\documentclass[aps,12pt]{revtex4}
\usepackage{dcolumn}
\usepackage{graphicx}
\usepackage{amsmath}
\usepackage{amsfonts}
\usepackage{amssymb}
\usepackage{psfrag}
\usepackage{wrapfig}
\usepackage{subfigure}
\usepackage{makeidx}
\usepackage{bm}
\usepackage{epsf}
\usepackage{float}
\usepackage{color}

\usepackage{cases}
\usepackage{booktabs} 
\usepackage{makecell} 
\usepackage{multirow}
\usepackage{bm}
\usepackage{enumitem}
\usepackage{hyperref}

\makeatletter

\newcommand{\Rmnum}[1]{\uppercase\expandafter{\romannumeral #1}}
\usepackage{amsmath}

\begin{document} 

\vspace{12mm}

\title{Curvature-induced scalarization of charged AdS black holes}

\author{Chao-Ming Zhang}
\email{chaomingzhang70@gmail.com}
\affiliation{Department of Physics, Nanchang University, Nanchang, 330031, China}

\author{Yun Soo Myung}
\email{ysmyung@inje.ac.kr}
\affiliation{Center for Quantum Spacetime, Sogang University, Seoul 04107, Republic of Korea}

\author{Lina Zhang}
\email{linazhang@hnit.edu.cn}
\affiliation{College of Science, Hunan Institute of Technology, Hengyang 421002, China}

\author{De-Cheng Zou}
\email{dczou@jxnu.edu.cn}
\affiliation{College of Physics and Communication Electronics, Jiangxi Normal University, Nanchang 330022, China}

\author{Fu-Wen Shu}
\email{shufuwen@ncu.edu.cn}
\affiliation{Department of Physics, Nanchang University, Nanchang, 330031, China}

\date{\today}

\begin{abstract}
We investigate how a negative cosmological constant affects the Gauss-Bonnet (GB) scalarization in the Einstein-Maxwell-scalar-Gauss-Bonnet theory with  a scalar coupling constant $\eta$ to GB term. We focus on the instability of Reissner-Nordstr\"om-AdS (RN-AdS) black holes under a scalar perturbation governed by an effective mass $\mu^2_{\text{eff}}$ sourced by  the GB term.  Unlike the asymptotically flat spacetime case, the onset of scalarization  is not merely determined by $\mu^2_{\text{eff}} < 0$, but it is constrained by the Breitenlohner-Freedman (BF) bound. In  case that the BF bound is violated ($\eta>2.25$ with $\Lambda=-0.5$), one  may find AdS-tachyonic instability.
We find that for $0<\eta<2.25$, the GB$^+$ scalarization may be performed  through spontaneous scalarization, while for $\eta<0$ the GB$^-$ scalarization is found to give the single branch of scalarized AdS  black holes.
For the GB$^+$ scalarization in $\eta_{th}\le\eta<2.25$ with  $\eta_{th}$ threshold instability,  we obtain the single  branch ($n=0$ fundamental branch) of scalarized  AdS black holes, in contrast to the infinite branches in asymptotically flat spacetime.  
A bulk fixed-charge thermodynamic analysis is performed thoroughly for GB$^\pm$ scalarizations.  

\end{abstract}


\maketitle

\section{Introduction}

General Relativity (GR) has achieved remarkable success in weak-field regimes, yet unresolved theoretical inconsistencies and cosmological anomalies—such as dark matter and the accelerated cosmic expansion—strongly suggest it operates merely as an effective low-energy limit of a more fundamental theory. To address these fundamental puzzles, a broad range of modified gravity models has been developed, introducing new degrees of freedom or short-distance modifications that naturally reduce to GR locally while offering richer dynamics at extreme scales \cite{Clifton:2011jh,Sotiriou:2011dz,Joyce:2014kja,Berti:2015itd,Heisenberg:2018vsk,Barack:2018yly}. Concurrently, the recent dawn of multi-messenger astronomy has provided direct empirical access to the strong-gravity regime. Landmark detections of compact binary mergers by the LIGO-Virgo collaboration have revealed highly nonlinear spacetime dynamics \cite{LIGOScientific:2016aoc}, while complementary observations by the Event Horizon Telescope (EHT) have established stringent constraints on horizon-scale geometry through the shadow profiles of supermassive black holes \cite{EventHorizonTelescope:2019dse,EventHorizonTelescope:2022wkp}. By comparing observed waveform models and shadow properties against theoretical expectations, these high-precision data streams offer unprecedented channels to rigorously test the Kerr hypothesis and distinguish viable alternative theories from standard GR in previously inaccessible regimes \cite{Troja:2017nqp,LIGOScientific:2017bnn,LIGOScientific:2017ycc,LIGOScientific:2017vwq}.

Among the various theoretical avenues explored to probe such deviations, the search for ``hairy'' compact objects stands out as a prominent observational test. In standard GR, classical no-hair theorems dictate that stationary black holes are remarkably simple, being uniquely characterized by their mass, spin, and electric charge \cite{Carter:1971zc,Ruffini:1971bza,Bekenstein:1995un}. To evade these stringent restrictions, one of the most natural and extensively studied extensions of Einstein's gravity is the framework of scalar-tensor theories, which introduces an additional dynamical scalar degree of freedom alongside the spacetime metric \cite{Brans:1961sx,Horndeski:1974wa,Kobayashi:2011nu,Deffayet:2013lga}. While minimally coupled scalar fields are strictly forbidden from forming stable hair around black holes, introducing a non-minimal coupling between the scalar field and spacetime curvature, or other source fields, provides a robust physical mechanism to circumvent these no-hair theorems \cite{Sotiriou:2014pfa,Babichev:2017guv}.

Within such non-minimally coupled scalar-tensor theories, a prominent phenomenon is spontaneous scalarization~\cite{Damour:1993hw}. Originally proposed for neutron stars, this mechanism extends to black holes as a robust pathway to evade classical no-hair theorems. By coupling the scalar field to source terms that become negative near the horizon, the strong-field regime triggers a tachyonic instability that destabilizes the bald background and induces scalar hair. Following the seminal demonstrations of curvature-induced scalarization in asymptotically flat Einstein-scalar-Gauss-Bonnet (EsGB) gravity---driven by the Gauss-Bonnet invariant $\mathcal{G}$~\cite{Doneva:2017bvd,Silva:2017uqg,Antoniou:2017acq,Herdeiro:2021vjo}---this paradigm has experienced a rapid theoretical expansion (for a comprehensive review, see Ref.~\cite{Doneva:2022ewd}). More recently, fully nonlinear scalarization, its interplay with spontaneous scalarization, and the effects of scalar-field mass and self-interactions have also been explored~\cite{Doneva:2021tvn,Pombo:2023lxg,Belkhadria:2023ooc}, together with rotationally driven instabilities in Kerr backgrounds~\cite{Dima:2020yac,Herdeiro:2020wei}.

Furthermore, the tachyonic trigger is not restricted to purely geometric terms; it can also be catalyzed by matter invariants. This has been extensively explored in Einstein-Maxwell-Scalar (EMS) models triggered by the electromagnetic invariant $F^2$~\cite{Herdeiro:2018wub,Zhang:2021nnn,Liu:2022fxy}, or through non-minimal couplings to the Ricci scalar~\cite{Herdeiro:2019yjy}. Beyond standard Maxwell fields, this mechanism extends to regular Bardeen~\cite{Zhang:2024bfu} and nonlinear Einstein-Euler-Heisenberg~\cite{Zhang:2025msi} black holes, yielding radially stable fundamental branches. Furthermore, Belkhadria and Mignemi~\cite{Belkhadria:2025lev} unified these approaches within a generalized flat-spacetime framework incorporating both couplings.

Scalarization has  been generalized to black holes with non-vanishing cosmological constants~\cite{Bakopoulos:2018nui,Guo:2020sdu,Brihaye:2019gla,Brihaye:2019dck,Guo:2021zed}. In this AdS context, Ref.~\cite{Guo:2020sdu} investigated the holographic scalarization of planar Schwarzschild-AdS and RN-AdS black holes and interpreted the  scalar hair as symmetry-preserving condensation in the boundary theory. Ref.~\cite{Brihaye:2019dck} studied spherically symmetric asymptotically AdS black holes with the coupling $\phi^2(\alpha R+\gamma\mathcal{G})$. In that work, the authors derived the BF-bound restriction and showed that scalarization can occur for both positive and negative couplings. They also analyzed the holographic phase transition and showed the absence of  extremal scalarized black holes. Furthermore, Ref.~\cite{Guo:2021zed} has studied charged AdS scalarization with a nonminimal scalar-Maxwell coupling and found a rich canonical phase structure.

In this paper, we wish to study scalarized RN-AdS black holes with mass $M$, charge $Q$, and cosmological constant $\Lambda$ in the pure GB sector $\mathcal{G}$. The scalar field in AdS is fundamentally distinct due to the confining nature of the AdS spacetime. As demonstrated in previous scalarization studies for AdS black holes~\cite{Bakopoulos:2018nui}, the onset of instability is no longer determined simply by a negative effective mass squared ($\mu_{\text{eff}}^2 < 0$). Instead, if the scalar field violates the local BF bound ($\mu_{\text{eff}}^2 < m_{\text{BF}}^2$) in the asymptotic region~\cite{Breitenlohner:1982jf}, it leads to AdS-tachyonic instability.

Compared with previous studies for  AdS scalarization, we focus on regular non-extremal scalarized RN-AdS black holes and analyze their thermodynamics thoroughly. To avoid AdS-tachyonic instability, we impose the condition $\eta<2.25$ for $\Lambda=-0.5$. Then, two scalarization channels appear: GB$^+$ for $0<\eta<2.25$ and GB$^-$ for $\eta<0$. Here, we find that GB$^-$ scalarization is not allowed for the $\Lambda$-$\eta$ branch. By solving the fully nonlinear field equations for $\eta_{\rm th}\leq \eta<2.25$  with $\eta_{\rm th}$ the threshold coupling of instability, we obtain the single branch of scalarized AdS black holes. For $\eta<0$, we also find a single branch of scalarized AdS black holes through GB$^-$ scalarization.

The remainder of this paper is organized as follows. In Section~\ref{sec:framework}, we outline the theoretical framework, including the action and the  field equations. Section~\ref{sec:probe_limit} discusses the  stability analysis in the probe limit, analyzing  onset  scalarization. We present our numerical results  known as GB$^+$ and GB$^-$ scalarizations by solving full equations in Section~\ref{sec:results}. 
Section~\ref{sec:TP} is devoted to analyzing thermodynamics and phase transitions thoroughly. 
Finally, we  conclude in Section~\ref{sec:conclusion}.

\section{Theoretical Framework}
\label{sec:framework}

We consider the Einstein-Maxwell-scalar-Gauss-Bonnet (EMsGB) theory with a negative cosmological constant $\Lambda$. The action is given by
\begin{equation}
    S = \frac{1}{16\pi} \int d^4x \sqrt{-g} \left[ R - 2\Lambda - 2\nabla_\mu \phi \nabla^\mu \phi + f(\phi)\mathcal{G} - F_{\mu\nu}F^{\mu\nu} \right],
    \label{eq:action}
\end{equation}
this action includes the pure GB sector of the model studied in Ref.~\cite{Brihaye:2019dck} with $\Lambda = -3/L^2$. In the scalar perturbation equation, this normalization amounts to matching $9+32\gamma\Lambda$ in~\cite{Brihaye:2019dck} to $9+8\eta\Lambda$ with  the present model for a coupling function $f(\phi)=\eta \phi^2/2$.
The GB term $\mathcal{G}$ is defined as
\begin{equation}
    \mathcal{G} = R^2 - 4R_{\mu\nu}R^{\mu\nu} + R_{\mu\nu\alpha\beta}R^{\mu\nu\alpha\beta}.
\end{equation}
Here, $\phi$ denotes  a scalar field and $F_{\mu\nu} = \partial_\mu A_\nu - \partial_\nu A_\mu$ represents   the Maxwell  field strength.

Varying the action \eqref{eq:action} with respect to the metric tensor $g_{\mu\nu}$ yields the Einstein equation
\begin{equation}
    G_{\mu\nu} + \Lambda g_{\mu\nu} = T_{\mu\nu}^{\phi} + T_{\mu\nu}^{M},
    \label{eq:einstein_eq1}
\end{equation}
where the  energy-momentum tensor for scalar $T_{\mu\nu}^{(\phi)}$ is expressed by
\begin{equation}
    T_{\mu\nu}^{\phi} = 2 ( \nabla_\mu \phi \nabla_\nu \phi - \frac{1}{2} g_{\mu\nu} \nabla_\alpha \phi \nabla^\alpha \phi) - 4 P_{\mu\alpha\nu\beta} \nabla^\beta \nabla^\alpha f(\phi),
    \label{eq:einstein_eq2}
\end{equation}
with  $P_{\mu\alpha\nu\beta}$ tensor defined by
\begin{equation}
    P_{\mu\alpha\nu\beta} = R_{\mu\alpha\nu\beta} + g_{\mu\beta}R_{\alpha\nu} - g_{\mu\nu}R_{\alpha\beta} + g_{\alpha\nu}R_{\mu\beta} - g_{\alpha\beta}R_{\mu\nu} + \frac{1}{2}\left(g_{\mu\nu}g_{\alpha\beta} - g_{\mu\beta}g_{\alpha\nu}\right)R.
\end{equation}
Here, the energy-momentum tensor for Maxwell term is given by
\begin{equation}
    T_{\mu\nu}^{M} = 2 ( F_{\mu\alpha}F_\nu{}^\alpha - \frac{1}{4} g_{\mu\nu} F_{\alpha\beta}F^{\alpha\beta} ).
\end{equation}
The equation of motion for  $\phi$ takes the form
\begin{equation}
    \Box \phi + \frac{1}{4} \frac{df(\phi)}{d\phi} \mathcal{G}= 0,
    \label{eq:scalar_eq}
\end{equation}
where $\Box = \nabla_\mu \nabla^\mu$ is the d'Alembertian.
Finally, the  Maxwell equation is given simply  by
\begin{equation}
    \nabla_\mu \left( F^{\mu\nu} \right) = 0.
    \label{eq:maxwell_eq}
\end{equation}

\section{Scalarization in the Probe Limit}
\label{sec:probe_limit}

To elucidate the physical mechanism triggering onset scalarization, we first examine the system in the probe limit. By treating the scalar field as a test perturbation on a fixed black hole  background, we decouple its dynamics from the metric backreaction. This linearized approximation provides a tractable framework to examine the threshold of instability and to understand how the scalar effective mass is modified by the curvature and cosmological constant. Specifically, we investigate the stability of the RN-AdS black hole against scalar perturbation driven by the interplay of a non-minimal coupling to the GB term.

\subsection{Background geometry and analysis of GB term}

We  may define the background spacetime using a spherically symmetric metric ansatz
\begin{equation}
    ds^2 = -A(r) dt^2 + \frac{dr^2}{B(r)}  + r^2 \left(d\theta^2 + \sin^2\theta d\varphi^2\right).
    \label{eq:metric_ansatz}
\end{equation}
For a specific case of RN-AdS black hole, the metric functions $A(r)$ and $B(r)$ coincide and are given by
\begin{equation}
    A(r) = B(r) = 1 - \frac{2M}{r} + \frac{Q^2}{r^2} - \frac{\Lambda}{3} r^2,
\end{equation}
where $M$ is the ADM mass, $Q$ is the electric charge, and $\Lambda$ is the negative cosmological constant ($\Lambda < 0$).  The background vector potential is  given by $A_t = -Q/r$, which yields the Maxwell term  $F_{\mu\nu}F^{\mu\nu} = -2Q^2/r^4$.  We find four roots from $A(r)=0$. One is the outer horizon $r_+(M,Q,\Lambda)$ and the other is the inner horizon $r_-(M,Q,\Lambda)$ with two more complex solutions. Even though their forms are complicated, one can always access their explicit forms.

\begin{figure}[H]
\centering
\subfigure[~$\Lambda-r$ ]
{\label{Lr} 
\includegraphics[width=3.1in]{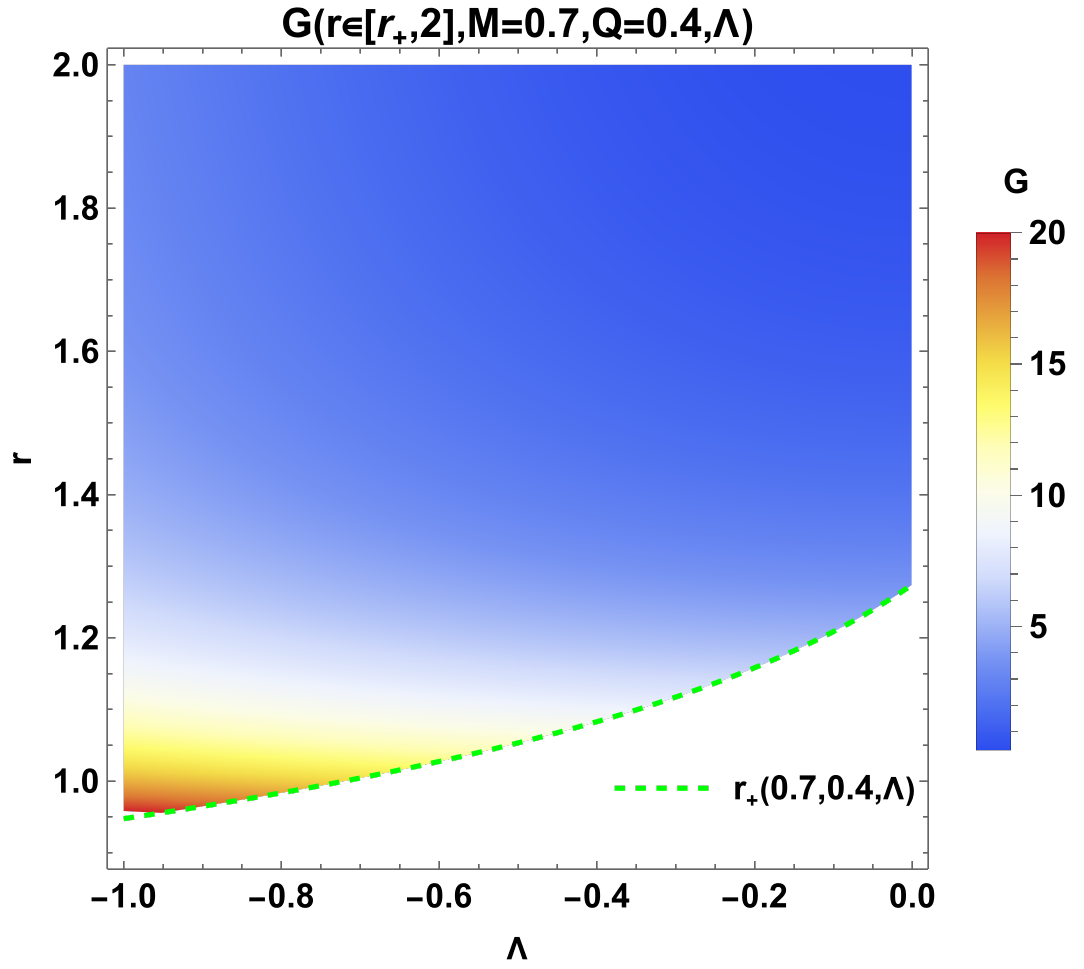}}
\hfill
\subfigure[~$Q-M$]
{\label{QM} 
\includegraphics[width=2.6in]{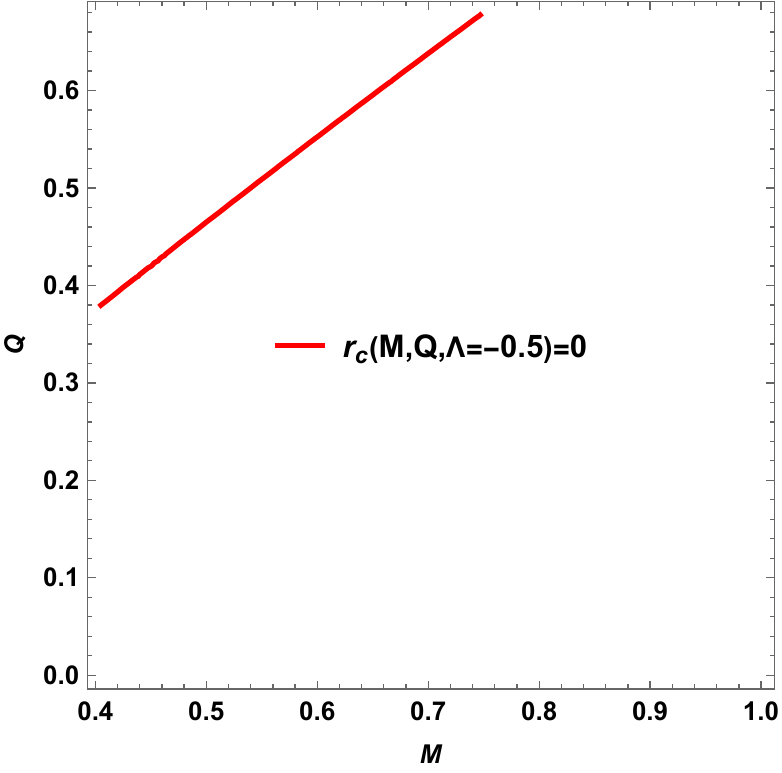}}
\caption{Profiles of GB term in the near-horizon and critical onset curve. (a) $\mathcal{G}(r\in[r_+,2.0],M=0.7,Q=0.4,\Lambda)$. One finds that $\mathcal{G}>0$  for any $\Lambda$. (b) 
Critical onset curve $r_c(M,Q,\Lambda=-0.5)=0$ is a set of red dots. }\label{fig:GB2}
\end{figure}

The instability mechanism is sourced by the GB term. For  the RN-AdS black hole, it  takes the explicit form
\begin{equation}
    \mathcal{G} = \frac{48M^2}{r^6} - \frac{96MQ^2}{r^7} + \frac{40Q^4}{r^8} + \frac{8}{3}\Lambda^2,
\end{equation}
whose $Q=0$ limit leads to the GB term for Schwarzschild-AdS black hole. In this case, its GB$^+$ scalarization was discussed in~\cite{Zou:2022hjj}.
It is worth noting that the cosmological constant contributes a positive constant background shift $8\Lambda^2/3$ to the curvature term of RN black hole. This term is crucial as it modifies the asymptotic behavior of the effective mass of scalar compared to the asymptotically flat case, affecting the scalar's fall-off behavior at infinity.
\begin{figure}[H]
\centering
\subfigure[~$M-r$ ]
{\label{Mr} 
\includegraphics[width=3.1in]{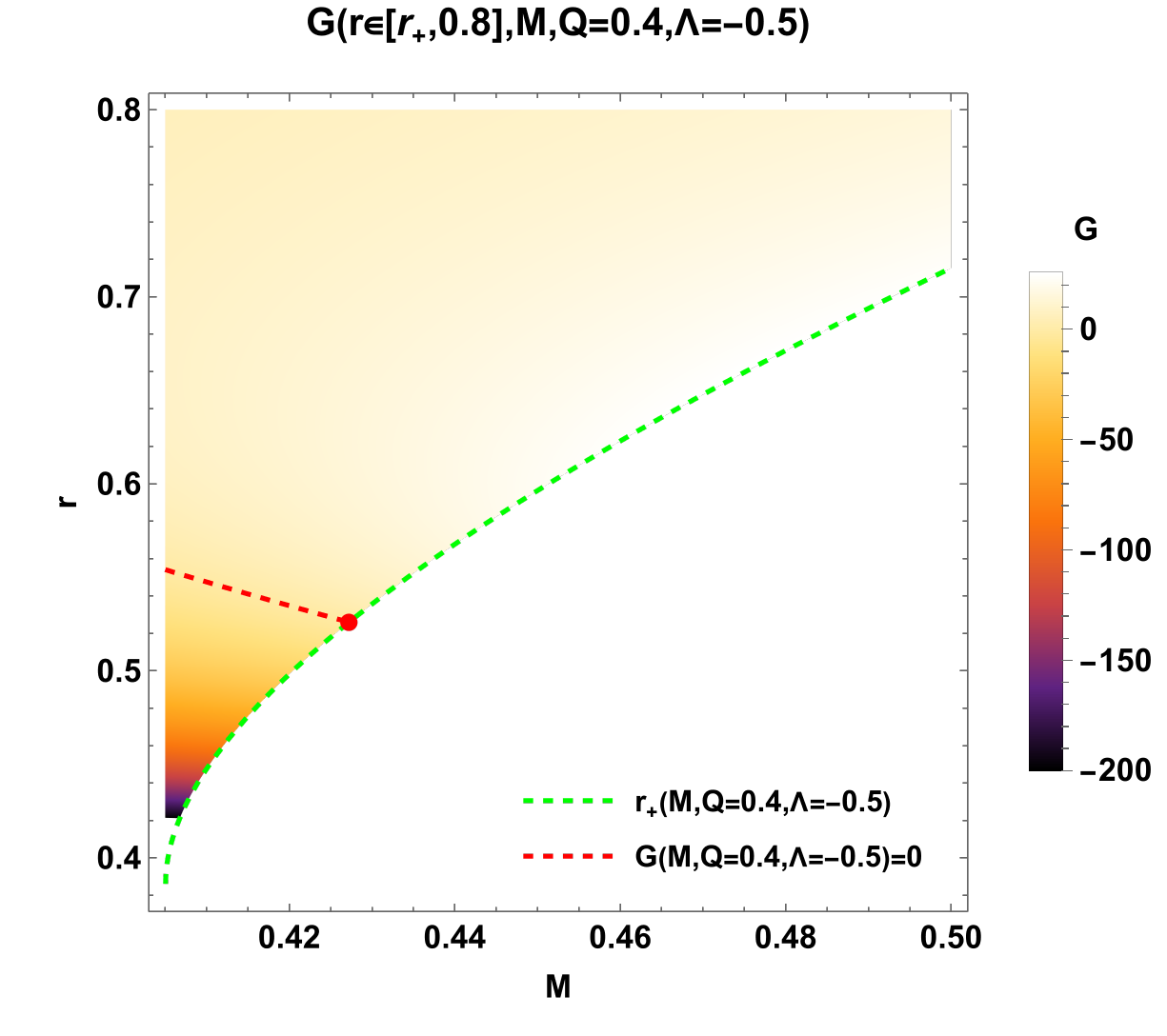}}
\hfill
\subfigure[~$Q-r$]
{\label{Qr} 
\includegraphics[width=3.1in]{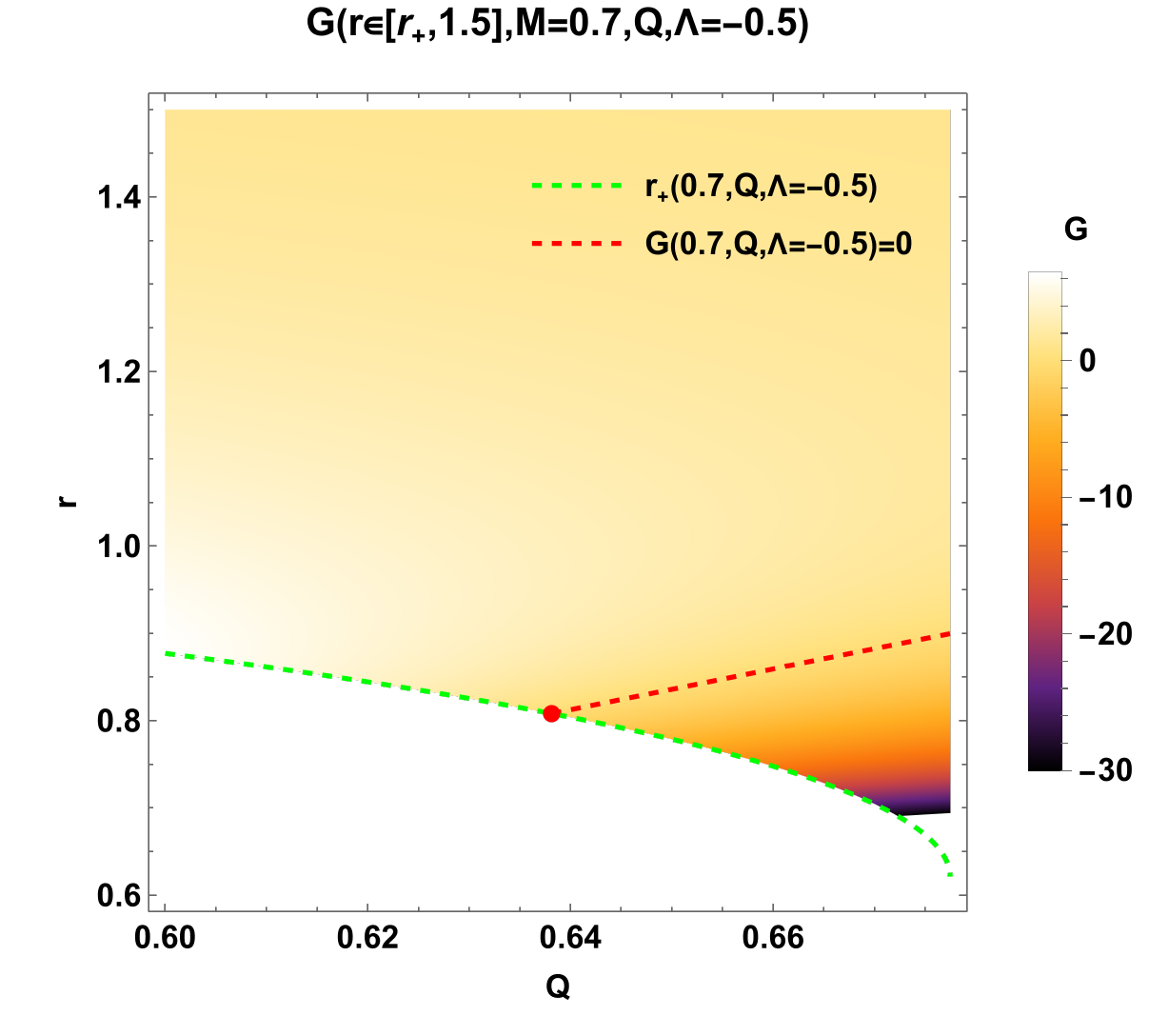}}
\caption{Profiles of GB term and outer horizon $r_+$ in the near-horizon. (a) $\mathcal{G}(r\in[r_+,0.8],M,Q=0.4,\Lambda=-0.5)$. For  $M_{\rm min}(=0.405)<M<M_c(=0.4272$, red dot), $\mathcal{G}<0$, while  one finds $\mathcal{G}>0$ for $M>M_c$. (b)  $\mathcal{G}(r\in[r_+,1.5],M=0.7,Q,\Lambda=-0.5)$. For  $0<Q<Q_c(=0.6381$, red dot), $\mathcal{G}>0$, while  one finds $\mathcal{G}<0$ for $Q_c<Q<Q_e(=0.6775)$.  }\label{fig:GB1}
\end{figure}
To systematically determine the geometric prerequisites for triggering the tachyonic instability, we display the profiles of the GB term $\mathcal{G}(M,Q,\Lambda)$ and the outer horizon $r_+(M,Q,\Lambda)$ in the near-horizon region in Figs.~\ref{fig:GB1} and \ref{fig:GB2}. The sign of $\mathcal{G}$ at the horizon is of paramount importance, as it critically determines whether a tachyonic instability can be triggered in the near-horizon region.

In Fig.~\ref{Mr}, we observe a sign reversal of $\mathcal{G}$ at the critical mass $M=M_c=0.4272$. Specifically, for $M_{\rm min}(=0.405) < M < M_c$, the horizon value of $\mathcal{G}$ becomes negative. This provides the essential geometric condition for the GB$^-$ scalarization. Conversely, Fig.~\ref{Qr} reveals that the charge $Q$ exerts an opposite regulatory effect. The GB term turns negative when the charge exceeds a critical value $Q=Q_c=0.6381$, meaning that the GB$^-$ scalarization is available within the parameter window $Q_c < Q < Q_e(=0.677)$. Thus, $M_c$ serves as an upper mass bound for GB$^-$ scalarization, while $Q_c$ acts as a lower charge bound for triggering the GB$^-$ scalarization.

Furthermore, Fig.~\ref{Lr} illustrates the effect of the cosmological constant $\Lambda$. For the chosen background configuration ($M=0.7$, $Q=0.4$), $\mathcal{G}$ remains strictly positive at the horizon across the varied range of $\Lambda$. This indicates that the sign of the near-horizon GB term is highly robust against $\Lambda$. Consequently, it is important to note that the GB$^-$ scalarization is forbidden for this specific $(M, Q)$ setup regardless of any magnitude of $\Lambda$. 

Finally, Fig.~\ref{QM} synthesizes these critical onset points into a unified phase boundary curve in the $Q-M$ parameter space. The red critical curve distinctly delineates the geometric parameter domains for scalarization: the parameter region above and to the left of the curve corresponds to $\mathcal{G} < 0$ , whereas the region below and to the right corresponds to $\mathcal{G} > 0$ .

\subsection{Linearized theory and effective mass term}

We study the evolution of a scalar perturbation $\delta\phi$ around the RN-AdS black hole. Considering the quadratic coupling function 
\begin{equation}
 f(\phi) =\frac{\eta}{2}\phi^2,   \label{coup-f}
\end{equation}
 the linearized scalar equation reads as
\begin{equation}
    \left( \Box - \mu_{\text{eff}}^2(r) \right) \delta\phi = 0
    \label{eq:linearized_kg}
\end{equation}
with  the effective mass term 
\begin{equation}
    \mu_{\text{eff}}^2(r) = -\frac{1}{4} \frac{d^2f}{d\phi^2}\bigg|_0 \mathcal{G}= -\frac{\eta}{4} \mathcal{G}.
    \label{eq:eff_mass}
\end{equation}
Here, $\eta$ characterizes the strength of the scalar-curvature interaction. In the context of AdS spacetime, the instability condition possesses twofold.  A negative squared mass term in the near-horizon  may imply a tachyonic instability. As well, the scalar field becomes unstable if $\mu_{\text{eff}}^2(r)$ violates the Breitenlohner-Freedman (BF) bound when $\mu_{\text{eff}}^2 < m_{\text{BF}}^2 = \frac{3}{4}\Lambda$  in the far-region. It is called as AdS-tachyonic  instability.
Here, we consider the former case of near-horizon instability  mainly  because we are interested in carrying out  GB$^\pm$ scalarizations. 

\subsection{Schr\"odinger formulation and  Potential analysis}

To perform onset scalarization, we decompose the scalar perturbation into spherical harmonics $Y_{lm}(\theta, \varphi)$ with  the radial dependence $u_{lm}(r)$
\begin{equation}
    \delta\phi(t, r, \theta, \varphi) = \sum_{l,m} \frac{u_{lm}(r)}{r} Y_{lm}(\theta, \varphi) e^{-i\omega t}.
\end{equation}
By introducing the tortoise coordinate $r_*$ defined by $dr_* = dr/A(r)$, the radial part of the scalar equation can be recast into a Schrödinger-like form:
\begin{equation}
    \frac{d^2 u_{lm}}{dr_*^2} + \Big[\omega^2 - V_{\text{eff}}(r) \Big] u_{lm} = 0
    \label{eq:schrodinger}
\end{equation}
where the effective potential $V_{\text{eff}}(r)$ is given by
\begin{equation}
    V_{\text{eff}}(r) = A(r) \Big[ \frac{l(l+1)}{r^2} + \frac{2M}{r^3}-\frac{2Q^2}{r^4}-\frac{2\Lambda}{3} + \mu_{\text{eff}}^2(r) \Big].
    \label{eq:potential}
\end{equation}
From now on, we consider the $s(l=0)$-mode of the scalar perturbation.
For $\eta>0$ and  $u_{00}(t,r_*)\sim u(r_*)e^{-i\omega t}$, GB$^+$ scalarization of Schwarzschild-AdS black hole was found in Ref.~\cite{Bakopoulos:2018nui}.
For $\eta<0$, one expects to find  GB$^-$ scalarization.
As is shown in Fig.~\ref{fig:GB2}, the sign of $\mu_{\rm eff}^2(r)=-\eta \mathcal{G}/4$ changes at $M=M_c$ and $Q=Q_c$. Considering GB$^-$ scalarization, we choose negative $\eta$, leading to $\mu_{\rm eff}^2(r)<0$ in the near-horizon. 

\begin{figure}[H]
\centering
\subfigure[~$V_{\rm eff}^a=\frac{2\Lambda^2}{9}(1+\eta\Lambda)$ ]
{\label{veff1-1} 
\includegraphics[width=2.5in]{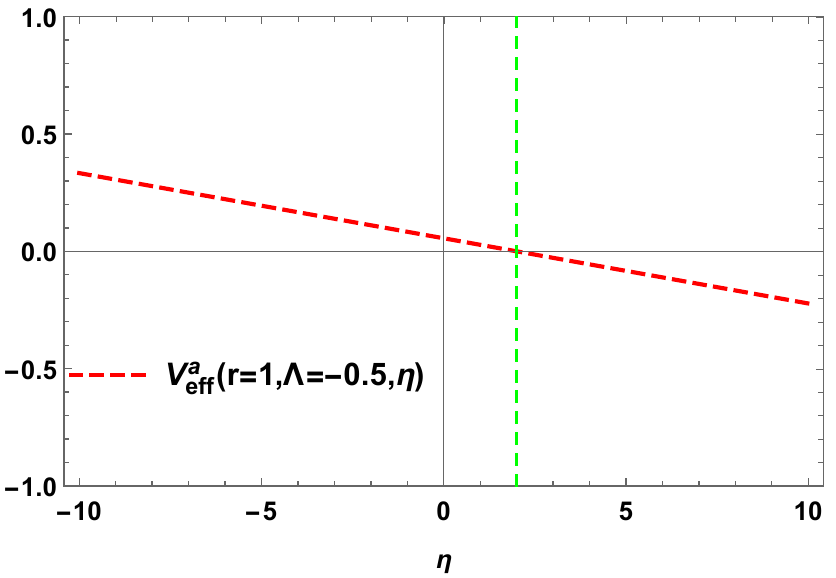}}
\hfill
\subfigure[~$\eta_b=-\frac{9}{8\Lambda}$]
{\label{veff1-2} 
\includegraphics[width=2.5in]{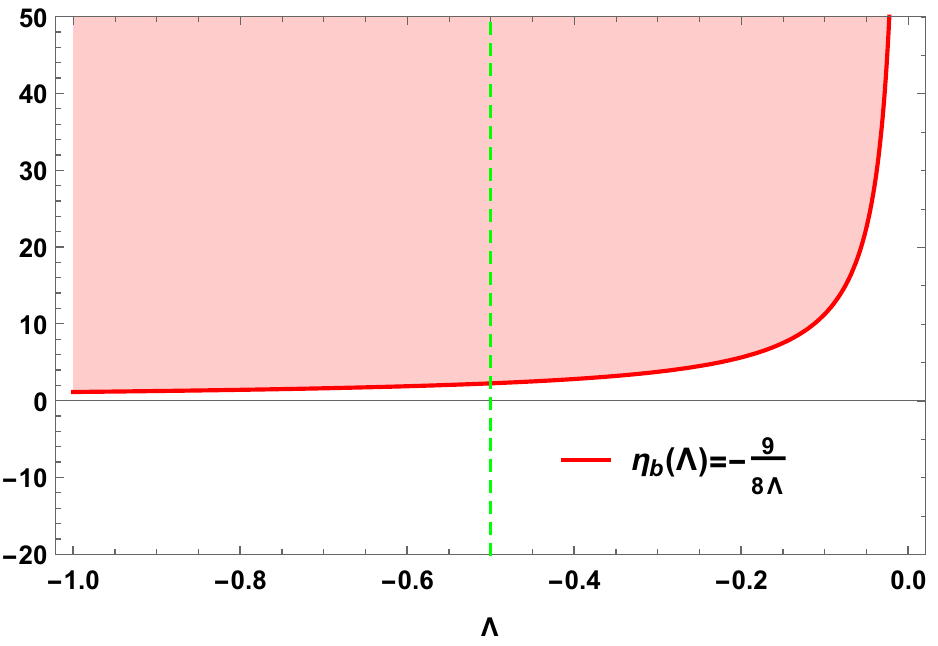}}
\caption{(a) Coefficient of asymptotic potential $V_{\rm eff}^a(r=1,\Lambda=-0.5,\eta)$. At $\eta=2$ (dashed line), the sign of $V^a_{\rm eff}$ changes.  (b) BF boundary $\eta_b(\Lambda)$. Shaded region respects  violation of the BF bound (AdS-tachyonic instability), like as $\eta>-\frac{9}{8\Lambda}$ and the dashed line is at $\Lambda=-0.5$.} \label{vef1}
\end{figure}

On the other hand, its asymptotic potential takes the form 
\begin{equation} \label{a-pot}
V_{\rm eff}^{a}=\frac{2\Lambda^2}{9}(1+\eta \Lambda)r^2. 
\end{equation}
Sign change of its coefficient  around $\eta=2$ in Fig.~\ref{veff1-1} shows the AdS-tachyonic instability.
Applying the violation of the BF boundary condition to here,
one finds an inequality of
\begin{equation} \label{bf-con}
\eta >-\frac{9}{8\Lambda},
\end{equation}
leading to $\eta>2.25$ for $\Lambda=-0.5$. Here, one finds  AdS-tachyonic instability.
This is the $\alpha=0$ version of the BF-bound restriction derived in~\cite{Brihaye:2019dck}. Taking into account the coupling normalization, the condition $\Delta\geq0$ becomes $9+8\eta\Lambda\geq0$, which provides a bound of  $\eta\leq2.25$ for $\Lambda=-0.5$.
So, we do not  worry about the violation of the BF bound when choosing $\eta<2.25$, being free from  AdS-tachyonic instability. 
Also, it is allowable to  focus on the near-horizon instability  when choosing  negative $\eta$ for GB$^-$ scalarization (see Fig.~\ref{veff1-2} ). 

Now, we wish to find the  critical onset parameter mass $M_c$ and charge $Q_c$, which determine the lower and upper bounds for the  onset of GB$^-$ scalarization  by making use of Hod's approach~\cite{Hod:2020jjy}.
To get the critical onset parameters, it is enough to consider the potential term: $V_{\rm eff}(r)u_{00}(t,r_*)=0$.
The onset of  scalarization  is defined by critical black hole which  denotes   the boundary between RN-AdS  and  scalarized AdS black holes  in the limit of $\eta \to -\infty$.
In this limit, it is characterized by  the presence of a degenerate  binding potential well  whose two turning points
merge at the outer horizon at $r=r_+(M,Q,\Lambda)$ as
\begin{eqnarray}
 \mu^2_{\rm eff}(r_+)u_{00}(t,r_*)=0. \label{crit-cond}
\end{eqnarray}
From Eq.(\ref{crit-cond}), we find the resonance condition as
\begin{equation} \label{c-cond}
r_c(M,Q,\Lambda)\equiv \frac{12M^2}{ r^6_+}-\frac{24 M Q^2}{r_+^7}+\frac{10Q^4}{ r^8_+} +\frac{2}{3}\Lambda^2=0.
\end{equation}
The critical onset mass and charge  are  determined by the resonance condition.
Solving Eq.(\ref{c-cond}) for $M_c$ with $Q=0.4$ and $\Lambda=-0.5$, and for $Q=Q_c$ with $M=0.7$ and $\Lambda=-0.5$ leads to
the critical onset  mass and charge   for  GB$^-$ scalarization as
\begin{eqnarray}
M_c=0.4272 \quad Q_c=0.6381.
\end{eqnarray}
 Fig.~\ref{QM}  shows  the critical onset curve $\{r_c(M,Q,\Lambda=-0.5)=0\}$ for any $M_c$ and $ Q_c$.
  We note that this curve  is regarded  as the same  in the critical-coupling curves found from Ref.~\cite{Brihaye:2019dck} expressed in the $(M,Q,\Lambda,\gamma)$ parametrization.

\begin{figure}[H]
\centering
\subfigure[~$V_{\rm eff}-\eta=3,2.25,2,1.5,1$ ]
{\label{veff3} 
\includegraphics[width=2.8in]{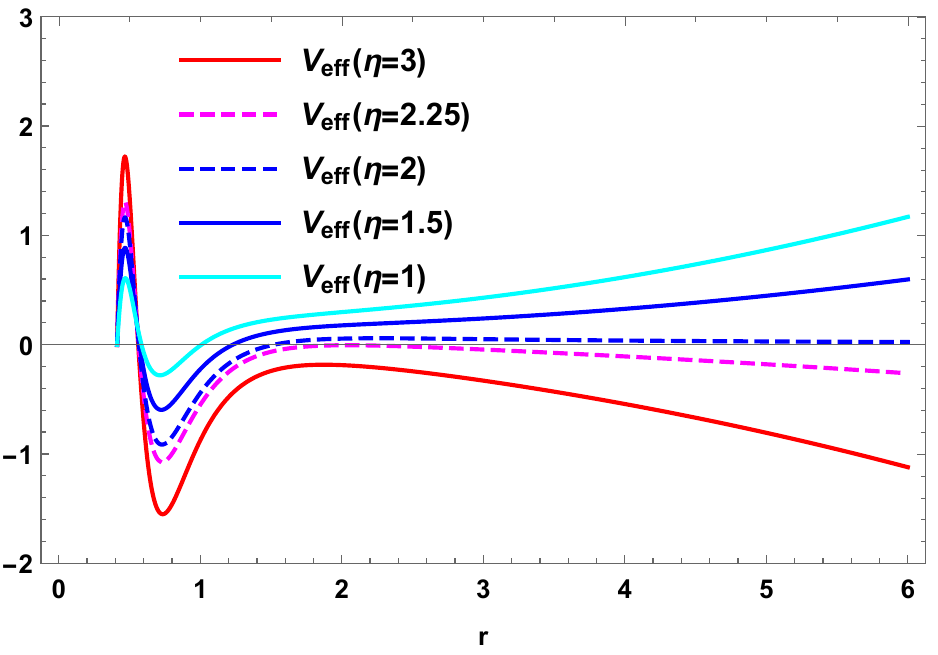}}
\hfill
\subfigure[~$V_{\rm eff}-\eta=-1,-2,-3,-4,-5$]
{\label{veff4} 
\includegraphics[width=2.8in]{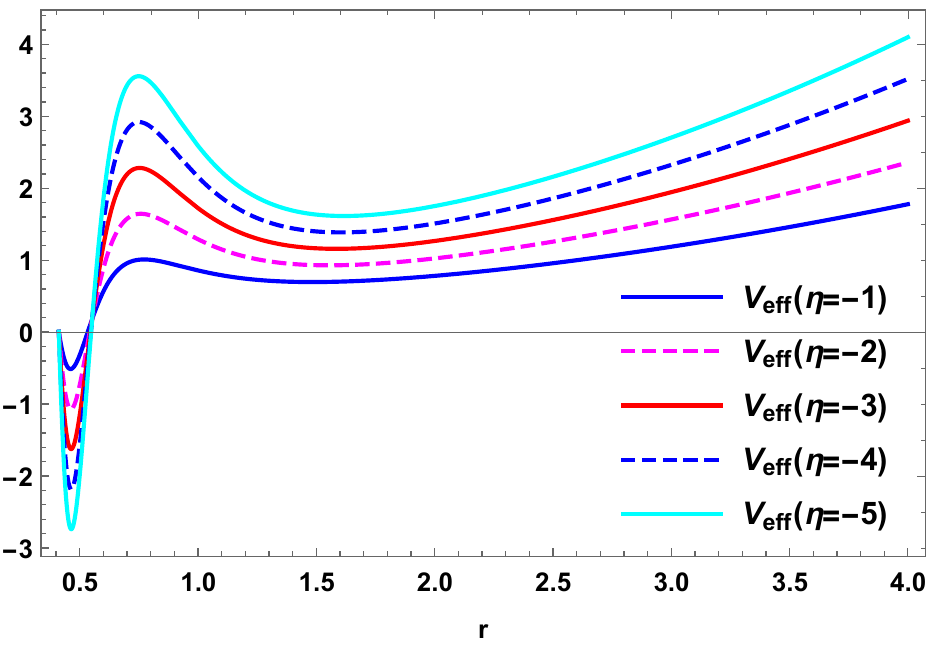}}
\caption{Effective potentials $V_{\rm eff}(r, M=0.406, Q=0.4, \Lambda=-0.5, \eta)$ as functions of the radial coordinate $r \ge r_+ = 0.4119$. It belongs to Fig.~\ref{Mr}.
(a) Potentials for   $\eta=3,2.25, 2,1.5,1$. For $\eta=2$, its asymptotic slop is zero. For $\eta>2.25$, it predicts  AdS-tachyonic instability.
(b) Potentials for  $\eta=-1,-2,-3,-4,-5$. In this case,  GB$^-$ scalarization is triggered by  negative potential in the near-horizon but there is no AdS-tachyonic instability.}\label{vef2-1}
\end{figure}

We are in a position to discuss how instability comes out from the shape of the effective potentials. As is shown in Fig.~\ref{vef2-1}  with $M-\eta$ branch (Fig.~\ref{Mr} ), it depends on the value $\eta$ for choosing $M=0.406,Q=0.4,\Lambda=-0.5$. For $\eta>2.25$, there may exist AdS-tachyonic instability. However, there is  GB$^+$ spontaneous scalarization for $\eta>0$  triggered by tachyon, giving finite branches of scalarized AdS black holes. 
For $\eta<0$, the negative region appears in the near-horizon, leading to tachyonic instability for GB$^-$ scalarization without AdS-tachyonic instability.

\begin{figure}[H]
\centering
\subfigure[~$V_{\rm eff}-\eta=3,2.25,2,1.5,1$ ]
{\label{veff2-2-1} 
\includegraphics[width=2.8in]{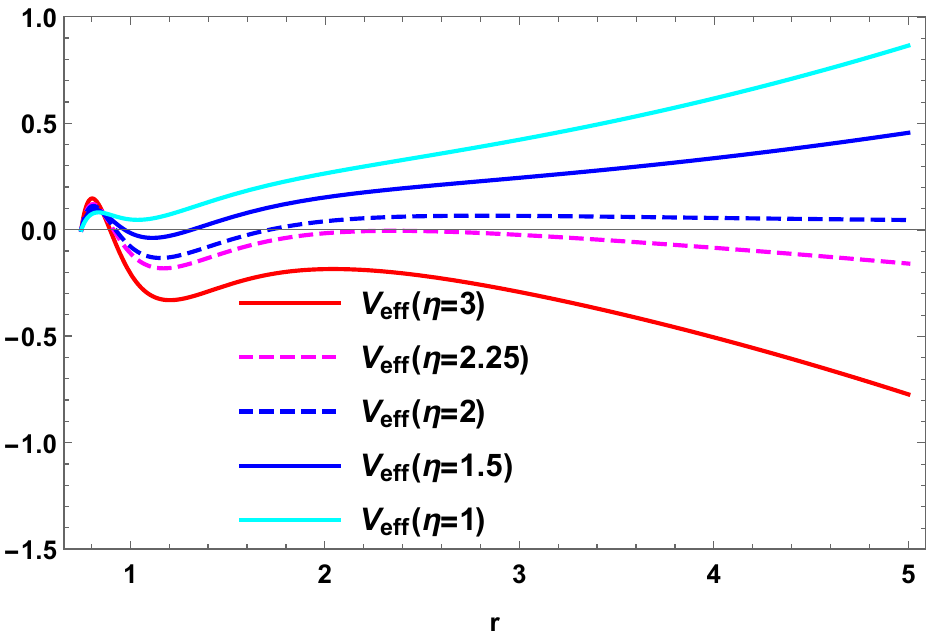}}
\subfigure[~$V_{\rm eff}-\eta=-1,-2,-3,-4,-5$]
{\label{veff2-2-2} 
\includegraphics[width=2.8in]{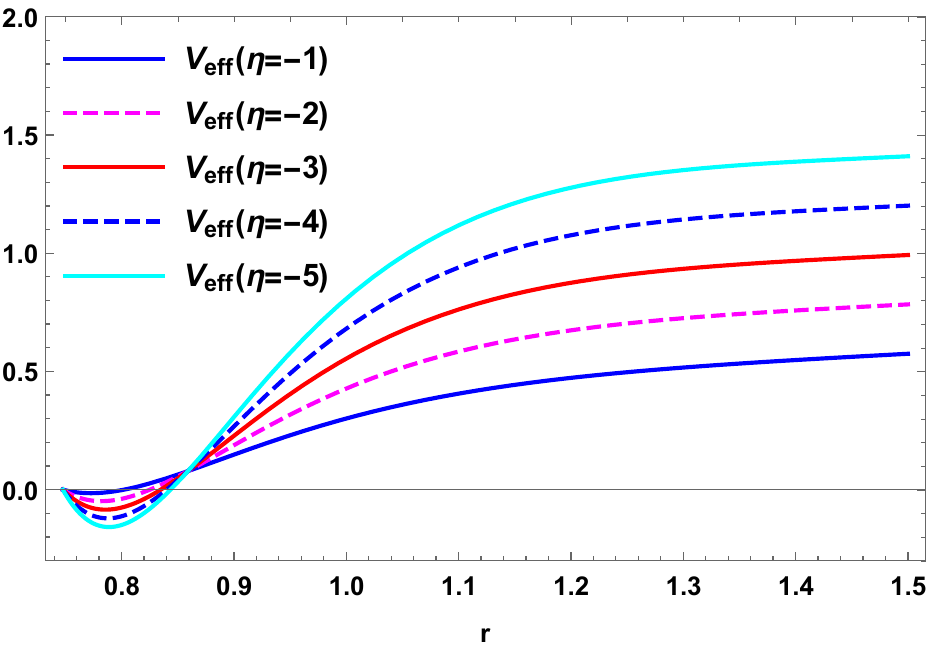}}
\caption{Effective potentials $V_{\rm eff}(r,M=0.7,Q=0.66,\Lambda=-0.5,\eta)$ as  functions of the radial coordinate $r \ge r_+ = 0.7478$.  It belongs to Fig.~\ref{Qr}.
(a) Potentials for   $\eta=3,2.25,2,1.5,1$. For $\eta=2$, its asymptotic slop is zero. For $\eta>2.25$, it predicts  AdS-tachyonic instability.
(b) Potentials for  $\eta=-1,-2,-3,-4,-5$. In this case,  GB$^-$ scalarization is triggered by  negative potential in the near-horizon, but there is no AdS-tachyonic instability.}\label{vef2-2}
\end{figure}

Showing in Fig.~\ref{vef2-2} with $Q-\eta$ branch (Fig.~\ref{Qr} ), it depends on the value $\eta$ for choosing $M=0.7,Q=0.66,\Lambda=-0.5$. For $\eta>2.25$, there may exist AdS-tachyonic instability.
One expects that  GB$^+$ scalarization appears with finite branches. 
For $\eta<0$, the negative region always appears in the near-horizon, leading to GB$^-$ scalarization. Here, there is no AdS-tachyonic instability. 

\begin{figure}[H]
\centering
\subfigure[~$V_{\rm eff}-\eta=3,2.25,2,1.5,1$ ]
{\label{veff2-3-1} 
\includegraphics[width=2.8in]{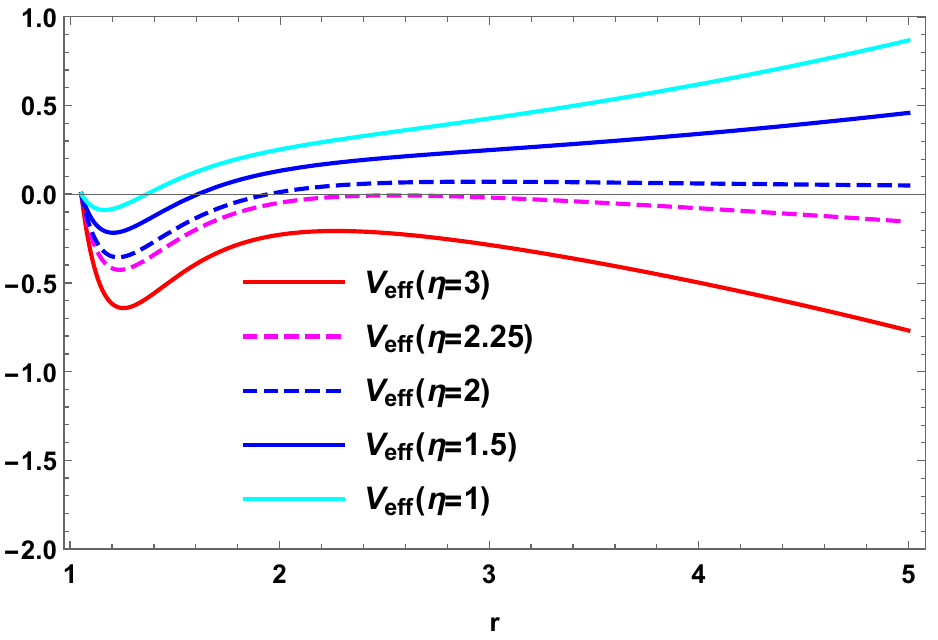}}
\hfill
\subfigure[~$V_{\rm eff}-\eta=-1,-2,-3,-4,-5$]
{\label{veff2-3-2} 
\includegraphics[width=2.8in]{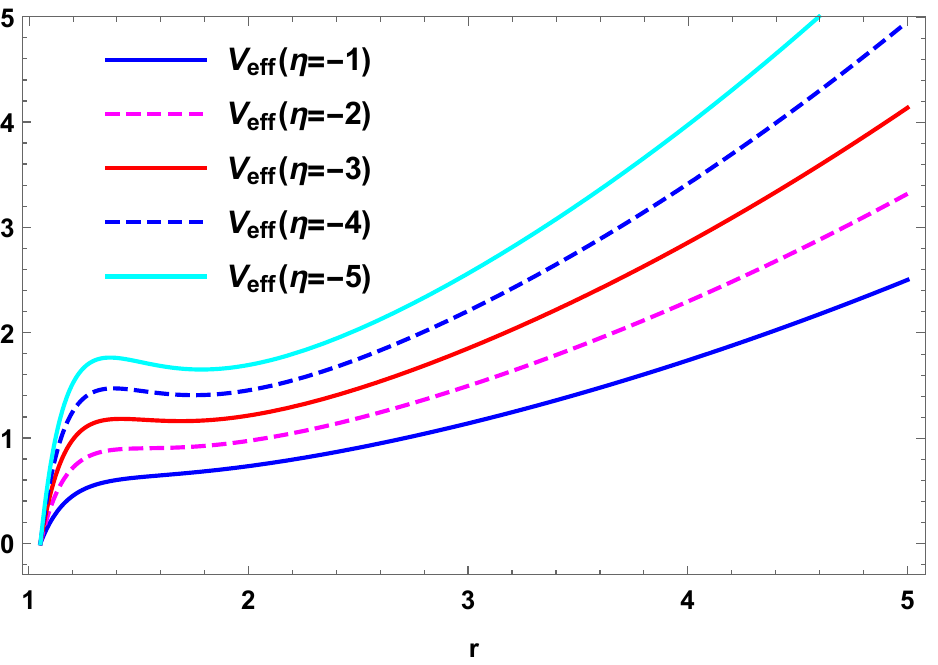}}
\caption{Effective potentials $V_{\rm eff}(r,M=0.7,Q=0.4,\Lambda=-0.5,\eta)$ as  functions of the radial coordinate $r \ge r_+ = 1.0533$.  It belongs to Fig.~\ref{Lr}.
(a) Potentials for   $\eta=3,2.25,2,1.5,1$. For $\eta=2$, its asymptotic slop is zero. For $\eta>2.25$, it predicts  AdS-tachyonic instability. In this case, we expect to have onset of  GB$^+$ scalarization. 
(b) Potentials for  $\eta=-1,-2,-3,-4,-5$. In this case,  GB$^-$ scalarization is not allowed because negative potential is absent in the near-horizon. Also, there is no AdS-tachyonic instability.}\label{vef2-3}
\end{figure}

Fig.~\ref{vef2-3} with $\Lambda-\eta$ branch Fig.~\ref{Lr} shows potentials  depending  on the value $\eta$ for choosing $M=0.7,Q=0.4,\Lambda=-0.5$. For $\eta>2.25$, there may exist AdS-tachyonic instability. GB$^+$ scalarization may occur for  $0<\eta<2.25$. 
For $\eta<0$, the negative region is not allowed  outside the horizon, leading to no GB$^-$ scalarization. Also, there is no AdS-tachyonic instability.


\section{Onset scalarization and single branch}
\par
To  determine the instability threshold quantitatively, we  impose appropriate boundary conditions. Near the AdS boundary ($r \to \infty$), the scalar field exhibits the asymptotic behavior as 
\begin{equation}
    \phi(r) \sim \phi_{1\infty} r^{-\Delta_-} + \phi_{2\infty} r^{-\Delta_+},
\label{eq-phi-inf}
\end{equation}
where the characteristic exponents are given by
\begin{equation}
    \Delta_{\pm} = \frac{1}{2} \pm \frac{1}{2}\sqrt{9 + 8\eta\Lambda}.
\label{eq-delta}
\end{equation}
\begin{figure}[H]
\centering
\includegraphics[width=0.6\textwidth]{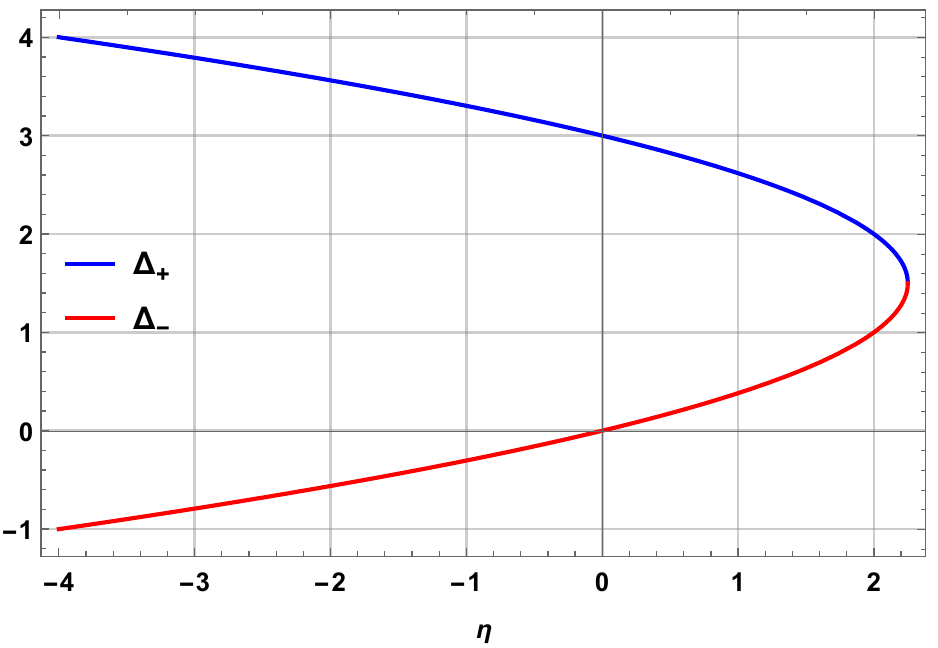} 
\caption{Characteristic exponents $\Delta_\pm$ with $\Lambda=-0.5$ as functions of coupling constant $\eta$. They are allowed for $0<\eta<2.25$ and $\eta<0$. For $\eta=2$, one finds that $\Delta_+=2,~\Delta_-=1$.}\label{figdel}
\end{figure}

From the positivity condition for square-root in Eq.(\ref{eq-delta}), one finds the violation of  BF boundary condition ($\eta>2.25$ for $\Lambda=-0.5$)  in Eq.(\ref{bf-con}).  As is shown in Fig.~\ref{figdel}, two allowed regions for $\Delta_\pm$ are $0<\eta<2.25$ and $\eta<0$. For the latter case, one observes that $\Delta_-<0$. However, we could not define $\Delta_\pm$ for $\eta>2.25$. 
According to the AdS/CFT correspondence, the coefficient $\phi_{1\infty}$, associated with the slower fall-off ($r^{-\Delta_-}$), represents the non-normalizable mode and acts as a source for the dual scalar operator on the conformal boundary.  Conversely, the coefficient $\phi_{2\infty}$, associated with the faster fall-off ($r^{-\Delta_+}$), corresponds to the normalizable mode and yields the expectation value  of the dual operator.  Since (spontaneous) scalarization describes a dynamical phase transition without external fields, the scalar field is usually required to be unsourced. 
Here, we may set the source term to be zero ($\phi_{1\infty} = 0$), ensuring that the bound state solution is purely normalizable at asymptotic infinity.
We note that this scalar asymptotic structure Eq.(\ref{eq-phi-inf}) is quite different from  asymptotically flat case of $\Lambda=0$,  where it takes the form of $\phi(r\to\infty)=\frac{Q_s}{r}+\cdots$ with scalar charge $Q_s$. This corresponds to a primary scalar hair. Hence,  all scalars  belong to secondary hair in AdS spacetime. 

However, if one considers the case of $\eta=2$ whose exponents are given by $\Delta_+=2$ and $\Delta_-=1$, we may  obtain two scalarized AdS black holes~\cite{Guo:2025flg}.

\par
The onset of scalarization implies  the existence of  static bound states. To find it, one  obtains  solutions that are regular at the horizon and normalizable at asymptotic infinity. 
Also,  we determine the bifurcation points ($\{\eta_n\}, n=0,1,\cdots$) in the parameter space $(\eta, Q, M)$ by numerically solving Eq. \eqref{eq:linearized_kg} for $\omega=0$ subject to these boundary conditions. In the fundamental ($n=0$) branch, it is important to  note that $\eta_{0}$ is identified  with threshold value $\eta_{th}$ ($\eta_0=\eta_{th}$). 

\begin{figure}[H]
\centering
\subfigure[~$M$ vs $\eta_{th}$]
{\label{fig-clouds1-1} 
\includegraphics[width=2.05in]{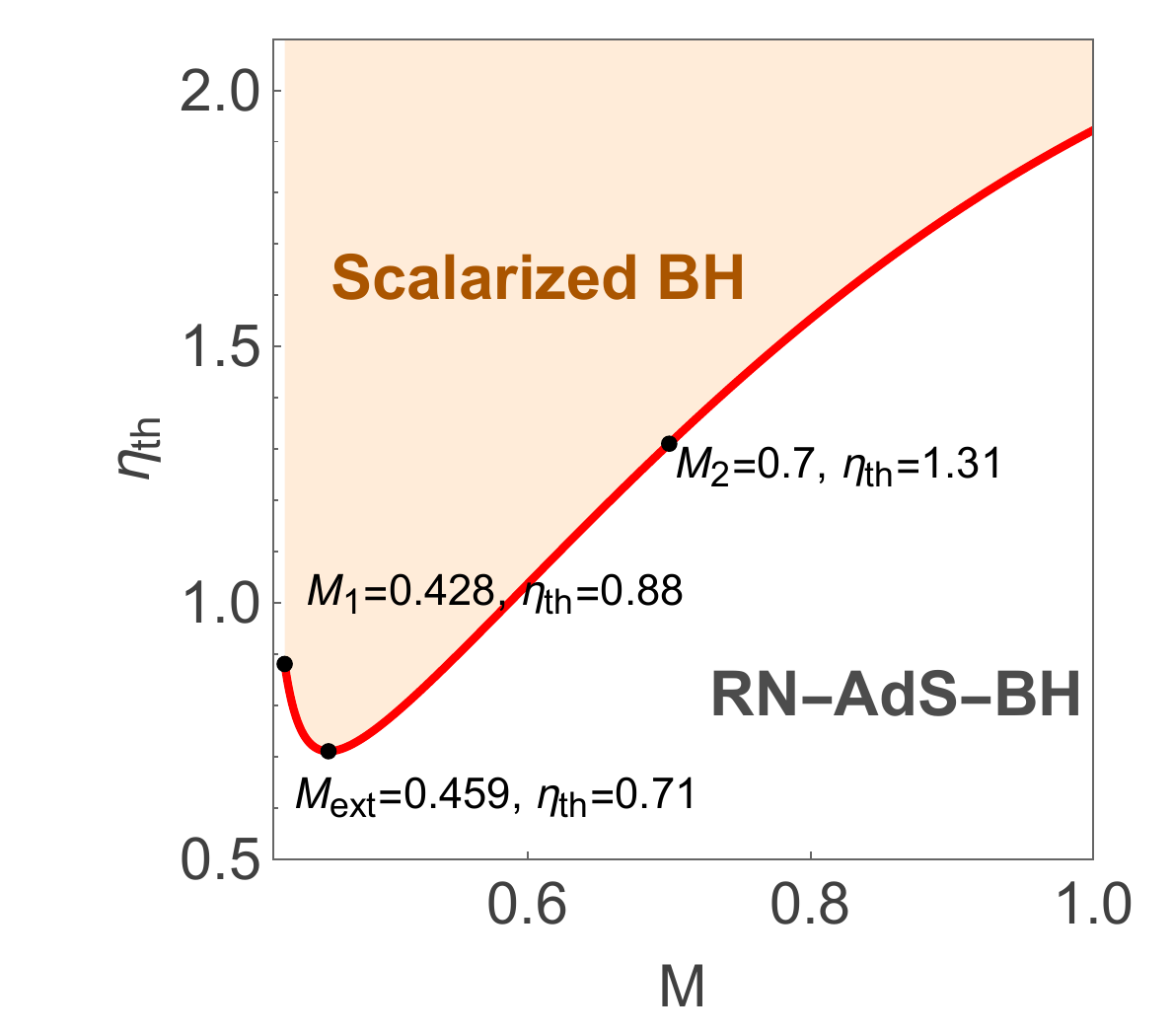}}
\hfill
\subfigure[~$Q$ vs $\eta_{th}$]
{\label{fig-clouds1-2}
\includegraphics[width=2.05in]{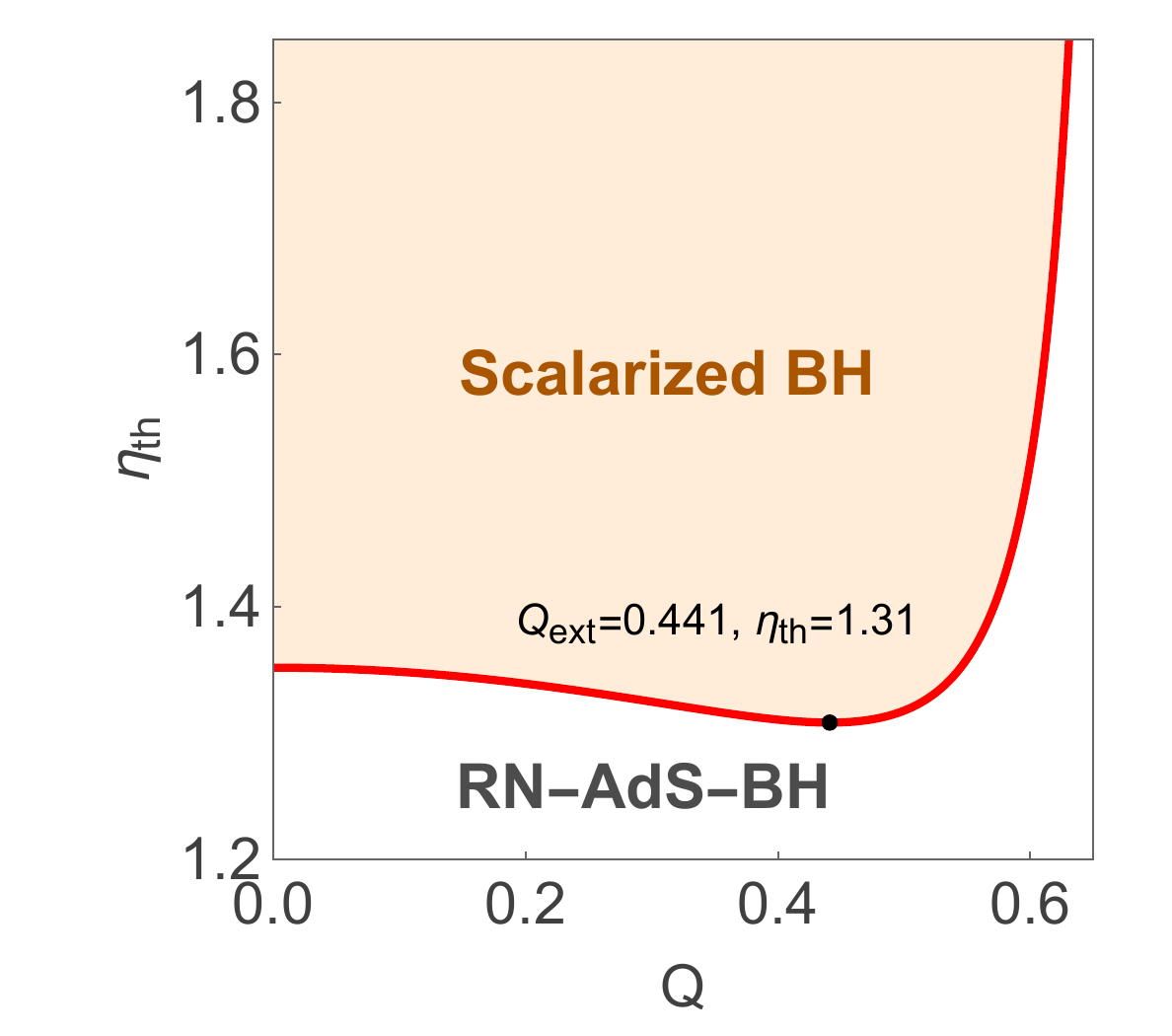}}
\hfill
\subfigure[~$-\Lambda$ vs $\eta_{th}$]
{\label{fig-clouds1-3}
\includegraphics[width=2.05in]{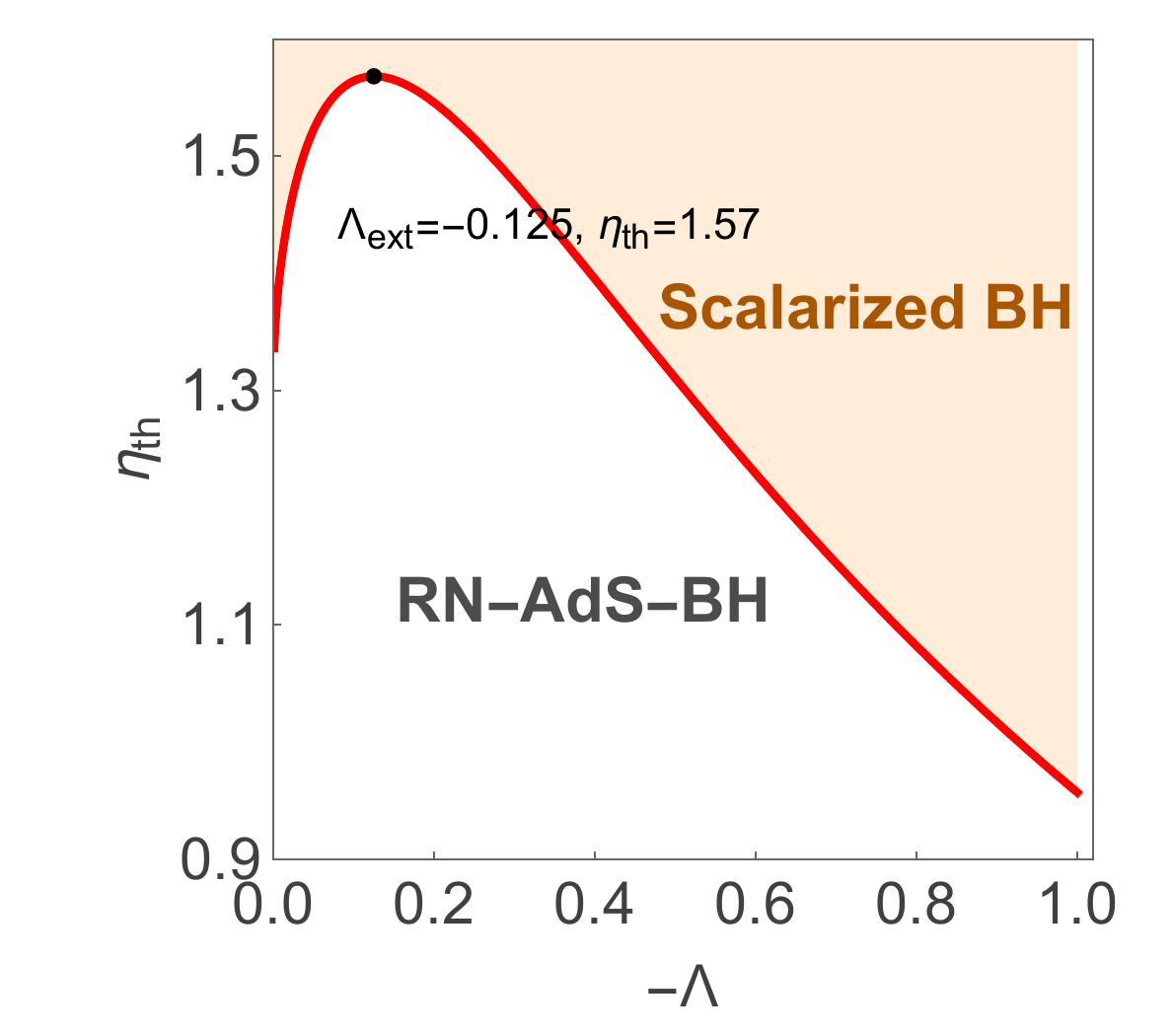}}
\hfill
\caption{Threshold coupling constant $\eta_{th}>0$ from scalar cloud configurations for GB$^+$ scalarization.  (a) $\eta_{th} (M)$ for fixed charge $Q=0.4$ and cosmological constant $\Lambda=-0.5$. (b) $\eta_{th}(Q)$ for  $M=0.7$ and $\Lambda=-0.5$. (c) $\eta_{th}(-\Lambda)$ for  $M=0.7$ and $Q=0.4$. Here, the red lines denote boundaries  between  RN-AdS and  scalarized AdS black holes . The shaded regions represent unstable regions where scalarized AdS black holes can be found.}
\label{fig-clouds1}
\end{figure}

As is shown in Fig. \ref{fig-clouds1} for $\mathcal{G}>0$ (GB$^+$ scalarization), we display the threshold value of coupling constant $\eta_{th}$ as functions of $M, Q, -\Lambda$ by computing  scalar clouds.
The threshold constant $\eta_{th}(M)$ for $Q=0.4$ and  $\Lambda=-0.5$ starts with $\eta_{th}=0.88$, decreases and  increases as $M$ increases.
 Similarly, $\eta_{th}(Q)$ for  $M=0.7$ and $\Lambda=-0.5$ arrives at minimum value  $\eta_{th}=1.31$ at $Q=0.441$ and it increases rapidly  as $Q$ increases  . On the other hand, $\eta_{th}(-\Lambda)$ for  $M=0.7$ and $Q=0.4$ increases, arrives at maximum value $\eta=1.57$ at $-\Lambda=0.125$ , and decreases as $-\Lambda$ increases. 
 In all panels, the red lines denote boundaries  between   RN-AdS and scalarized black holes. This may imply  that there are infinite branches of scalarized black holes for $\eta>0$ through spontaneous scalarization.
 
However, considering the bound of  $\eta_{th}\le\eta <2.25$ to avoid the AdS-tachyonic instability and to keep unstable region, one expects to find finite branches. 
In our work on the spontaneous scalarization of RN-AdS black holes, the number of branches depends on the number of branch points for scalar clouds.  Generating more higher branches requires a deeper effective potential well in the near-horizon, which corresponds to a larger coupling constant $\eta$.   Finding branch points is essentially the process of solving for the static scalar equation for the perturbed scalar $\phi(r)$ with  requiring  the scalar field to smoothly converge at $r=\infty$ (satisfying the boundary condition $\phi(\infty)$ = 0). We may obtain  the number of branches  within a given range of $\eta_{th}\le\eta<2.25$  when  determining  the number of nodes which describes how many times  the $\phi(r)$ crosses the $r$-axis.  That is, the number of branches is the number of nodes.

\begin{figure}[H]
\centering
\subfigure[~Appearance of $n=0$ branch with $\eta=\eta_0=1.31$]
{\label{fig-phicheck-1} 
\includegraphics[width=3in]{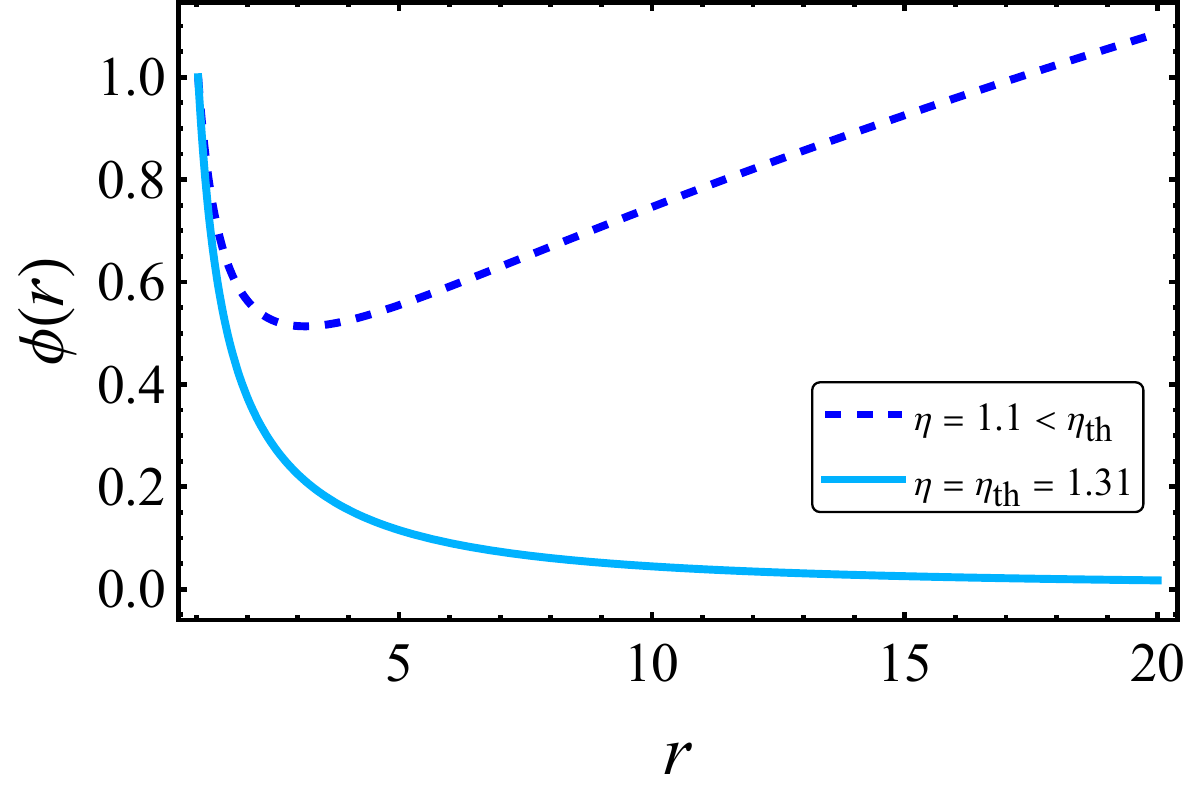}}
\hfill
\subfigure[~Appearance of $n=1$ branch with $\eta=\eta_1\approx2.2756$]
{\label{fig-phicheck-2}
\includegraphics[width=3in]{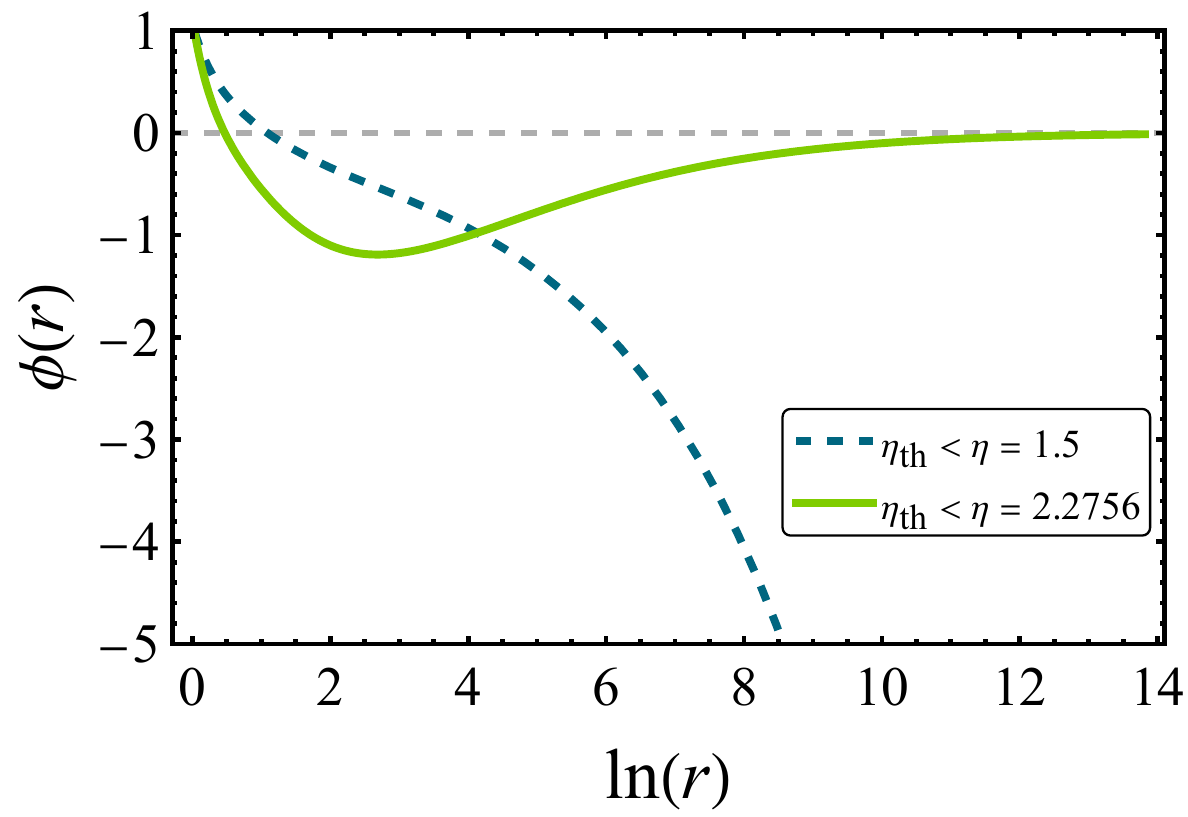}}
\caption{Radial profiles of the scalar cloud $\phi(r)$ under different parameter choices. (a) For $\eta=1.1$, the scalar diverges without crossing the $r$-axis, while for $\eta=\eta_{th}$ it converges at infinity without crossing. That is, $\eta=\eta_{th}=\eta_0$ is the branch point of $n=0$ branch ($\eta_0\le \eta<\eta_1)$. (b) For $\eta=1.5$, the scalar exhibits one node and diverges at infinity. However,  the curve for $\eta=\eta_1$ exhibits a local minimum (turning point) before approaching to zero, indicating the branch point of $n=1$ branch  ($\eta_1\le \eta<\eta_2$). }
\label{fig-phicheck}
\end{figure}

In Fig.~\ref{fig-phicheck}, we display several plots under the choice of different parameters. We have already determined the threshold coupling $\eta_{th} = 1.31$ for $M = 0.7$, $Q = 0.441$, and $\Lambda = -0.5$ (see Fig.~\ref{fig-clouds1-2}). As shown in Fig.~\ref{fig-phicheck-1}, when $\eta=1.1 < \eta_{th}$, it corresponds to the stable region and onset scalarization does not occur. In this case, $\phi(r)$ has no intersections with the $r$-axis (it is always greater than $0$) and it diverges as $r$ increases. For $\eta = \eta_{th}$, $\phi(r)$ converges at infinity and has no intersections with the $r$-axis. It determines that $\eta_{th}$ is the first branch point (that is, the starting point of the $n = 0$ fundamental branch).  This implies that the $n=0$ fundamental branch is confined to be $\eta_{th}\le \eta<\eta_1(=2.2756)$.

As depicted in Fig.~\ref{fig-phicheck-2}, continuing to increase $\eta$ beyond the theoretical limit of $2.25$ up to $\eta_1=2.2756$, $\phi(r)$ crosses the $r$-axis, exhibiting one node. The asymptotic behavior splits into two distinct types. For $\eta = 1.5$, $\phi(r)$ monotonically decreases after passing through the node and ultimately diverges downwards. In contrast, for $\eta =\eta_1$, the scalar field reaches a local minimum (a turning point) and then increases, eventually converging to zero at infinity. The emergence of this converging behavior with one node signifies the existence of another branch of solutions (such as the $n = 1$ excited branch).  Consequently, we prove that there exists  the single branch ($n=0$ fundamental branch) within our range of $\eta_0\le \eta<2.25$ (GB$^+$ scalarization) because the $n=1$ excited branch starts with $\eta=\eta_1=2.2756$.

Finally, for $\eta<0$, we do not make such boundaries between  RN-AdS and scalarized AdS black holes, even though this scalarization avoids AdS-tachyonic instability automatically. 
 For the negative coupling channel, we do not use the same scalar-cloud shooting diagnostic as in the GB$^+$ case. Instead, we will construct the nonlinear scalarized branch directly from regular horizon data in the next section. This corresponds to the channel in which GB$^-$ scalarization is triggered by negative near-horizon potentials.

\section{Numerical Solutions for Scalarized AdS Black Holes}
\label{sec:results}

We consider static and spherically symmetric space-times as well as static and a radial-dependent scalar as 
\begin{equation}
ds^2=-A(r)dt^2+\frac{1}{B(r)}dr^2+r^2\left(d\theta^2+\sin^2\theta d\varphi^2\right),\quad \phi=\phi(r).\label{ms}
\end{equation}
Now,  we are in a position to find the numerical solutions for scalarized AdS black hole in the ESGB theory.  For this purpose,
we first introduce a coordinate transformation of $z=\frac{r_+}{r}$ so that the metric functions can be derived in the compact region of $0 \leq z\leq 1$, and $A(r)$ and $B(r)$ become $A=A(z)$ and $B=B(z)$.
{Therefore, $z = 0$ denotes  infinity ($r\rightarrow\infty$),
and $z=1$ corresponds to the event horizon $r=r_+$ of the black hole. The metric functions $A(r)$ and $B(r)$ in Eq.\eqref{ms} approach $r^2$ as $r\rightarrow\infty$. In this case,  $A(z)$ and $B(z)$ of $1/z^2$ are divergent at $z=0$.  Then, we can  define new metric functions as 
\begin{eqnarray}
A_z(z)\rightarrow  z^2 A(z),\quad B_{z}(z)\rightarrow z^2B(z) \label{metricz}
\end{eqnarray}
so that $A_z(z)$ and $B_{z}(z)$ are always regular in the whole region of $0 \leq z\leq 1$.
Fortunately, the scalar field $\phi(z)$ and the electromagnetic potential
 $V(z)$ are always regular in the whole region under $z\to \frac{r_+}{r}$. Here, we reset
\begin{eqnarray}
\phi_{z}(z)\rightarrow \phi(z),V_{z}(z)\rightarrow V(z). \label{pvz}
\end{eqnarray}

Substituting  Eqs.~\eqref{metricz} and \eqref{pvz} into Eqs.~\eqref{eq:einstein_eq1} and \eqref{eq:einstein_eq2} , we have three equations:
\begin{eqnarray}
eq_1&=&3z^4 - \frac{z^6}{B_z} + \frac{r_+^2 z^4 \Lambda}{B_z} - \frac{z^5 A_z'}{A_z} + \frac{z^8 (V_z')^2}{A_z} \frac{4 z^7 \eta \phi_z \phi_z'}{r_+^2} + \frac{12 z^5 \eta B_z \phi_z \phi_z'}{r_+^2} \nonumber \\
&&+ \frac{2 z^8 \eta \phi_z A_z' \phi_z'}{r_+^2 A_z} - \frac{6 z^6 \eta B_z \phi_z A_z' \phi_z'}{r_+^2 A_z} - z^6 (\phi_z')^2 = 0 ,\label{eq1z}
\end{eqnarray}
\begin{eqnarray}
eq_2&=&\frac{z^4 A_z}{r_+^2} - z^2 \Lambda A_z - \frac{3 z^2 A_z B_z}{r_+^2} + \frac{z^3 A_z B_z'}{r_+^2} - \frac{z^6 B_z (V_z')^2}{r_+^2}  - \frac{4 z^5 \eta A_z B_z \phi_z \phi_z'}{r_+^4}\nonumber \\
&&- \frac{4 z^3 \eta A_z B_z^2 \phi_z \phi_z'}{r_+^4} - \frac{2 z^6 \eta A_z \phi_z B_z' \phi_z'}{r_+^4} + \frac{6 z^4 \eta A_z B_z \phi_z B_z' \phi_z'}{r_+^4} \nonumber \\
&&- \frac{z^4 A_z B_z (\phi_z')^2}{r_+^2} - \frac{4 z^6 \eta A_z B_z (\phi_z')^2}{r_+^4} + \frac{4 z^4 \eta A_z B_z^2 (\phi_z')^2}{r_+^4} \nonumber \\
&&- \frac{4 z^6 \eta A_z B_z \phi_z \phi_z''}{r_+^4} + \frac{4 z^4 \eta A_z B_z^2 \phi_z \phi_z''}{r_+^4}=0,\label{eq2z}
\end{eqnarray}
\begin{eqnarray}
eq_3&=&\frac{4 r_+^2 \Lambda A_z^2}{z^2} + \frac{12 A_z^2 B_z}{z^2} - \frac{4 A_z B_z A_z'}{z} - B_z (A_z')^2 - \frac{4 A_z^2 B_z'}{z} + A_z A_z' B_z' - 4 z^2 A_z B_z (V_z')^2 \nonumber \\
&& + \frac{16 \eta A_z^2 B_z^2 \phi_z \phi_z'}{r_+^2 z} - \frac{8 \eta A_z B_z^2 \phi_z A_z' \phi_z'}{r_+^2} - \frac{4 z \eta B_z^2 \phi_z (A_z')^2 \phi_z'}{r_+^2} - \frac{24 \eta A_z^2 B_z \phi_z B_z' \phi_z'}{r_+^2} \nonumber \\
&& + \frac{12 z \eta A_z B_z \phi_z A_z' B_z' \phi_z'}{r_+^2} + 4 A_z^2 B_z (\phi_z')^2 - \frac{16 \eta A_z^2 B_z^2 (\phi_z')^2}{r_+^2} + \frac{8 z \eta A_z B_z^2 A_z' (\phi_z')^2}{r_+^2} \nonumber \\
&& + 2 A_z B_z A_z'' + \frac{8 z \eta A_z B_z^2 \phi_z \phi_z' A_z''}{r_+^2} - \frac{16 \eta A_z^2 B_z^2 \phi_z \phi_z''}{r_+^2} + \frac{8 z \eta A_z B_z^2 \phi_z A_z' \phi_z''}{r_+^2}=0,\label{eq3z}
\end{eqnarray}
where primes denote derivatives with respect to $z$.

In order to obtain  scalarized AdS black hole solutions, we solve three Eqs. \eqref{eq1z}--\eqref{eq3z} numerically via a shooting method.
Spherically symmetric black holes have an event horizon $(z=1)$, where  $A_z$, $B_z$, and $V_z$ vanish, while $\phi_z$ takes a horizon scalar $\phi_0$:
\begin{eqnarray}\label{expansion}
&&A_{z}(z\approx 1)=A_{1}(1-z)+A_{2}(1-z)^2+\cdots, \\
&&B_{z}(z\approx 1)=B_{1}(1-z)+B_{2}(1-z)^2+\cdots,\\
&&V_{z}(z\approx 1)=V_{1}(1-z)+V_{2}(1-z)^2+\cdots,\\
&&\phi_{z}(z\approx 1)=\phi_{0}+\phi_{1}(1-z)+\cdots.\label{eqexpandh}
\end{eqnarray}
Here, $A_1$ and $\phi_0$ are specified parameters evaluated at the event horizon. The remaining  coefficients are determined by algebraic expressions depending on these two quantities and their explicit forms will be  presented in Appendix~\ref{app:expansion_coefficients}. 

It is worth pointing out that the regularity of the scalar, and its first
and second derivatives on the horizon give an additional  condition
\begin{eqnarray}
r_+^8 + 8 r_+^4 \eta^2 \left(-3 + 2 Q^2 + 2 r_+^2 \Lambda\right) \phi_0^2 + 16 \eta^4 \left[Q^4 + r_+^2 \Lambda \left(-6 + r_+^2 \Lambda\right) + 2 Q^2 \left(3 + r_+^2 \Lambda\right)\right] \phi_0^4 >0,\label{horzcond2}
\end{eqnarray}
which recovers the RN black hole when imposing $\Lambda=0$.\par

In Ref.~\cite{Guo:2025flg}, AdS scalarization was investigated in the context of primary hair,  where the scalar field carries an independent charge. In the present work, we are interested in the case of secondary hair. To describe a spontaneous phase transition without introducing external sources, we require the non-normalizable mode in Eq.~\eqref{eq-phi-inf} to vanish ($\phi_{1\infty} = 0$). This boundary condition ensures that the scalar does not possess any primary  charge  and the asymptotic form takes the form
\begin{eqnarray}
A_z=B_z=-\frac{\Lambda r_+^2}{3}, \quad \phi_z= 0, \quad V(r) = \mu - \frac{Q}{r} \quad {\rm when} \quad z\rightarrow 0 \quad (r\to\infty)
\label{infcond}
\end{eqnarray}
with $\mu$ chemical potential. 

\subsection{GB\texorpdfstring{$^+$}{+} scalarization}

With the threshold of instability established in the probe limit, we are willing to  construct scalarized AdS black hole solutions for  their existence domain  of $\eta_{th} \le  \eta < 2.25$ with $\Lambda = -0.5$. 
The core objective here is to understand how the backreaction of the scalar cloud  modifies the spacetime geometry.  By solving the coupled field equations numerically, we track the evolution of the scalar hair from its onset at the bifurcation points deep into the non-linear regime.

The scalarized AdS black hole solutions emerge as the $n=0$ fundamental  branch  bifurcating from the  RN-AdS black holes. We construct these solutions using a numerical shooting method by integrating the coupled field equations from the event horizon out to spatial infinity. The physical solutions are singled out by imposing regularity at the horizon and ensuring that the scalar field respects the  asymptotic AdS boundary conditions. We would like to  mention that backreacted positive-coupling scalarized RN-AdS solutions in the pure GB sector were  constructed in~\cite{Brihaye:2019dck}. Our GB$^+$ scalarized solutions  therefore serve as benchmarks in our conventions and provide the data  for the fixed-charge  thermodynamic analysis of Sec.~\ref{sec:TP}.

\begin{figure}[H]
\centering
\subfigure[~Metric function $B(r)$]
{\label{br} 
\includegraphics[width=3in]{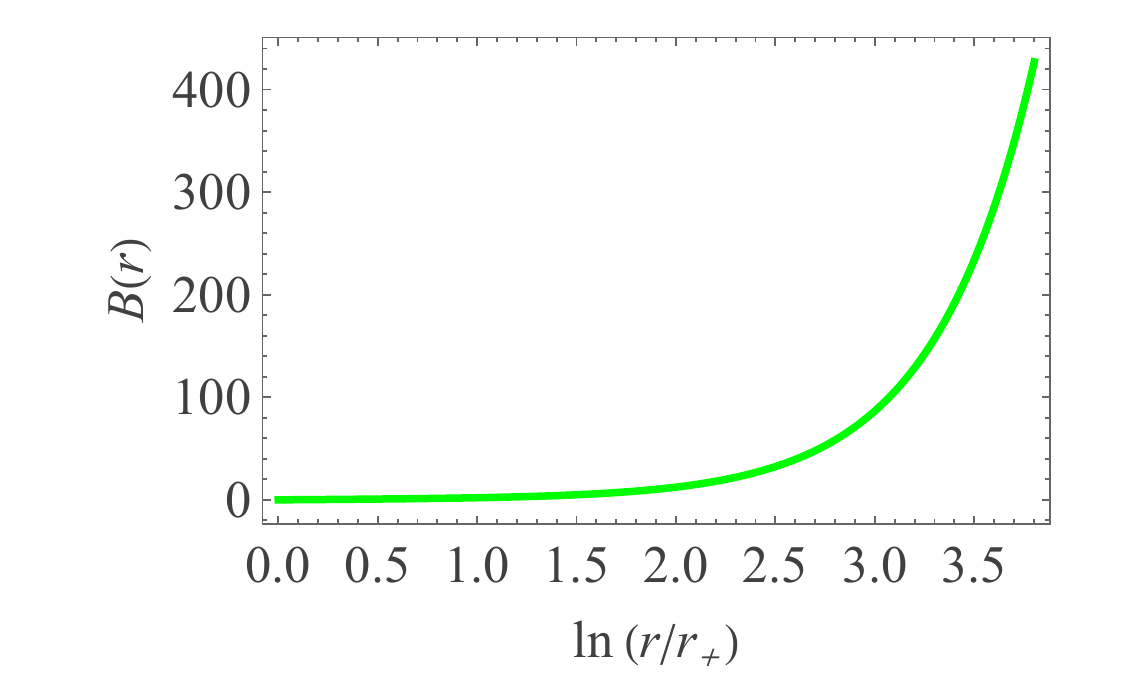}}
\hfill
\subfigure[~Ratio of metric functions $\Delta$]
{\label{deta}
\includegraphics[width=3in]{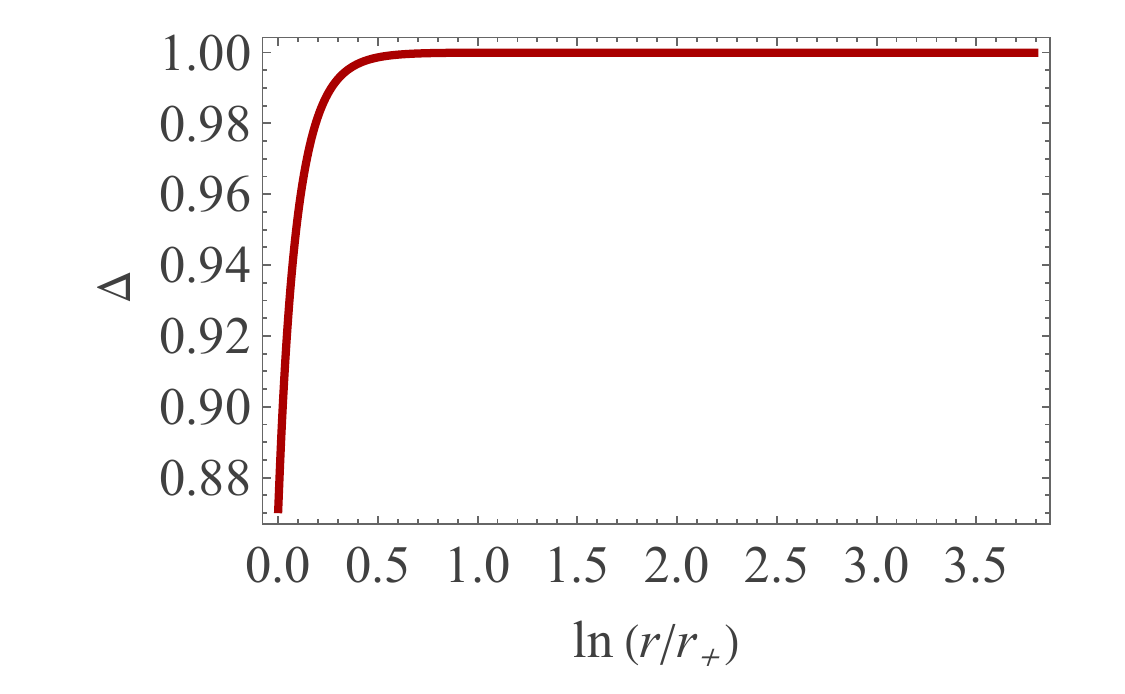}}
\hfill
\subfigure[~Scalar field $\phi(r)$]
{\label{phir}
\includegraphics[width=3in]{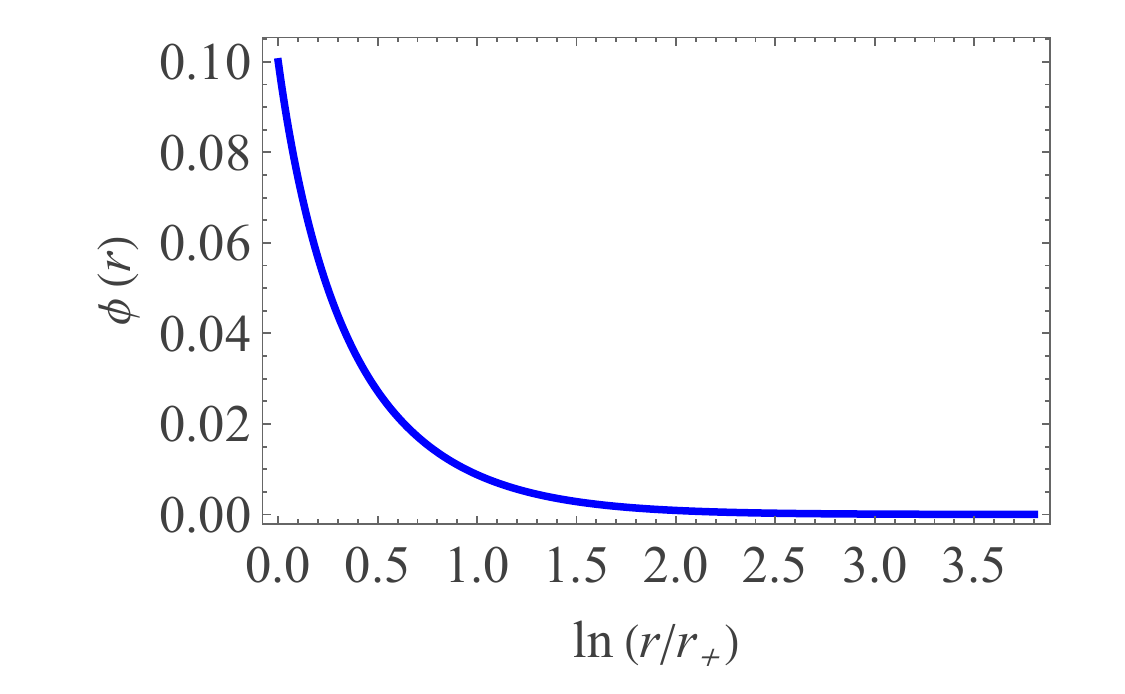}}
\hfill
\subfigure[~Electromagnetic potential $V(r)$]
{\label{vr}
\includegraphics[width=3in]{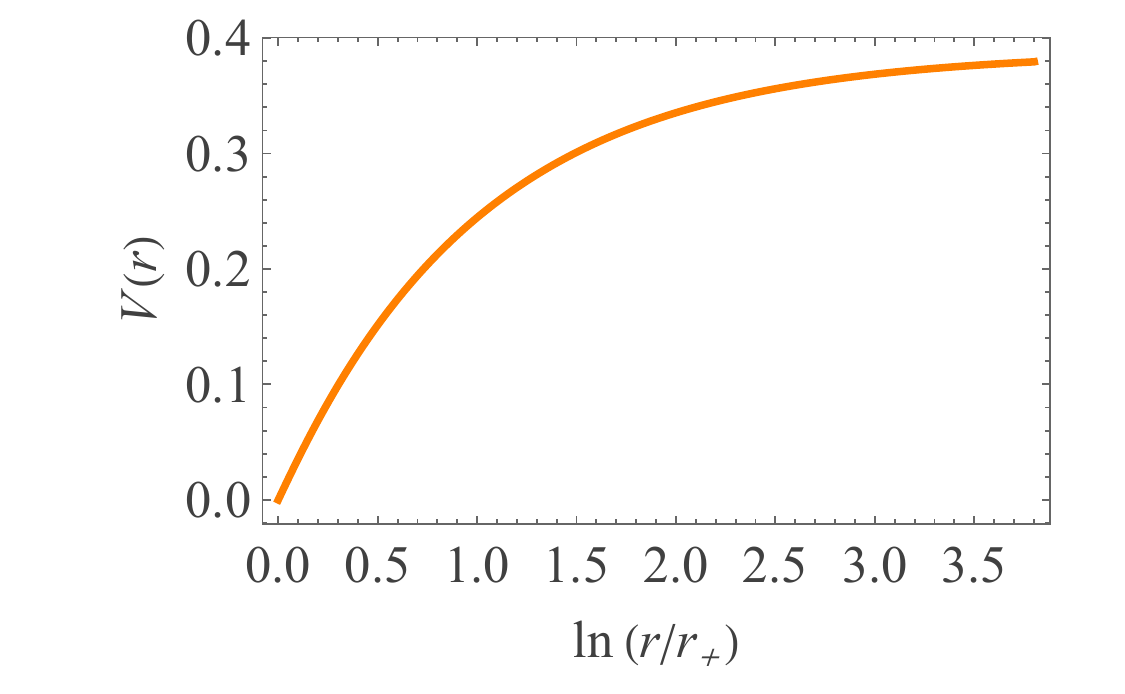}}
\caption{Radial profiles of  GB$^+$ scalarized AdS black hole solution. (a) The metric function $B(r)$ as a function of the logarithmic radial coordinate. (b) Ratio of the metric functions ($\Delta = A(r)/B(r)$),  illustrating the local geometric deformation caused by the scalar backreaction near the horizon. (c) Scalar profile $\phi(r)$. (d) Electromagnetic potential $V(r)$.  Parameters are set to  $Q=0.441$, $\Lambda=-0.5$, $\phi_0=0.1$ and $r_+=1.129$, yielding a required GB coupling $\eta=1.447$  between $\eta_{th}(=1.31)$ and 2.25.}
\label{fig-matric} 
\end{figure}

Four panels in Fig.~\ref{fig-matric} illustrate the  physical and geometric characteristics of the numerical solutions for the GB$^+$ scalarized AdS black holes. 
Fig.~\ref{br} shows the radial profile of the metric function $B(r)$. It vanishes at the horizon $r=r_+$ and approaches the  AdS asymptotic limit at large $r$, ensuring a regular event horizon and proper boundary conditions.
Fig.~\ref{deta} illustrates the scalar backreaction by plotting the ratio of the metric functions, $\Delta \equiv A(r)/B(r)$. The deviation of $\Delta$ from unity  in the near-horizon reflects the strong  geometric deformation induced by the scalar hair. Asymptotically, $\Delta$  approaches $1$, satisfying the AdS boundary condition.
Fig.~\ref{phir} shows the scalar  profile $\phi(r)$. The scalar  peaks at the horizon ($\phi_0 = 0.1$) and decays monotonically to zero at infinity.  This shows  that the scalar hair is tightly bound by the near-horizon effective potential, satisfying both regularity and normalizability conditions.
Fig.~\ref{vr} plots the electromagnetic potential $V(r)$.  It is  consistent with gauge regularity of  $V(r_+) = 0$. The potential rises monotonically to a constant at infinity, corresponding to the chemical potential of the scalarized  AdS black hole. Despite the non-trivial scalar backreaction, $V(r)$ retains the  Coulombic profile of a charged AdS background.

\subsection{GB\texorpdfstring{$^-$}{-} scalarization}

Next, we consider the $\eta < 0$ regime to investigate scalarized AdS black holes named  by GB$^-$ scalarization, where the Gauss-Bonnet invariant $\mathcal{G} < 0$. In this case, we also find a single branch  of scalarized black holes.

\begin{figure}[H]
\centering
\subfigure[~Metric function $B(r)$]
{\label{brgb-} 
\includegraphics[width=3in]{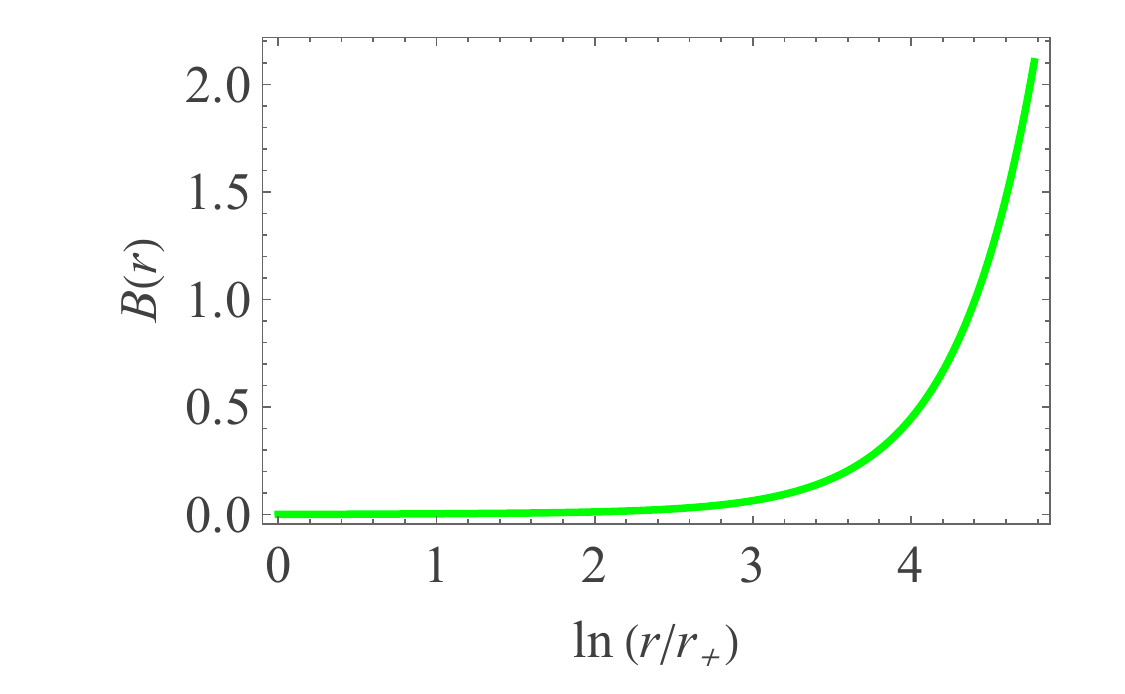}}
\hfill
\subfigure[~Ratio of metric functions $\Delta$]
{\label{detagb-}
\includegraphics[width=3in]{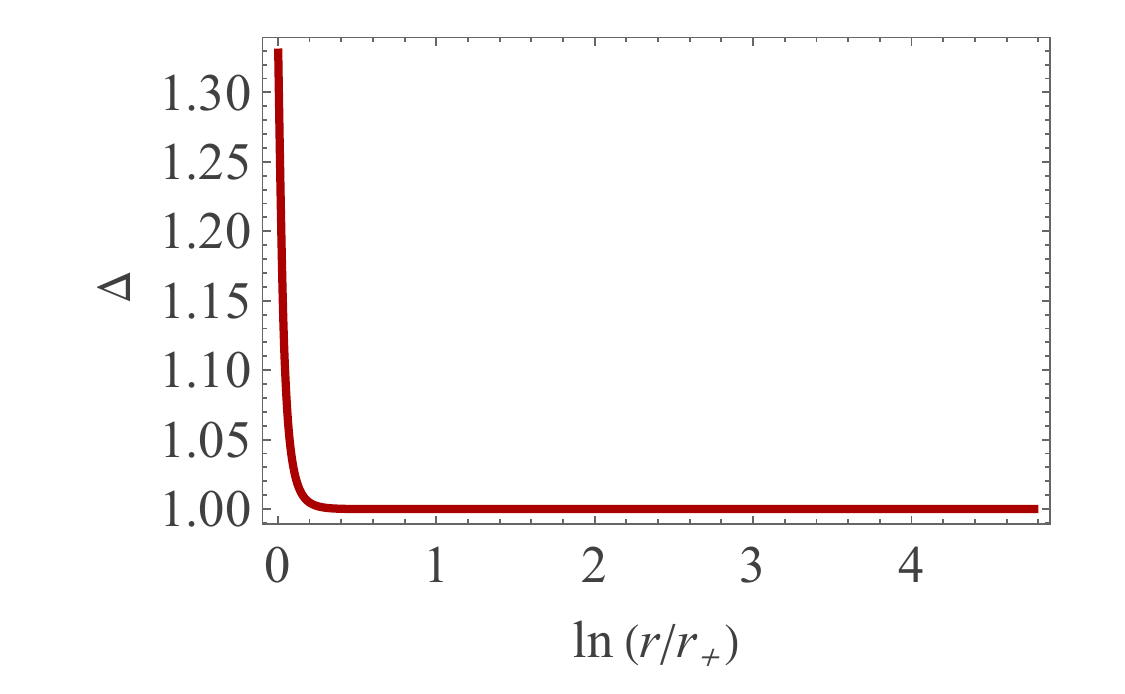}}
\hfill
\subfigure[~Scalar field $\phi(r)$]
{\label{phirgb-}
\includegraphics[width=3in]{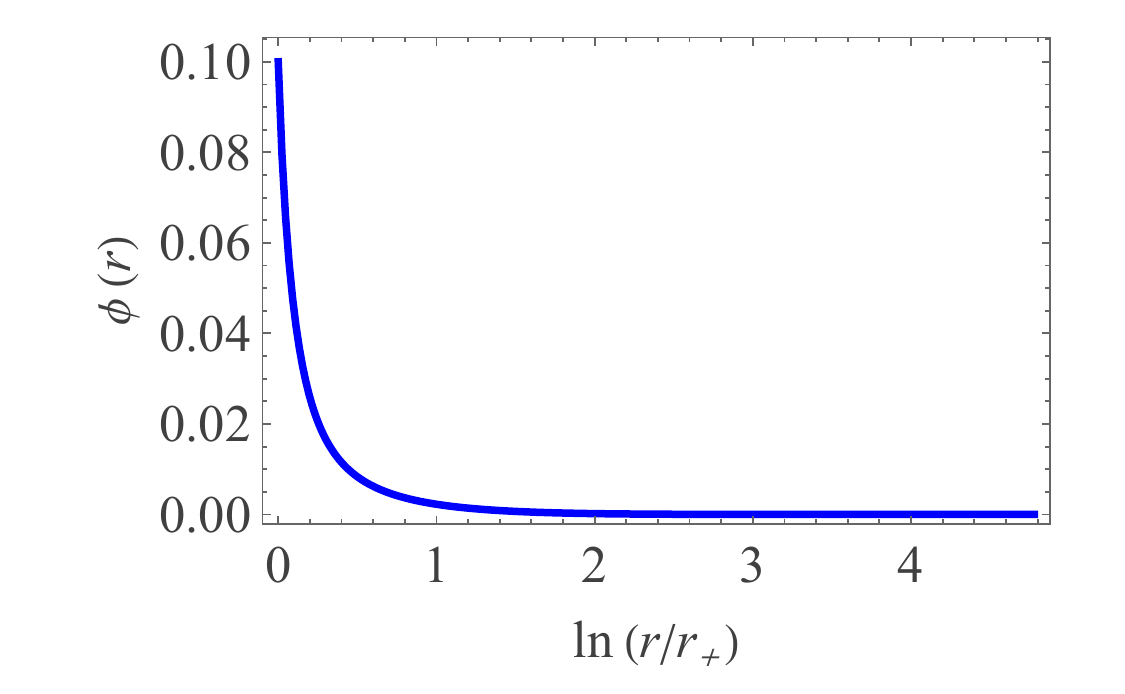}}
\hfill
\subfigure[~Electromagnetic potential $V(r)$]
{\label{vrgbb-}
\includegraphics[width=3in]{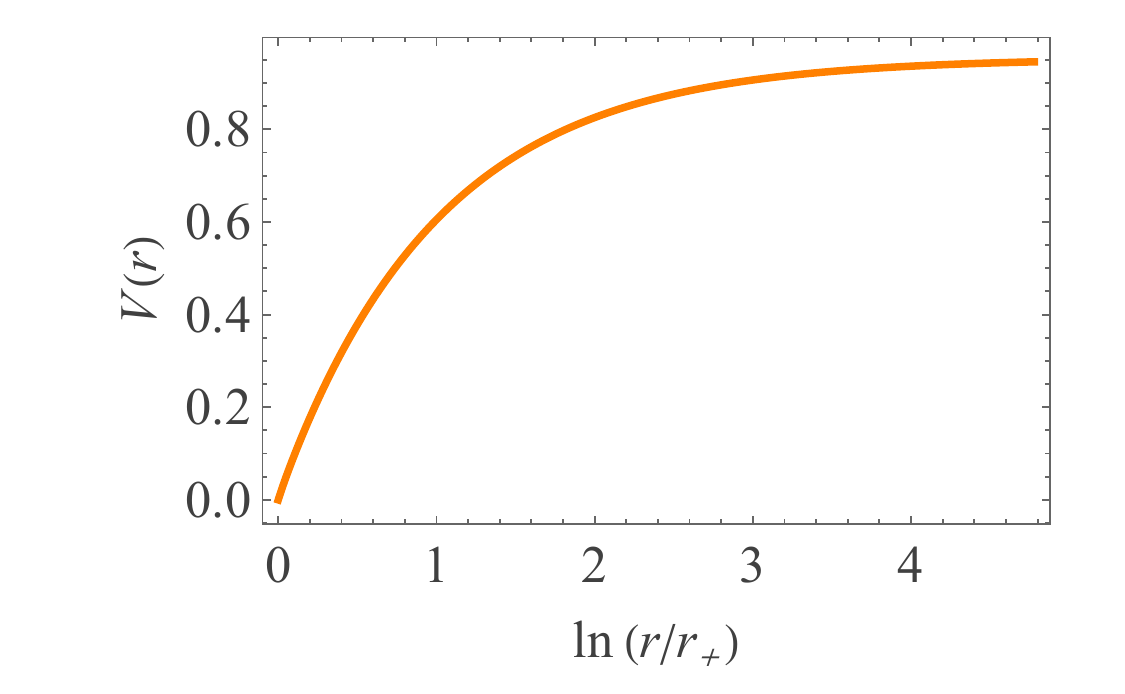}}
\caption{Characteristic behaviors of the GB$^-$ scalarized AdS black hole ($\eta < 0$). Panels display: (a) Radial dependence of the metric component $B(r)$. (b) The ratio of the metric functions, $\Delta = A(r)/B(r)$, showing a monotonically decreasing behavior. (c) Configuration of the highly compact scalar hair $\phi(r)$ decaying towards the boundary. (d) Electrostatic potential $V(r)$. The parameters are chosen as  $Q = 0.4$, $\Lambda = -0.5$, $\phi_0 = 0.1$ and $r_+=0.422$ corresponding to a negative GB coupling of $\eta = -0.182$.}
\label{fig-matric-GB-}
\end{figure} 

The four panels in Fig.~\ref{fig-matric-GB-} illustrate the numerical solutions for the GB$^-$ case. Since the global topology and asymptotic boundary conditions are similar to the GB$^+$ case (Fig.~\ref{fig-matric}), we focus here on the distinct local features induced by the negative Gauss-Bonnet coupling and the smaller horizon radius.
Fig.~\ref{brgb-} shows the metric function $B(r)$. Its primary distinction from the GB$^+$ case is a visibly smaller horizon radius $r_+$, while  asymptotic AdS behavior is  preserved.
Fig.~\ref{detagb-} depicts the ratio of the metric functions, $\Delta \equiv A(r)/B(r)$. In stark contrast to the GB$^+$ case where $\Delta$ is suppressed near the horizon and increases towards unity, here $\Delta$ begins at an elevated value of $1.35$ in the near-horizon and monotonically decreases to $1$. This striking difference highlights how the negative Gauss-Bonnet coupling  alters the local geometric deformation backreacted by the scalar hair,  while it recovers the asymptotic AdS boundary condition.
Fig.~\ref{phirgb-} displays the scalar field $\phi(r)$. Due to the smaller $r_+$ and the correspondingly stronger near-horizon curvature, the scalar field decays much more sharply towards zero than in the GB$^+$ case. The negative coupling tightly confines  the scalar hair to the immediate vicinity of the horizon, resulting in a highly compact scalar hair.
Fig.~\ref{vrgbb-} plots the electromagnetic potential $V(r)$. While maintaining gauge regularity at the horizon, its asymptotic value at spatial infinity is significantly higher than that of the GB$^+$ case. This elevation is a direct consequence of the smaller horizon radius.

\section{Thermodynamics and phase transition}
\label{sec:TP}
To explore the physical relevance of scalarized AdS black hole solutions, it is useful to address their thermodynamic stability and phase transition. We work in the canonical ensemble by keeping the electric charge $Q$ fixed. The Hawking temperature is calculated from the surface gravity at the horizon as
\begin{equation}
    T = \frac{1}{4\pi} \sqrt{A'(r_+) B'(r_+)}.
\end{equation}

Due to the presence of the higher-curvature GB term, we compute the Wald entropy, which interprets entropy as a Noether charge associated with the diffeomorphism invariance of the theory. It is given by
\begin{equation}
    S_{\text{Wald}} = -2\pi \int_{\mathcal{H}} \frac{\partial \mathcal{L}}{\partial R_{\mu\nu\rho\sigma}} \epsilon_{\mu\nu}\epsilon_{\rho\sigma} \sqrt{h} d\Omega
    = \frac{\mathcal{A}_H}{4} + 4\pi f(\phi_0)  \,,
    \label{eq:Wald_def}
\end{equation}
where $\mathcal{A}_H = 4\pi r_+^2$ is the horizon area. 
To distinguish the mass parameter $M_{\rm RN}$ of the bald RN-AdS solution from the mass of a hairy solution, we denote the latter by $M_{\rm hair}$ below. 

Being consistent with the boundary conditions shown in Eq.~(\ref{infcond}), $A_z(z) \equiv z^2 A(r)$ admits the asymptotic expansion near the AdS boundary ($z \to 0$) as 
\begin{equation}
    A_z(z) = -\frac{\Lambda r_+^2}{3} + z^2 - \frac{2M_{\rm hair}}{r_+} z^3 + \frac{Q^2}{r_+^2} z^4 + \mathcal{O}(z^5).
    \label{eq:Azexpansion}
\end{equation}
To extract the physical mass $M_{\rm hair}$ from our numerical solutions without introducing high-order derivative terms, we choose a near-boundary polynomial fit to the numerical data as $A_z^{\text{num}}(z) = a_0 + a_2 z^2 + a_3 z^3 + a_4 z^4$. 
Comparing this fit with Eq.~(\ref{eq:Azexpansion}), the mass parameter is encoded in the cubic coefficient $a_3$. We note that Eq.~(\ref{infcond}) demands the constant term to be $-\Lambda r_+^2/3$. Integrating from the horizon via the shooting method inevitably introduces a slight global scaling offset. Since the field equations possess a scaling symmetry and we have already chosen the coordinate gauge of $A_z = B_z$ at infinity, we can obtain the physical mass by compensating for this numerical scaling drift. Thus, $M_{\rm hair}$ is evaluated as
\begin{equation}
    M_{\rm hair} = \frac{a_3 r_+^3\Lambda }{6a_0}.
\end{equation}

To validate the first law of thermodynamics for our scalarized AdS black hole solutions numerically, we generate sequences of configurations by varying the horizon scalar $\phi_0$. We consider three representative fixed-charge cases: the GB$^+$ scalarization with $(M_{\rm RN},Q)=(0.7,0.4)$, the small charge  GB$^-$ scalarization with $(M_{\rm RN},Q)=(0.406,0.4)$, and the high-charge GB$^-$ scalarization with $(M_{\rm RN},Q)=(0.7,0.66)$. The extracted numerical mass $M_{\rm hair}$  together with the temperature $T$ and the Wald entropy $S$  are summarized in Table~\ref{table1}. To make the comparison transparently, the three data sets are arranged in one table with a common $\phi_0$ column. For a $Q=0.66$ case, the absence of entries beyond $\phi_0=0.04$ has a physical meaning such that  increasing $\phi_0$ further makes the near-horizon initial data cease to be real and prevents the AdS-boundary condition from being satisfied at the same time. Accordingly, the thermodynamic conclusions drawn from Table~\ref{table1} and Fig.~\ref{figgt0} should be understood as statements on the scanned regular segments, but not as a complete  phase diagram.

\begin{table}[htbp]
\centering
\caption{Thermodynamic quantities of the scalarized RN-AdS black holes for the three representative fixed-charge cases. We set $\Lambda=-0.5$. The symbol $\backslash$ means that any tabulated regular data is not allowed  for that parameter slice at the corresponding value of $\phi_0$. In the GB$^-$ case with $Q=0.66$, we note that the entries stop at the endpoint of the regular solution family.}
\label{tab:thermo_data_combined}
\renewcommand{\arraystretch}{1.12}
\setlength{\tabcolsep}{2.0mm}
\resizebox{\textwidth}{!}{%
\begin{tabular}{c c c c c c c c c c c c c c c c}
\hline\hline
$\phi_0$ & \multicolumn{5}{c}{GB$^+$: $M_{\rm RN}=0.7$, $Q=0.4$} & \multicolumn{5}{c}{GB$^-$: $M_{\rm RN}=0.406$, $Q=0.4$} & \multicolumn{5}{c}{GB$^-$: $M_{\rm RN}=0.7$, $Q=0.66$} \\
\cline{2-6} \cline{7-11} \cline{12-16}
 & $r_+$ & $\eta$ & $M_{\rm hair}$ & $T$ & $S$ & $r_+$ & $\eta$ & $M_{\rm hair}$ & $T$ & $S$ & $r_+$ & $\eta$ & $M_{\rm hair}$ & $T$ & $S$ \\
\hline
0.01 & 1.063 & 1.32866 & 0.707154 & 0.106552 & 3.55290 & 0.413 & -0.14452 & 0.406074 & 0.028359 & 0.53567 & 0.758 & -4.20017 & 0.702288 & 0.055436 & 1.80141 \\
0.02 & 1.073 & 1.34603 & 0.714596 & 0.106566 & 3.62257 & 0.414 & -0.14934 & 0.406140 & 0.029198 & 0.53798 & 0.768 & -5.15977 & 0.704288 & 0.057238 & 1.83901 \\
0.03 & 1.083 & 1.36268 & 0.722281 & 0.106595 & 3.69465 & 0.415 & -0.15407 & 0.406203 & 0.029995 & 0.54009 & 0.778 & -5.64475 & 0.706199 & 0.058932 & 1.86861 \\
0.04 & 1.093 & 1.37860 & 0.730221 & 0.106640 & 3.76919 & 0.416 & -0.15869 & 0.406256 & 0.030749 & 0.54198 & 0.789 & -5.73268 & 0.707970 & 0.071316 & 1.89704 \\
0.05 & 1.103 & 1.39378 & 0.738434 & 0.106700 & 3.84623 & 0.417 & -0.16316 & 0.406322 & 0.031460 & 0.54363 & $\backslash$ & $\backslash$ & $\backslash$ & $\backslash$ & $\backslash$ \\
0.06 & 1.113 & 1.40819 & 0.746914 & 0.106775 & 3.92583 & 0.418 & -0.16745 & 0.406348 & 0.032125 & 0.54503 & $\backslash$ & $\backslash$ & $\backslash$ & $\backslash$ & $\backslash$ \\
0.07 & 1.123 & 1.42184 & 0.755685 & 0.106866 & 4.00802 & 0.419 & -0.17153 & 0.406385 & 0.032744 & 0.54616 & $\backslash$ & $\backslash$ & $\backslash$ & $\backslash$ & $\backslash$ \\
0.08 & 1.133 & 1.43472 & 0.764733 & 0.106972 & 4.09283 & 0.420 & -0.17537 & 0.406386 & 0.033316 & 0.54703 & $\backslash$ & $\backslash$ & $\backslash$ & $\backslash$ & $\backslash$ \\
0.09 & 1.143 & 1.44682 & 0.774074 & 0.107093 & 4.18029 & 0.421 & -0.17894 & 0.406428 & 0.033843 & 0.54761 & $\backslash$ & $\backslash$ & $\backslash$ & $\backslash$ & $\backslash$ \\
0.10 & 1.153 & 1.45814 & 0.783712 & 0.107227 & 4.27043 & 0.422 & -0.18221 & 0.406444 & 0.034323 & 0.54792 & $\backslash$ & $\backslash$ & $\backslash$ & $\backslash$ & $\backslash$ \\
0.11 & 1.163 & 1.46868 & 0.793647 & 0.107375 & 4.36325 & 0.423 & -0.18518 & 0.406453 & 0.034759 & 0.54795 & $\backslash$ & $\backslash$ & $\backslash$ & $\backslash$ & $\backslash$ \\
0.12 & 1.173 & 1.47844 & 0.803876 & 0.107535 & 4.45876 & 0.424 & -0.18783 & 0.406415 & 0.035151 & 0.54769 & $\backslash$ & $\backslash$ & $\backslash$ & $\backslash$ & $\backslash$ \\
0.13 & 1.183 & 1.48744 & 0.814413 & 0.107706 & 4.55698 & 0.425 & -0.19015 & 0.406393 & 0.035502 & 0.54716 & $\backslash$ & $\backslash$ & $\backslash$ & $\backslash$ & $\backslash$ \\
0.14 & 1.193 & 1.49570 & 0.825255 & 0.107884 & 4.65789 & 0.426 & -0.19214 & 0.406367 & 0.035814 & 0.54636 & $\backslash$ & $\backslash$ & $\backslash$ & $\backslash$ & $\backslash$ \\
0.15 & 1.203 & 1.50326 & 0.836412 & 0.108066 & 4.76151 & 0.427 & -0.19381 & 0.406327 & 0.036090 & 0.54531 & $\backslash$ & $\backslash$ & $\backslash$ & $\backslash$ & $\backslash$ \\
0.16 & 1.213 & 1.51021 & 0.847877 & 0.108238 & 4.86783 & 0.428 & -0.19516 & 0.406280 & 0.036335 & 0.54400 & $\backslash$ & $\backslash$ & $\backslash$ & $\backslash$ & $\backslash$ \\
0.17 & 1.223 & 1.51680 & 0.859658 & 0.108349 & 4.97689 & 0.429 & -0.19621 & 0.406222 & 0.036554 & 0.54245 & $\backslash$ & $\backslash$ & $\backslash$ & $\backslash$ & $\backslash$ \\
0.18 & $\backslash$ & $\backslash$ & $\backslash$ & $\backslash$ & $\backslash$ & 0.430 & -0.19697 & 0.406153 & 0.036756 & 0.54068 & $\backslash$ & $\backslash$ & $\backslash$ & $\backslash$ & $\backslash$ \\
\hline\hline
\end{tabular}
}
\label{table1}
\end{table}

Since the electric charge is fixed ($dQ = 0$), the first law of thermodynamics requires $dM_{\rm hair} = T dS$. To verify this relation by making use of our discrete numerical data, we employ the central finite difference method to evaluate the variation between adjacent configurations. Explicitly, we compute the absolute deviation $|\Delta M_{\rm hair} - T \Delta S|$ and the corresponding relative error defined by
\begin{equation}
    \epsilon = \left| \frac{\Delta M_{\rm hair} - T \Delta S}{\Delta M_{\rm hair}} \right|.
\end{equation}
Our calculations reveal that the absolute deviations are remarkably small, staying within the order of $\mathcal{O}(10^{-7})$ to $\mathcal{O}(10^{-5})$. Hence, the relative error $\epsilon$ remains well below $1\%$ for the majority of the parameter space, typically fluctuating between $0.1\%$ and $0.8\%$. 
The excellent agreement between $dM_{\rm hair}$ and $T dS$ confirms that our data extraction method eliminates numerical artifacts efficiently. Furthermore, it proves that the scalar field is a secondary hair and its backreaction on the thermodynamic phase space has been accurately captured by the geometrically corrected Wald entropy.

Having established the validity of the first law of thermodynamics, we now turn to the thermodynamic stability and the phase transition. In the canonical ensemble where the temperature $T$ and the electric charge $Q$ are fixed parameters, the thermodynamic preference of the system is governed by the Gibbs free energy defined by
\begin{equation}
    G = M_{\rm hair} - T S.
\end{equation}
Here, within a fixed parameter slice, the thermodynamically preferred state is obtained  when  minimizing $G$.

\begin{figure}[H]
\centering
\subfigure[~GB$^+$: $M_{\rm RN}=0.7$, $Q=0.4$]
{\label{gtgb+}
\includegraphics[width=2.05in]{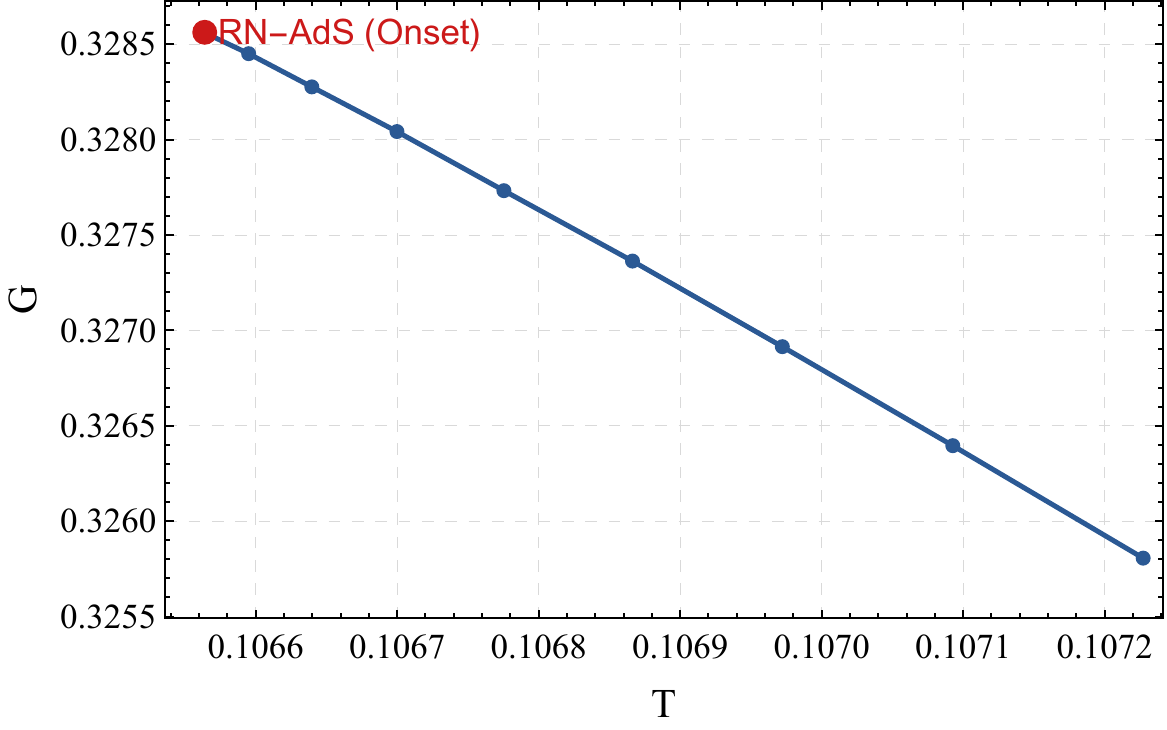}}
\hfill
\subfigure[~GB$^-$: $M_{\rm RN}=0.406$, $Q=0.4$]
{\label{gtgb-}
\includegraphics[width=2.05in]{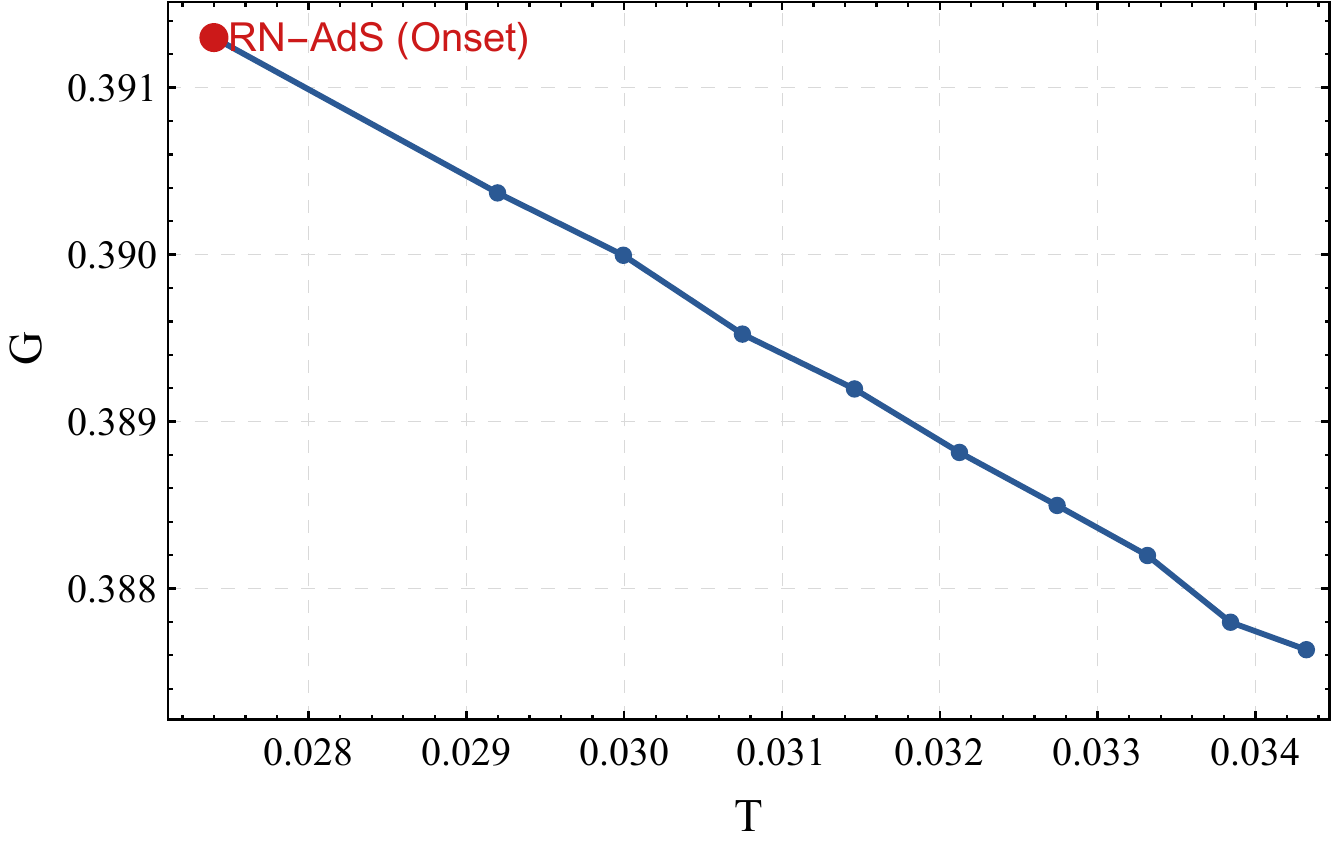}}
\hfill
\subfigure[~GB$^-$: $M_{\rm RN}=0.7$, $Q=0.66$]
{\label{gtgb-q066}
\includegraphics[width=2.05in]{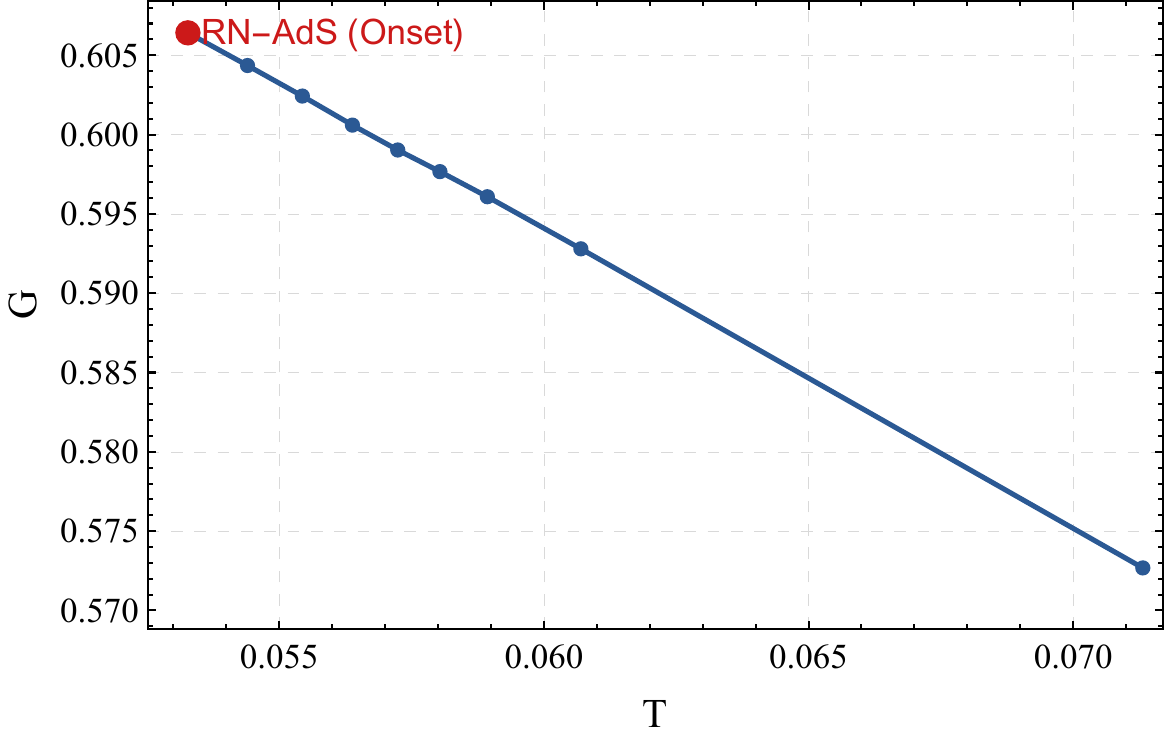}}

\caption{Gibbs free energy $G$ versus temperature $T$ diagram for the three fixed-charge cases listed in Table~\ref{tab:thermo_data_combined}. Panels (a) and (b) show the regular $Q=0.4$ sequences. Panel (c) shows the terminal regular sequence of the high charge GB$^-$ scalarization: the solution family ends near $\phi_0\simeq0.04$ because beyond this point the near-horizon initial data cannot remain real while the AdS-boundary condition is imposed. For panel (c), the curve is drawn by making use of  the denser verified data with $\Delta\phi_0=0.005$, while Table~\ref{tab:thermo_data_combined} lists the $\Delta\phi_0=0.01$ subset for a compact comparison. In all three displayed regular segments, the scalarized RN-AdS configurations lie below the corresponding  RN-AdS  configurations in free energy.}
\label{figgt0}
\end{figure}

In Fig.~\ref{figgt0}, utilizing the thermodynamic data from Table~\ref{tab:thermo_data_combined}, we present the Gibbs free energy $G$ as a function of the temperature $T$ for the scalarized AdS black holes. The three panels compare  GB$^+$ scalarization at $(M_{\rm RN},Q)=(0.7,0.4)$, the small charge  GB$^-$ scalarization at $(M_{\rm RN},Q)=(0.406,0.4)$, and the higher charge GB$^-$ scalarization at $(M_{\rm RN},Q)=(0.7,0.66)$. The red dots denote the corresponding RN-AdS reference points used for comparison.

As illustrated in Fig.~\ref{gtgb+}, the GB$^+$ scalarized configurations emerge from the RN-AdS onset at $(T, G) \approx (0.10656, 0.32856)$. As the horizon scalar $\phi_0$ increases, the hairy black hole evolves along a trajectory where the temperature $T$ increases and the free energy $G$ decreases. Within the scanned regular segment, these scalarized RN-AdS black holes possess lower free energy than the  RN-AdS black holes and they are therefore thermodynamically preferred in this parameter slice.

The small charge  GB$^-$ scalarization shown in Fig.~\ref{gtgb-} shows  a slightly different internal thermodynamic behavior. Along the sequence with $(M_{\rm RN},Q)=(0.406,0.4)$, the temperature increases monotonically, but $M_{\rm hair}$ and the product $TS$ first increase and then decrease, with a shallow turning point around $\phi_0=0.11$. This indicates  a physical feature of the small charge family of solutions. Since $\eta<0$ and $f(\phi_0)=\eta\phi_0^2/2$, the GB correction of $4\pi f(\phi_0)$ in the Wald entropy becomes increasingly negative as the scalar hair grows. For this small charge black hole, the mild increase of the horizon area $\pi r_+^2$ cannot compensate this negative GB contribution at larger $\phi_0$, so the entropy turns over because $T$ continues to increase  moderately, the product of  $TS$ also turns over. The asymptotic mass $M_{\rm hair}$ displays the same shallow nonmonotonicity, indicating that this fixed-charge sequence approaches a turning/end-point region. Nevertheless, the free energy of the regular scalarized configurations remains below the corresponding  RN-AdS value.

For the higher charge GB$^-$ case in Fig.~\ref{gtgb-q066}, the regular scalarized sequence is intrinsically short. As $\phi_0$ increases toward $\phi_0=0.04$, the required negative coupling grows rapidly in magnitude and the regular solution family reaches its endpoint beacuse the near-horizon expansion loses real-valued initial data and the AdS-boundary condition fails simultaneously. Therefore, the  panel (c) represents the completed terminal segment of this high charge scalarization, but not a truncated long sequence. Within this terminal segment, the scalarized RN-AdS configurations also lie below the  RN-AdS  family in $G$.

Finally, we analyzed the fixed-charge bulk thermodynamics of the regular scalarized solutions by using the Wald entropy, the asymptotic mass extraction, and the Gibbs free energy. This canonical ensemble analysis is in line with the charge-sensitive thermodynamic behavior found in related AdS scalarizations~\cite{Guo:2021zed}, while here the thermodynamic preference is established directly for the GB-coupled scalarized black holes. For  GB$^+$ and GB$^-$ cases studied here with $\Lambda=-0.5$, the scalarized RN-AdS configurations have lower free energy than the corresponding RN-AdS references. They connect smoothly to the RN-AdS onset without a jump in $G$ or swallowtail behavior, supporting a continuous second-order phase transition in these fixed-charge sectors.

\section{Conclusion}
\label{sec:conclusion}

In this work, we have  investigated  two different  scalarizations in the EMsGB theory with a negative cosmological constant by focusing on RN-AdS black hole. By combining analytical insights in the probe limit with full numerical solutions, we have  clarified both the dynamical origin and thermodynamic aspect of scalarized  AdS black holes.
In AdS spacetime, the tachyonic  instability is strictly constrained by the BF bound, leading to a restriction on number of branches. We demonstrated that the GB coupling $\eta$ provides the dominant destabilizing effect in the near-horizon. In the strong curvature regime, the cosmological constant $\Lambda$ may  promote the instability through higher-order $\Lambda^2$ contribution from GB term to the effective mass.

Combining  the tachyonic instability with the  BF bound, we divided coupling constant $\eta$ into two cases with $\Lambda=-0.5$:   $0<\eta<2.25$ (GB$+$ scalarization) and  $\eta<0$ (GB$^-$ scalarization). 
For $\eta>2.25$,  we could  not analyze its scalarization explicitly.

At the linearized level of GB$^+$ with $0<\eta<2.25$, scalarization mechanism is governed by a tachyonic instability encoded in the effective mass squared $\mu_{\text{eff}}^2$.  
Through the Schr\"odinger-like reformulation, we  demonstrated how the effective potential $V_{\text{eff}}(r)$ develops  negative wells in the near-horizon, enabling the formation of bound state scalars (scalar clouds) that signal the onset of scalarization. 
The numerical construction of these scalar clouds allowed us to  map the threshold curves into the parameter space as depicted in Fig.~\ref{fig-clouds1} and identify the number of branch points.
Because of the GB$^+$ condition ($\eta_{th}\le \eta<2.25$) arising from the absence of AdS-tachyonic instability and the presence of unstable region, we proved that there exists the single branch of scalarized AdS black holes (see Fig.~\ref{fig-phicheck}). 
However, it is interesting to  note that there exist  infinite branches of scalarized black holes through spontaneous scalarization triggered by tachyon in asymptotically flat spacetime~\cite{Doneva:2017bvd,Myung:2018vug}. 
Beyond the probe approximation, we constructed fully backreacted scalarized black hole solutions belonging to the single branch and thoroughly analyzed their physical and geometric properties, as shown in Fig.~\ref{fig-matric}.  The backreaction effect significantly modifies the near-horizon geometry, while strictly preserving the AdS asymptotics. 
 
On the other hand, for $\eta<0$, we found GB$^-$ scalarization, providing the single branch of scalarized AdS black holes as is shown in Fig.~\ref{fig-matric-GB-}. Because of the absence of  scalar clouds, these solutions were directly constructed. 

Finally, we have analyzed the fixed-charge  thermodynamics of the regular scalarized RN-AdS solutions by using the Wald entropy, the asymptotic mass extraction, and the Gibbs free energy. For the representative GB$^+$ and GB$^-$ cases with $\Lambda=-0.5$, the scalarized configurations have lower free energy than the corresponding RN-AdS references.  They connect smoothly to the RN-AdS onset without a jump in $G$ or swallowtail behavior, implying  a continuous second-order phase transition.  In addition, the high charge GB$^-$ case with $(M_{\rm RN},Q)=(0.7,0.66)$ exhibits a nearby endpoint of the regular scalarized sequence, while remaining a thermodynamically preferred state along the displayed segment.

\vspace{1cm}

{\bf Acknowledgments}

\vspace{1cm}

 F. W. Shu was supported by the
National Natural Science Foundation of China (Grant No. 12375049), Key Program of the Natural Science Foundation of Jiangxi Province under Grant No. 20232ACB201008, and the Ganpo High-Level Innovative Talent Program. D. C. Zou was  supported by the National Natural Science Foundation of China (NNSFC) (Grant No.12365009). Y. S. Myung was supported by the National Research Foundation of Korea (NRF) grant funded by the Korea government(MSIT) (RS-2022-NR069013). 
 
 \vspace{1cm}

\newpage 

\appendix
\section{Explicit forms of the expansion coefficients}
\label{app:expansion_coefficients}

In this appendix, we provide the explicit algebraic expressions for the remaining expansion coefficients introduced in the near-horizon expansion, Eq.~\eqref{eqexpandh}. As discussed in the main text, these quantities are completely determined by the independently specified horizon parameters $A_1$ and $\phi_0$. 

We  substitute   Eq.~\eqref{eqexpandh} into the equations of motion and solve them  order by order. To simplify the presentation of the expansion coefficients, it is convenient to  define  a quantity $K$, which is determined by the horizon radius $r_+$, the cosmological constant $\Lambda$, the charge parameter $Q$, the coupling constant $\eta$, and the scalar at the horizon $\phi_0$ as 
\begin{equation}
\label{eq:K_def}
\begin{aligned}
K =& \, r_+^8 + 8 r_+^4 \eta^2 \left(-3 + 2 Q^2 + 2 r_+^2 \Lambda\right) \phi_0^2 + 16 \eta^4 \left[Q^4 + r_+^2 \Lambda \left(-6 + r_+^2 \Lambda\right) + 2 Q^2 \left(3 + r_+^2 \Lambda\right)\right] \phi_0^4 \,.
\end{aligned}
\end{equation}
Using  $K$, the explicit forms of $B_1$, $V_1$, $\phi_1$, $A_2$, $B_2$, $V_2$, and $\phi_2$ can be expressed as follows:

\begin{widetext}
\begin{eqnarray}
B_1 &=& -\frac{-r_+^4 + \sqrt{K} + 4 Q^2 \eta^2 \phi_0^2 + 4 r_+^2 \eta^2 \Lambda \phi_0^2}{12 \eta^2 \phi_0^2} \,, \label{eq:B1} \\[12pt]
\phi_1 &=& -\frac{r_+^2}{4 \eta \phi_0 \left[ (-1 + Q^2) r_+^4 + r_+^6 \Lambda + 4 Q^2 \eta^2 \phi_0^2 - 4 r_+^2 \eta^2 \Lambda \phi_0^2 \right]} \nonumber \\
&& \times \Bigg[ r_+^6 \Lambda + (1 - Q^2) \sqrt{K} + 4 Q^2 \eta^2 \phi_0^2 (1 + Q^2) \nonumber \\
&& \qquad\quad - r_+^2 \Lambda \left(\sqrt{K} + 4 (3 - 2 Q^2) \eta^2 \phi_0^2\right) + r_+^4 \left(-1 + Q^2 + 4 \eta^2 \Lambda^2 \phi_0^2\right) \Bigg] \,, \label{eq:phi1} \\[12pt]
V_1 &=& -\frac{\sqrt{-Q^2 A_1 \left(r_+^4 + \sqrt{K} - 4 Q^2 \eta^2 \phi_0^2 - 4 r_+^2 \eta^2 \Lambda \phi_0^2\right)}}{\sqrt{2} \sqrt{(-1 + Q^2) r_+^4 + r_+^6 \Lambda + 4 Q^2 \eta^2 \phi_0^2 - 4 r_+^2 \eta^2 \Lambda \phi_0^2}} \,. \label{eq:V1}
\end{eqnarray}

\begin{eqnarray}
A_2 &=& -\frac{A_1}{48 \left[ r_+^4 \eta \phi_0 (-1 + Q^2 + r_+^2 \Lambda) + 4 \eta^3 \phi_0^3 (Q^2 - r_+^2 \Lambda) \right]^2 \sqrt{K}} \nonumber \\
&& \times \Bigg[ -9 r_+^{20} \Lambda^2 + r_+^{14} \Big( -9 (-1 + Q^2)^2 \eta + 18 \sqrt{K} (-1 + Q^2) \Lambda + 9 \sqrt{K} \eta \Lambda^2 + 36 (11 - 9 Q^2) \eta^3 \Lambda^2 \phi_0^2 \nonumber \\
&& \qquad\quad + 4 \eta^2 \Lambda \left(-113 + 197 Q^2 - 96 Q^4 + 14 \sqrt{K} \Lambda^2\right) \phi_0^2 + 40 (33 - 16 Q^2) \eta^4 \Lambda^3 \phi_0^4 - 144 \eta^5 \Lambda^4 \phi_0^4 \Big) \nonumber \\
&& \qquad + r_+^{10} \eta \Big( 9 \sqrt{K} (-1 + Q^2)^2 + 4 \sqrt{K} \left(5 - 23 Q^2 + 42 Q^4\right) \eta \Lambda \phi_0^2 \nonumber \\
&& \qquad\quad - 36 \eta^2 \left(-3 + 4 \sqrt{K} \Lambda^2 + Q^2 \left(7 - 7 Q^2 + 3 Q^4 - 3 \sqrt{K} \Lambda^2\right)\right) \phi_0^2 \nonumber \\
&& \qquad\quad - 144 \left(11 - 11 Q^2 + 6 Q^4\right) \eta^4 \Lambda^2 \phi_0^4 \nonumber \\
&& \qquad\quad - 8 \eta^3 \Lambda \left(-111 + 70 Q^2 - 7 Q^4 + 80 Q^6 + 35 \sqrt{K} \Lambda^2\right) \phi_0^4 + 32 \left(-47 + 44 Q^2\right) \eta^5 \Lambda^3 \phi_0^6 \Big) \nonumber \\
&& \qquad + 2 r_+^8 \eta^2 \phi_0^2 \Big( \sqrt{K} \left(33 - 86 Q^2 + 25 Q^4 + 28 Q^6\right) + 18 \sqrt{K} \left(3 - 4 Q^2 + 3 Q^4\right) \eta \Lambda \nonumber \\
&& \qquad\quad - 72 \left(-3 - 2 Q^2 - Q^4 + 4 Q^6\right) \eta^3 \Lambda \phi_0^2 \nonumber \\
&& \qquad\quad - 4 \eta^2 \left(9 - 8 \sqrt{K} \Lambda^2 + Q^2 \left(57 - 159 Q^2 + 79 Q^4 + 20 Q^6 + 41 \sqrt{K} \Lambda^2\right)\right) \phi_0^2 \nonumber \\
&& \qquad\quad + 48 \left(1 - 5 Q^2 + 6 Q^4\right) \eta^4 \Lambda^2 \phi_0^4 + 256 \eta^6 \Lambda^4 \phi_0^6 \Big) \nonumber \\
&& \qquad + r_+^{12} \Big( 9 \sqrt{K} (-1 + Q^2)^2 + 18 \sqrt{K} (-1 + Q^2) \eta \Lambda + 36 \eta^3 \Lambda \left(-11 + 18 Q^2 - 9 Q^4 + \sqrt{K} \Lambda^2\right) \phi_0^2 \nonumber \\
&& \qquad\quad - 2 \eta^2 \left(-69 + 71 \sqrt{K} \Lambda^2 + Q^2 \left(178 - 173 Q^2 + 64 Q^4 - 84 \sqrt{K} \Lambda^2\right)\right) \phi_0^2 \nonumber \\
&& \qquad\quad - 8 \left(253 - 251 Q^2 + 120 Q^4\right) \eta^4 \Lambda^2 \phi_0^4 + 144 (7 - 4 Q^2) \eta^5 \Lambda^3 \phi_0^4 + 608 \eta^6 \Lambda^4 \phi_0^6 \Big) \nonumber \\
&& \qquad - 128 Q^2 r_+^2 \eta^6 \Lambda \phi_0^6 \left(12 \sqrt{K} + Q^2 \left(\sqrt{K} + 126 \eta^2 \phi_0^2\right)\right) \nonumber \\
&& \qquad + 64 Q^4 \eta^6 \phi_0^6 \left(9 \sqrt{K} + 2 Q^2 \left(\sqrt{K} + 2 \left(27 + 2 Q^2\right) \eta^2 \phi_0^2\right)\right) \nonumber \\
&& \qquad- r_+^{18} \Lambda \left(-18 + 18 Q^2 + \eta \Lambda \left(9 + 128 \eta \Lambda \phi_0^2\right)\right) \nonumber \\
&& \qquad + 4 r_+^6 \eta^3 \phi_0^2 \Big( -9 \sqrt{K} Q^2 + 9 \sqrt{K} Q^6 + 114 \sqrt{K} \eta \Lambda \phi_0^2 \nonumber \\
&& \qquad\quad - 120 \sqrt{K} Q^2 \eta \Lambda \phi_0^2 + 46 \sqrt{K} Q^4 \eta \Lambda \phi_0^2 \nonumber \\
&& \qquad\quad + 32 \sqrt{K} \eta^3 \Lambda^3 \phi_0^4 + 4 \eta^2 \phi_0^2 \big( -9 Q^2 \left(3 - 5 Q^2 + 3 Q^4 + Q^6\right) \nonumber \\
&& \qquad\quad - 2 (-3 + Q^2) \left(-3 + 13 Q^2 + 20 Q^4\right) \eta \Lambda \phi_0^2 - 336 \eta^3 \Lambda^3 \phi_0^4 \big) \Big) \nonumber \\
&& \qquad - 8 r_+^4 \eta^4 \phi_0^4 \Big( 52 Q^8 \eta^2 \phi_0^2 - 120 \sqrt{K} \eta^2 \Lambda^2 \phi_0^2 - Q^6 \left(29 \sqrt{K} + 60 \eta^2 \phi_0^2\right) \nonumber \\
&& \qquad\quad + 4 Q^4 \left(-13 \sqrt{K} + 57 \eta^2 \phi_0^2 + 32 \eta^4 \Lambda^2 \phi_0^4\right) \nonumber \\
&& \qquad\quad + Q^2 \left(51 \sqrt{K} - 4 \eta^2 \phi_0^2 \left(9 - 4 \Lambda^2 \left(\sqrt{K} - 114 \eta^2 \phi_0^2\right)\right)\right) \Big) \nonumber \\
&& \qquad - r_+^{16} \Big( 9 + 9 Q^4 - 18 \eta \Lambda + 6 Q^2 \big( -3 + \eta \Lambda (3 + 64 \eta \Lambda \phi_0^2) \big) \nonumber \\
&&\qquad\quad + \Lambda^2 \big(-9 \sqrt{K} + 2 \eta^2 \phi_0^2 ( -221 + 2 \eta \Lambda (27 + 40 \eta \Lambda \phi_0^2) ) \big) \Big) \Bigg] \,, \label{eq:A2}
\end{eqnarray}

\begin{eqnarray}
B_2 &=& \frac{1}{288 \eta^4 \phi_0^4 \left[ r_+^4 (-1 + Q^2 + r_+^2 \Lambda) + 4 \eta^2 (Q^2 - r_+^2 \Lambda) \phi_0^2 \right] \sqrt{K}} \nonumber \\
&& \times \Bigg[ -11 r_+^{18} \Lambda + 128 Q^4 \eta^6 \phi_0^6 \left(\sqrt{K} (-3 + Q^2) + 2 (63 + 2 Q^4) \eta^2 \phi_0^2\right) \nonumber \\
&& \qquad + 64 Q^2 r_+^2 \eta^6 \Lambda \phi_0^6 \left(\sqrt{K} (15 + 2 Q^2) + 4 \left(-72 + 3 Q^2 + 4 Q^4\right) \eta^2 \phi_0^2\right) \nonumber \\
&& \qquad + r_+^{10} \eta \Big( 27 \sqrt{K} (-1 + Q^2) + 2 \sqrt{K} (-59 + 62 Q^2) \eta \Lambda \phi_0^2 \nonumber \\
&& \qquad\quad - 54 \eta^2 \left(9 + 7 Q^2 (-2 + Q^2) - 3 \sqrt{K} \Lambda^2\right) \phi_0^2 + 216 (16 - 9 Q^2) \eta^4 \Lambda^2 \phi_0^4 \nonumber \\
&& \qquad\quad + 8 \eta^3 \Lambda \left(-147 + 177 Q^2 - 99 Q^4 + 4 \sqrt{K} \Lambda^2\right) \phi_0^4 + 16 (97 + 32 Q^2) \eta^5 \Lambda^3 \phi_0^6 \Big) \nonumber \\
&& \qquad + 2 r_+^8 \eta^2 \phi_0^2 \Big( \sqrt{K} \left(6 - 15 Q^2 + 31 Q^4\right) + 54 \sqrt{K} (-4 + 3 Q^2) \eta \Lambda \nonumber \\
&& \qquad\quad - 324 \left(5 - 4 Q^2 + 3 Q^4\right) \eta^3 \Lambda \phi_0^2 \nonumber \\
&& \qquad\quad - 2 \eta^2 \left(-9 + 61 \sqrt{K} \Lambda^2 + 3 Q^2 \left(-18 + Q^2 + 22 Q^4 - 8 \sqrt{K} \Lambda^2\right)\right) \phi_0^2 \nonumber \\
&& \qquad\quad + 24 \left(-90 + 49 Q^2 + 16 Q^4\right) \eta^4 \Lambda^2 \phi_0^4 - 256 \eta^6 \Lambda^4 \phi_0^6 \Big) \nonumber \\
&& \qquad + r_+^{12} \Big( 11 \sqrt{K} (-1 + Q^2) + 27 \sqrt{K} \eta \Lambda + 108 (9 - 7 Q^2) \eta^3 \Lambda \phi_0^2 \nonumber \\
&& \qquad\quad + 2 \eta^2 \left(-72 + 125 Q^2 - 75 Q^4 + 31 \sqrt{K} \Lambda^2\right) \phi_0^2 \nonumber \\
&& \qquad\quad + 12 (119 - 66 Q^2) \eta^4 \Lambda^2 \phi_0^4 - 648 \eta^5 \Lambda^3 \phi_0^4 + 128 \eta^6 \Lambda^4 \phi_0^6 \Big) \nonumber \\
&& \qquad + 4 r_+^4 \eta^4 \phi_0^4 \Big( -27 \sqrt{K} + 66 \sqrt{K} Q^2 + 32 Q^8 \eta^2 \phi_0^2 - 4 Q^2 \eta^2 \left(261 + 8 \sqrt{K} \Lambda^2\right) \phi_0^2 \nonumber \\
&& \qquad\quad + 4 Q^6 \left(2 \sqrt{K} - 47 \eta^2 \phi_0^2\right) - 144 \eta^2 \Lambda^2 \phi_0^2 \left(\sqrt{K} - 4 \eta^2 \phi_0^2\right) + 3 Q^4 \left(\sqrt{K} + 408 \eta^2 \phi_0^2\right) \Big) \nonumber \\
&& \qquad - r_+^{16} \left(-11 + 11 Q^2 + 3 \eta \Lambda \left(9 + 50 \eta \Lambda \phi_0^2\right)\right) \nonumber \\
&& \qquad - 2 r_+^6 \eta^3 \phi_0^2 \Big( -81 \sqrt{K} - 60 \sqrt{K} \eta \Lambda \phi_0^2 + 4 Q^6 \eta^2 \phi_0^2 \left(81 - 64 \eta \Lambda \phi_0^2\right) \nonumber \\
&& \qquad\quad + 8 \eta^3 \Lambda \phi_0^4 \left(-99 + 8 \Lambda^2 \left(\sqrt{K} + 6 \eta^2 \phi_0^2\right)\right) \nonumber \\
&& \qquad\quad - 3 Q^4 \left(27 \sqrt{K} + 8 \eta \phi_0^2 \left(-18 \eta + 2 \sqrt{K} \Lambda + \eta^2 \Lambda \phi_0^2\right)\right) \nonumber \\
&& \qquad\quad + 4 Q^2 \left(27 \sqrt{K} + \eta \phi_0^2 \left(-243 \eta + 29 \sqrt{K} \Lambda - 312 \eta^2 \Lambda \phi_0^2 + 128 \eta^4 \Lambda^3 \phi_0^4\right)\right) \Big) \nonumber \\
&& \qquad + r_+^{14} \Big( 11 \sqrt{K} \Lambda + \eta \Big(27 - 3 Q^2 \left(9 + 100 \eta \Lambda \phi_0^2\right) \nonumber \\
&& \qquad\quad - 2 \eta \Lambda \phi_0^2 \left(-169 + 3 \eta \Lambda \left(63 + 44 \eta \Lambda \phi_0^2\right)\right)\Big)\Big) \Bigg] \,. \label{eq:B2}
\end{eqnarray}

\begin{eqnarray}
V_2 &=& -\frac{Q^2 A_1}{2 \sqrt{2K} \left[ r_+^4 (-1 + Q^2 + r_+^2 \Lambda) + 4 \eta^2 (Q^2 - r_+^2 \Lambda) \phi_0^2 \right]^{3/2}} \nonumber \\
&& \times \frac{1}{\left(-\sqrt{K} + r_+^4 - 4 Q^2 \eta^2 \phi_0^2 - 4 r_+^2 \eta^2 \Lambda \phi_0^2\right) \sqrt{-Q^2 A_1 \left(\sqrt{K} + r_+^4 - 4 Q^2 \eta^2 \phi_0^2 - 4 r_+^2 \eta^2 \Lambda \phi_0^2\right)}} \nonumber \\
&& \times \Bigg[ 5 r_+^{20} \Lambda^2 + 6912 Q^4 r_+^2 \eta^8 \Lambda \phi_0^8 + 192 Q^4 \eta^6 \phi_0^6 \left(\sqrt{K} - 20 Q^2 \eta^2 \phi_0^2\right) \nonumber \\
&& \qquad - 2 r_+^8 \eta^2 \phi_0^2 \Big( \sqrt{K} \left(9 - 26 Q^2 + 5 Q^4 + 12 Q^6\right) + 18 \sqrt{K} \left(3 - 4 Q^2 + 3 Q^4\right) \eta \Lambda \nonumber \\
&& \qquad\quad - 72 \left(-3 - 2 Q^2 - Q^4 + 4 Q^6\right) \eta^3 \Lambda \phi_0^2 \nonumber \\
&& \qquad\quad - 4 \eta^2 \left(9 - 27 Q^4 + 31 Q^6 + 12 Q^8 - 16 \sqrt{K} \Lambda^2 + 3 Q^2 \left(-5 + 7 \sqrt{K} \Lambda^2\right)\right) \phi_0^2 \nonumber \\
&& \qquad\quad + 16 \left(3 - 19 Q^2 + 18 Q^4\right) \eta^4 \Lambda^2 \phi_0^4 \Big) \nonumber \\
&& \qquad + r_+^{14} \Big( -10 \sqrt{K} (-1 + Q^2) \Lambda + 9 \eta \left((-1 + Q^2)^2 - \sqrt{K} \Lambda^2\right) + 36 (-11 + 9 Q^2) \eta^3 \Lambda^2 \phi_0^2 \nonumber \\
&& \qquad\quad - 4 \eta^2 \Lambda \left(-47 + 85 Q^2 - 48 Q^4 + 6 \sqrt{K} \Lambda^2\right) \phi_0^2 \nonumber \\
&& \qquad\quad + 8 (-85 + 48 Q^2) \eta^4 \Lambda^3 \phi_0^4 + 144 \eta^5 \Lambda^4 \phi_0^4 \Big) \nonumber \\
&& \qquad + r_+^{10} \eta \Big( -9 \sqrt{K} (-1 + Q^2)^2 - 4 \sqrt{K} \left(7 - 15 Q^2 + 18 Q^4\right) \eta \Lambda \phi_0^2 \nonumber \\
&& \qquad\quad + 36 \eta^2 \left(-3 + 4 \sqrt{K} \Lambda^2 + Q^2 \left(7 - 7 Q^2 + 3 Q^4 - 3 \sqrt{K} \Lambda^2\right)\right) \phi_0^2 \nonumber \\
&& \qquad\quad + 144 \left(11 - 11 Q^2 + 6 Q^4\right) \eta^4 \Lambda^2 \phi_0^4 \nonumber \\
&& \qquad\quad + 8 \eta^3 \Lambda \left(-39 + 46 Q^2 - 23 Q^4 + 48 Q^6 + 15 \sqrt{K} \Lambda^2\right) \phi_0^4 + 32 (35 - 36 Q^2) \eta^5 \Lambda^3 \phi_0^6 \Big) \nonumber \\
&& \qquad + r_+^{12} \Big( -5 \sqrt{K} (-1 + Q^2)^2 - 18 \sqrt{K} (-1 + Q^2) \eta \Lambda - 36 \eta^3 \Lambda \left(-11 + 18 Q^2 - 9 Q^4 + \sqrt{K} \Lambda^2\right) \phi_0^2 \nonumber \\
&& \qquad\quad + 2 \eta^2 \left(Q^2 \left(54 - 65 Q^2 + 32 Q^4 - 36 \sqrt{K} \Lambda^2\right) + 7 \left(-3 + 5 \sqrt{K} \Lambda^2\right)\right) \phi_0^2 \nonumber \\
&& \qquad\quad + 8 \left(113 - 139 Q^2 + 72 Q^4\right) \eta^4 \Lambda^2 \phi_0^4 + 144 (-7 + 4 Q^2) \eta^5 \Lambda^3 \phi_0^4 - 480 \eta^6 \Lambda^4 \phi_0^6 \Big) \nonumber \\
&& \qquad + r_+^{18} \Lambda \left(-10 + 10 Q^2 + \eta \Lambda \left(9 + 64 \eta \Lambda \phi_0^2\right)\right) \nonumber \\
&& \qquad - 8 r_+^4 \eta^4 \phi_0^4 \Big( -36 Q^8 \eta^2 \phi_0^2 + 24 \sqrt{K} \eta^2 \Lambda^2 \phi_0^2 + 4 Q^4 \left(\sqrt{K} - 57 \eta^2 \phi_0^2\right) \nonumber \\
&& \qquad\quad + Q^6 \left(9 \sqrt{K} + 76 \eta^2 \phi_0^2\right) + Q^2 \left(-3 \sqrt{K} + 36 \eta^2 \phi_0^2 + 672 \eta^4 \Lambda^2 \phi_0^4\right) \Big) \nonumber \\
&& \qquad + 4 r_+^6 \eta^3 \phi_0^2 \Big( 9 \sqrt{K} Q^2 - 9 \sqrt{K} Q^6 - 18 \sqrt{K} \eta \Lambda \phi_0^2 + 40 \sqrt{K} Q^2 \eta \Lambda \phi_0^2 - 6 \sqrt{K} Q^4 \eta \Lambda \phi_0^2 \nonumber \\
&& \qquad\quad + 4 \eta^2 \phi_0^2 \left(9 Q^2 \left(3 - 5 Q^2 + 3 Q^4 + Q^6\right) + 2 \left(9 - 42 Q^2 - 35 Q^4 + 12 Q^6\right) \eta \Lambda \phi_0^2 + 144 \eta^3 \Lambda^3 \phi_0^4\right) \Big) \nonumber \\
&& \qquad + r_+^{16} \Big( 5 + 5 Q^4 - 18 \eta \Lambda + 2 Q^2 \left(-5 + 3 \eta \Lambda \left(3 + 32 \eta \Lambda \phi_0^2\right)\right) \nonumber \\
&& \qquad\quad + \Lambda^2 \left(-5 \sqrt{K} + 6 \eta^2 \phi_0^2 \left(-35 + 2 \eta \Lambda \left(9 + 8 \eta \Lambda \phi_0^2\right)\right)\right) \Big) \Bigg] \,. \label{eq:V2}
\end{eqnarray}

\begin{eqnarray}
\phi_2 &=& \frac{1}{16 \eta^3 \phi_0^3 \left(\sqrt{K} - r_+^4 + 4 Q^2 \eta^2 \phi_0^2 + 4 r_+^2 \eta^2 \Lambda \phi_0^2\right) \left[ r_+^4 (-1 + Q^2 + r_+^2 \Lambda) + 4 \eta^2 (Q^2 - r_+^2 \Lambda) \phi_0^2 \right]^2 \sqrt{K}} \nonumber \\
&& \times \Bigg[ r_+^2 \bigg( -r_+^{24} \Lambda^2 - 4 r_+^8 \eta^4 \phi_0^4 \Big( \sqrt{K} \left(15 - 49 Q^2 + 63 Q^4 - 39 Q^6 + 6 Q^8\right) \nonumber \\
&& \qquad\quad + 2 \sqrt{K} \left(21 + 26 Q^2 - 43 Q^4 + 12 Q^6\right) \eta \Lambda \nonumber \\
&& \qquad\quad + 8 \eta^3 \Lambda \left(45 - 156 Q^4 + 82 Q^6 - 15 Q^8 - 32 \sqrt{K} \Lambda^2 + 6 Q^2 \left(19 + 2 \sqrt{K} \Lambda^2\right)\right) \phi_0^2 \nonumber \\
&& \qquad\quad - 4 \eta^2 \left(-9 + 65 \sqrt{K} \Lambda^2 + 3 Q^2 \left(-16 - 12 Q^4 - 13 Q^6 + 2 Q^8 - 19 \sqrt{K} \Lambda^2 + 6 Q^2 \left(7 + \sqrt{K} \Lambda^2\right)\right)\right) \phi_0^2 \nonumber \\
&& \qquad\quad + 32 \left(-60 + 45 Q^2 - 41 Q^4 + 9 Q^6\right) \eta^4 \Lambda^2 \phi_0^4 - 192 \left(17 - 4 Q^2 + 2 Q^4\right) \eta^5 \Lambda^3 \phi_0^4 - 640 \eta^6 \Lambda^4 \phi_0^6 \Big) \bigg) \nonumber \\
&& \qquad - 2 r_+^{10} \eta^3 \phi_0^2 \Big( \sqrt{K} \left(9 - 15 Q^2 + 11 Q^4 - 5 Q^6\right) + 2 \sqrt{K} (-1 + Q) (1 + Q) \left(37 - 65 Q^2 + 24 Q^4\right) \eta \Lambda \phi_0^2 \nonumber \\
&& \qquad\quad + 4 \eta^2 \left(9 - 35 \sqrt{K} \Lambda^2 + Q^2 \left(36 - 11 \sqrt{K} \Lambda^2 + Q^2 \left(-78 + 36 Q^2 + Q^4 + 18 \sqrt{K} \Lambda^2\right)\right)\right) \phi_0^2 \nonumber \\
&& \qquad\quad + 32 \eta^4 \Lambda^2 \left(-78 + 5 Q^2 + 27 Q^4 - 15 Q^6 + 3 \sqrt{K} \Lambda^2\right) \phi_0^4 \nonumber \\
&& \qquad\quad - 8 \eta^3 \Lambda \left(120 - 31 \sqrt{K} \Lambda^2 + 2 Q^2 \left(-116 + 94 Q^2 - 68 Q^4 + 15 Q^6 + 12 \sqrt{K} \Lambda^2\right)\right) \phi_0^4 \nonumber \\
&& \qquad\quad + 64 \left(40 - 49 Q^2 + 21 Q^4\right) \eta^5 \Lambda^3 \phi_0^6 + 128 (26 - 9 Q^2) \eta^6 \Lambda^4 \phi_0^6 \Big) \nonumber \\
&& \qquad + 2 r_+^{18} \Big( -(-1 + Q^2)^2 \eta + \sqrt{K} (-1 + Q^2) \Lambda + \sqrt{K} \eta \Lambda^2 + \left(55 - 39 Q^2\right) \eta^3 \Lambda^2 \phi_0^2 \nonumber \\
&& \qquad\quad + 2 \eta^2 \Lambda \left(-11 + 18 Q^2 - 9 Q^4 + \sqrt{K} \Lambda^2\right) \phi_0^2 + 2 \left(9 + 16 Q^2\right) \eta^4 \Lambda^3 \phi_0^4 - 4 \eta^5 \Lambda^4 \phi_0^4 + 48 \eta^6 \Lambda^5 \phi_0^6 \Big) \nonumber \\
&& \qquad + r_+^{16} \Big( \sqrt{K} (-1 + Q^2)^2 + 4 \sqrt{K} (-1 + Q^2) \eta \Lambda + 2 \eta^3 \Lambda \left(-63 + 94 Q^2 - 39 Q^4 + 5 \sqrt{K} \Lambda^2\right) \phi_0^2 \nonumber \\
&& \qquad\quad - 4 \eta^2 \left(-3 + 4 \sqrt{K} \Lambda^2 + Q^2 \left(7 - 7 Q^2 + 3 Q^4 - 3 \sqrt{K} \Lambda^2\right)\right) \phi_0^2 \nonumber \\
&& \qquad\quad - 32 \left(-13 + Q^2\right) \eta^5 \Lambda^3 \phi_0^4 - 4 \eta^4 \Lambda^2 \left(1 + 57 Q^2 - 24 Q^4 + 6 \sqrt{K} \Lambda^2\right) \phi_0^4 \nonumber \\
&& \qquad\quad + 16 \left(-19 + 30 Q^2\right) \eta^6 \Lambda^4 \phi_0^6 + 96 \eta^7 \Lambda^5 \phi_0^6 \Big) \nonumber \\
&& \qquad + 2 r_+^{12} \eta^2 \phi_0^2 \Big( 2 \sqrt{K} Q^2 \left(-1 + Q^4\right) + \sqrt{K} \left(31 - 38 Q^2 + 15 Q^4\right) \eta \Lambda \nonumber \\
&& \qquad\quad - 4 \eta^3 \Lambda \left(-84 + 7 \sqrt{K} \Lambda^2 + 4 Q^2 \left(9 + 5 Q^2 + Q^4 + 3 \sqrt{K} \Lambda^2\right)\right) \phi_0^2 \nonumber \\
&& \qquad\quad + 2 \eta^2 \left(33 - 23 \sqrt{K} \Lambda^2 + Q^2 \left(-133 + 61 \sqrt{K} \Lambda^2 + Q^2 \left(167 - 75 Q^2 + 4 Q^4 - 36 \sqrt{K} \Lambda^2\right)\right)\right) \phi_0^2 \nonumber \\
&& \qquad\quad + 160 \left(-12 + Q^2 + 3 Q^4\right) \eta^5 \Lambda^3 \phi_0^4 + 8 \eta^4 \Lambda^2 \left(-142 + 268 Q^2 - 174 Q^4 + 60 Q^6 + 9 \sqrt{K} \Lambda^2\right) \phi_0^4 \nonumber \\
&& \qquad\quad + 32 \left(14 - 33 Q^2\right) \eta^6 \Lambda^4 \phi_0^6 + 384 \eta^7 \Lambda^5 \phi_0^6 \Big) \nonumber \\
&& \qquad - 2 r_+^{14} \eta \Big( -\sqrt{K} (-1 + Q^2)^2 - 2 \sqrt{K} \left(3 - 4 Q^2 + 3 Q^4\right) \eta \Lambda \phi_0^2 \nonumber \\
&& \qquad\quad + \eta^2 \left(Q^2 \left(47 + 13 Q^2 \left(-3 + Q^2\right) - 15 \sqrt{K} \Lambda^2\right) + 3 \left(-7 + 9 \sqrt{K} \Lambda^2\right)\right) \phi_0^2 \nonumber
\end{eqnarray}

\begin{eqnarray}
&& \qquad\quad + 4 \eta^4 \Lambda^2 \left(130 - 68 Q^2 + 6 Q^4 + 3 \sqrt{K} \Lambda^2\right) \phi_0^4 \nonumber \\
&& \qquad\quad + 2 \eta^3 \Lambda \left(49 + 141 Q^4 - 16 Q^6 - 11 \sqrt{K} \Lambda^2 + 2 Q^2 \left(-91 + 12 \sqrt{K} \Lambda^2\right)\right) \phi_0^4 \nonumber\\
&& \qquad\quad - 32 \left(11 - 24 Q^2 + 15 Q^4\right) \eta^5 \Lambda^3 \phi_0^6 - 16 \left(14 + 15 Q^2\right) \eta^6 \Lambda^4 \phi_0^6 + 288 \eta^7 \Lambda^5 \phi_0^8 \Big) \nonumber \\
&& \qquad - 128 Q^4 \eta^8 \phi_0^8 \left(15 \sqrt{K} + Q^2 \left(-11 \sqrt{K} + 4 \left(9 + 19 Q^2\right) \eta^2 \phi_0^2\right)\right) - 2 r_+^{22} \Lambda \left(-1 + Q^2 + \eta \Lambda \left(1 + 6 \eta \Lambda \phi_0^2\right)\right) \nonumber \\
&& \qquad - 64 Q^2 r_+^2 \eta^7 \phi_0^6 \Big( 12 Q^8 \eta^2 \phi_0^2 - 48 \sqrt{K} \eta \Lambda \phi_0^2 + Q^6 \left(-3 \sqrt{K} + 56 \eta^2 \phi_0^2\right) \nonumber \\
&& \qquad\quad - 4 Q^4 \left(\sqrt{K} + 3 \eta^2 \phi_0^2 \left(7 + 8 \eta \Lambda \phi_0^2\right)\right) + Q^2 \left(9 \sqrt{K} + 2 \eta \phi_0^2 \left(36 \eta + 5 \sqrt{K} \Lambda - 180 \eta^2 \Lambda \phi_0^2\right)\right) \Big) \nonumber \\
&& \qquad - 8 r_+^6 \eta^5 \phi_0^4 \Big( -12 Q^{10} \eta^2 \phi_0^2 + Q^8 \left(3 \sqrt{K} + 8 \eta^2 \phi_0^2 \left(16 - 3 \eta \Lambda \phi_0^2\right)\right) \nonumber \\
&& \qquad\quad + Q^4 \left(45 \sqrt{K} - 2 \eta \phi_0^2 \left(-108 \eta + 19 \sqrt{K} \Lambda + 752 \eta^2 \Lambda \phi_0^2 + 640 \eta^3 \Lambda^2 \phi_0^2\right)\right) \nonumber \\
&& \qquad\quad + Q^6 \left(-25 \sqrt{K} + 8 \eta^2 \phi_0^2 \left(-23 + 2 \eta \Lambda \left(37 + 12 \eta \Lambda\right) \phi_0^2\right)\right) \nonumber \\
&& \qquad\quad + Q^2 \left(-27 \sqrt{K} + 4 \eta \phi_0^2 \left(-45 \eta - 3 \sqrt{K} \Lambda + 8 \sqrt{K} \eta \Lambda^2 + 348 \eta^2 \Lambda \phi_0^2 + 552 \eta^3 \Lambda^2 \phi_0^2 + 192 \eta^4 \Lambda^3 \phi_0^4\right)\right) \nonumber \\
&& \qquad\quad + 2 \eta \Lambda \phi_0^2 \left(39 \sqrt{K} + 60 \sqrt{K} \eta \Lambda - 40 \sqrt{K} \eta^2 \Lambda^2 \phi_0^2 + 12 \eta^2 \phi_0^2 \left(3 + 8 \eta \Lambda \left(3 - 5 \eta \Lambda \phi_0^2\right)\right)\right) \Big) \nonumber \\
&& \qquad + r_+^{20} \Big( -1 - Q^4 + 4 \eta \Lambda + Q^2 \left(2 - 4 \eta \Lambda \left(1 + 9 \eta \Lambda \phi_0^2\right)\right) \nonumber \\
&& \qquad+ \Lambda^2 \left(\sqrt{K} + 2 \eta^2 \phi_0^2 \left(22 + \eta \Lambda \left(-13 + 8 \eta \Lambda \phi_0^2\right)\right)\right) \Big) \nonumber \\
&& \qquad + 16 r_+^4 \eta^6 \phi_0^6 \Big( 12 Q^{10} \eta^2 \phi_0^2 - 72 \sqrt{K} \eta^2 \Lambda^2 \phi_0^2 + 3 Q^8 \left(-\sqrt{K} - 48 \eta^2 (2 + \eta \Lambda) \phi_0^2\right) \nonumber \\
&& \qquad\quad + 3 Q^6 \left(15 \sqrt{K} + 8 \sqrt{K} \eta \Lambda + 8 \eta^2 \left(1 + 8 \eta \Lambda\right) \phi_0^2\right) \nonumber \\
&& \qquad\quad + Q^2 \left(45 \sqrt{K} + 96 \sqrt{K} \eta \Lambda - 88 \sqrt{K} \eta^2 \Lambda^2 \phi_0^2 + 12 \eta^2 \phi_0^2 \left(3 + 8 \eta \Lambda \left(6 - 17 \eta \Lambda \phi_0^2\right)\right)\right) \nonumber \\
&& \qquad\quad + Q^4 \left(-83 \sqrt{K} - 64 \sqrt{K} \eta \Lambda + 8 \eta^2 \phi_0^2 \left(21 + 2 \eta \Lambda \left(-3 + 28 \eta \Lambda \phi_0^2\right)\right)\right) \Big) \Bigg] \,. \label{eq:phi2}
\end{eqnarray}

\end{widetext}

\bibliography{GR}

@article{Clifton:2011jh,
    author = "Clifton, Timothy and Ferreira, Pedro G. and Padilla, Antonio and Skordis, Constantinos",
    title = "{Modified Gravity and Cosmology}",
    eprint = "1106.2476",
    archivePrefix = "arXiv",
    primaryClass = "astro-ph.CO",
    doi = "10.1016/j.physrep.2012.01.001",
    journal = "Phys. Rept.",
    volume = "513",
    pages = "1--189",
    year = "2012"
}

@article{Sotiriou:2011dz,
    author = "Sotiriou, Thomas P. and Faraoni, Valerio",
    title = "{Black holes in scalar-tensor gravity}",
    eprint = "1109.6324",
    archivePrefix = "arXiv",
    primaryClass = "gr-qc",
    doi = "10.1103/PhysRevLett.108.081103",
    journal = "Phys. Rev. Lett.",
    volume = "108",
    pages = "081103",
    year = "2012"
}

@article{Joyce:2014kja,
    author = "Joyce, Austin and Jain, Bhuvnesh and Khoury, Justin and Trodden, Mark",
    title = "{Beyond the Cosmological Standard Model}",
    eprint = "1407.0059",
    archivePrefix = "arXiv",
    primaryClass = "astro-ph.CO",
    doi = "10.1016/j.physrep.2014.12.002",
    journal = "Phys. Rept.",
    volume = "568",
    pages = "1--98",
    year = "2015"
}

@article{Berti:2015itd,
    author = "Berti, Emanuele and others",
    title = "{Testing General Relativity with Present and Future Astrophysical Observations}",
    eprint = "1501.07274",
    archivePrefix = "arXiv",
    primaryClass = "gr-qc",
    doi = "10.1088/0264-9381/32/24/243001",
    journal = "Class. Quant. Grav.",
    volume = "32",
    pages = "243001",
    year = "2015"
}

@article{Heisenberg:2018vsk,
    author = "Heisenberg, Lavinia",
    title = "{A systematic approach to generalisations of General Relativity and their cosmological implications}",
    eprint = "1807.01725",
    archivePrefix = "arXiv",
    primaryClass = "gr-qc",
    doi = "10.1016/j.physrep.2018.11.006",
    journal = "Phys. Rept.",
    volume = "796",
    pages = "1--113",
    year = "2019"
}

@article{Barack:2018yly,
    author = "Barack, Leor and others",
    title = "{Black holes, gravitational waves and fundamental physics: a roadmap}",
    eprint = "1806.05195",
    archivePrefix = "arXiv",
    primaryClass = "gr-qc",
    doi = "10.1088/1361-6382/ab0587",
    journal = "Class. Quant. Grav.",
    volume = "36",
    number = "14",
    pages = "143001",
    year = "2019"
}

@article{LIGOScientific:2016aoc,
    author = "Abbott, B. P. and others",
    collaboration = "LIGO Scientific, Virgo",
    title = "{Observation of Gravitational Waves from a Binary Black Hole Merger}",
    eprint = "1602.03837",
    archivePrefix = "arXiv",
    primaryClass = "gr-qc",
    reportNumber = "LIGO-P150914",
    doi = "10.1103/PhysRevLett.116.061102",
    journal = "Phys. Rev. Lett.",
    volume = "116",
    number = "6",
    pages = "061102",
    year = "2016"
}

@article{EventHorizonTelescope:2019dse,
    author = "Akiyama, Kazunori and others",
    collaboration = "Event Horizon Telescope",
    title = "{First M87 Event Horizon Telescope Results. I. The Shadow of the Supermassive Black Hole}",
    eprint = "1906.11238",
    archivePrefix = "arXiv",
    primaryClass = "astro-ph.GA",
    doi = "10.3847/2041-8213/ab0ec7",
    journal = "Astrophys. J. Lett.",
    volume = "875",
    pages = "L1",
    year = "2019"
}

@article{EventHorizonTelescope:2022wkp,
    author = "Akiyama, Kazunori and others",
    collaboration = "Event Horizon Telescope",
    title = "{First Sagittarius A* Event Horizon Telescope Results. I. The Shadow of the Supermassive Black Hole in the Center of the Milky Way}",
    eprint = "2311.08680",
    archivePrefix = "arXiv",
    primaryClass = "astro-ph.HE",
    doi = "10.3847/2041-8213/ac6674",
    journal = "Astrophys. J. Lett.",
    volume = "930",
    number = "2",
    pages = "L12",
    year = "2022"
}

@article{Troja:2017nqp,
    author = "Troja, E. and others",
    title = "{The X-ray counterpart to the gravitational wave event GW 170817}",
    eprint = "1710.05433",
    archivePrefix = "arXiv",
    primaryClass = "astro-ph.HE",
    doi = "10.1038/nature24290",
    journal = "Nature",
    volume = "551",
    pages = "71--74",
    year = "2017"
}

@article{LIGOScientific:2017bnn,
    author = "Abbott, Benjamin P. and others",
    collaboration = "LIGO Scientific, VIRGO",
    title = "{GW170104: Observation of a 50-Solar-Mass Binary Black Hole Coalescence at Redshift 0.2}",
    eprint = "1706.01812",
    archivePrefix = "arXiv",
    primaryClass = "gr-qc",
    reportNumber = "LIGO-P170104",
    doi = "10.1103/PhysRevLett.118.221101",
    journal = "Phys. Rev. Lett.",
    volume = "118",
    number = "22",
    pages = "221101",
    year = "2017",
    note = "[Erratum: Phys.Rev.Lett. 121, 129901 (2018)]"
}

@article{LIGOScientific:2017ycc,
    author = "Abbott, B. P. and others",
    collaboration = "LIGO Scientific, Virgo",
    title = "{GW170814: A Three-Detector Observation of Gravitational Waves from a Binary Black Hole Coalescence}",
    eprint = "1709.09660",
    archivePrefix = "arXiv",
    primaryClass = "gr-qc",
    doi = "10.1103/PhysRevLett.119.141101",
    journal = "Phys. Rev. Lett.",
    volume = "119",
    number = "14",
    pages = "141101",
    year = "2017"
}

@article{LIGOScientific:2017vwq,
    author = "Abbott, B. P. and others",
    collaboration = "LIGO Scientific, Virgo",
    title = "{GW170817: Observation of Gravitational Waves from a Binary Neutron Star Inspiral}",
    eprint = "1710.05832",
    archivePrefix = "arXiv",
    primaryClass = "gr-qc",
    reportNumber = "LIGO-P170817",
    doi = "10.1103/PhysRevLett.119.161101",
    journal = "Phys. Rev. Lett.",
    volume = "119",
    number = "16",
    pages = "161101",
    year = "2017"
}

@article{Carter:1971zc,
    author = "Carter, B.",
    title = "{Axisymmetric Black Hole Has Only Two Degrees of Freedom}",
    doi = "10.1103/PhysRevLett.26.331",
    journal = "Phys. Rev. Lett.",
    volume = "26",
    pages = "331--333",
    year = "1971"
}

@article{Ruffini:1971bza,
    author = "Ruffini, Remo and Wheeler, John A.",
    title = "{Introducing the black hole}",
    doi = "10.1063/1.3022513",
    journal = "Phys. Today",
    volume = "24",
    number = "1",
    pages = "30",
    year = "1971"
}

@article{Bekenstein:1995un,
    author = "Bekenstein, J. D.",
    title = "{Novel {\textquoteleft}{\textquoteleft}no-scalar-hair{\textquoteright}{\textquoteright} theorem for black holes}",
    doi = "10.1103/PhysRevD.51.R6608",
    journal = "Phys. Rev. D",
    volume = "51",
    number = "12",
    pages = "R6608",
    year = "1995"
}

@article{Brans:1961sx,
    author = "Brans, C. and Dicke, R. H.",
    editor = "Hsu, Jong-Ping and Fine, D.",
    title = "{Mach's principle and a relativistic theory of gravitation}",
    doi = "10.1103/PhysRev.124.925",
    journal = "Phys. Rev.",
    volume = "124",
    pages = "925--935",
    year = "1961"
}

@article{Horndeski:1974wa,
    author = "Horndeski, Gregory Walter",
    title = "{Second-order scalar-tensor field equations in a four-dimensional space}",
    doi = "10.1007/BF01807638",
    journal = "Int. J. Theor. Phys.",
    volume = "10",
    pages = "363--384",
    year = "1974"
}

@article{Kobayashi:2011nu,
    author = "Kobayashi, Tsutomu and Yamaguchi, Masahide and Yokoyama, Jun'ichi",
    title = "{Generalized G-inflation: Inflation with the most general second-order field equations}",
    eprint = "1105.5723",
    archivePrefix = "arXiv",
    primaryClass = "hep-th",
    reportNumber = "KUNS-2339, RESCEU-9-11",
    doi = "10.1143/PTP.126.511",
    journal = "Prog. Theor. Phys.",
    volume = "126",
    pages = "511--529",
    year = "2011"
}

@article{Deffayet:2013lga,
    author = "Deffayet, C{\'e}dric and Steer, Dani{\`e}le A.",
    title = "{A formal introduction to Horndeski and Galileon theories and their generalizations}",
    eprint = "1307.2450",
    archivePrefix = "arXiv",
    primaryClass = "hep-th",
    doi = "10.1088/0264-9381/30/21/214006",
    journal = "Class. Quant. Grav.",
    volume = "30",
    pages = "214006",
    year = "2013"
}

@article{Sotiriou:2014pfa,
    author = "Sotiriou, Thomas P. and Zhou, Shuang-Yong",
    title = "{Black hole hair in generalized scalar-tensor gravity: An explicit example}",
    eprint = "1408.1698",
    archivePrefix = "arXiv",
    primaryClass = "gr-qc",
    doi = "10.1103/PhysRevD.90.124063",
    journal = "Phys. Rev. D",
    volume = "90",
    pages = "124063",
    year = "2014"
}

@article{Babichev:2017guv,
    author = "Babichev, Eugeny and Charmousis, Christos and Leh{\'e}bel, Antoine",
    title = "{Asymptotically flat black holes in Horndeski theory and beyond}",
    eprint = "1702.01938",
    archivePrefix = "arXiv",
    primaryClass = "gr-qc",
    reportNumber = "LPT-ORSAY-17-03",
    doi = "10.1088/1475-7516/2017/04/027",
    journal = "JCAP",
    volume = "04",
    pages = "027",
    year = "2017"
}

@article{Damour:1993hw,
    author = "Damour, Thibault and Esposito-Farese, Gilles",
    title = "{Nonperturbative strong field effects in tensor - scalar theories of gravitation}",
    reportNumber = "IHES-P-93-1, CPT-93-PE-2868",
    doi = "10.1103/PhysRevLett.70.2220",
    journal = "Phys. Rev. Lett.",
    volume = "70",
    pages = "2220--2223",
    year = "1993"
}

@article{Doneva:2017bvd,
    author = "Doneva, Daniela D. and Yazadjiev, Stoytcho S.",
    title = "{New Gauss-Bonnet Black Holes with Curvature-Induced Scalarization in Extended Scalar-Tensor Theories}",
    eprint = "1711.01187",
    archivePrefix = "arXiv",
    primaryClass = "gr-qc",
    doi = "10.1103/PhysRevLett.120.131103",
    journal = "Phys. Rev. Lett.",
    volume = "120",
    number = "13",
    pages = "131103",
    year = "2018"
}

@article{Silva:2017uqg,
    author = "Silva, Hector O. and Sakstein, Jeremy and Gualtieri, Leonardo and Sotiriou, Thomas P. and Berti, Emanuele",
    title = "{Spontaneous scalarization of black holes and compact stars from a Gauss-Bonnet coupling}",
    eprint = "1711.02080",
    archivePrefix = "arXiv",
    primaryClass = "gr-qc",
    doi = "10.1103/PhysRevLett.120.131104",
    journal = "Phys. Rev. Lett.",
    volume = "120",
    number = "13",
    pages = "131104",
    year = "2018"
}

@article{Antoniou:2017acq,
    author = "Antoniou, G. and Bakopoulos, A. and Kanti, P.",
    title = "{Evasion of No-Hair Theorems and Novel Black-Hole Solutions in Gauss-Bonnet Theories}",
    eprint = "1711.03390",
    archivePrefix = "arXiv",
    primaryClass = "hep-th",
    doi = "10.1103/PhysRevLett.120.131102",
    journal = "Phys. Rev. Lett.",
    volume = "120",
    number = "13",
    pages = "131102",
    year = "2018"
}

@article{Herdeiro:2021vjo,
    author = "Herdeiro, Carlos A. R. and Pombo, Alexandre M. and Radu, Eugen",
    title = "{Aspects of Gauss-Bonnet Scalarisation of Charged Black Holes}",
    eprint = "2111.06442",
    archivePrefix = "arXiv",
    primaryClass = "gr-qc",
    doi = "10.3390/universe7120483",
    journal = "Universe",
    volume = "7",
    number = "12",
    pages = "483",
    year = "2021"
}

@article{Bakopoulos:2018nui,
    author = "Bakopoulos, Athanasios and Antoniou, Georgios and Kanti, Panagiota",
    title = "{Novel Black-Hole Solutions in Einstein-Scalar-Gauss-Bonnet Theories with a Cosmological Constant}",
    eprint = "1812.06941",
    archivePrefix = "arXiv",
    primaryClass = "hep-th",
    doi = "10.1103/PhysRevD.99.064003",
    journal = "Phys. Rev. D",
    volume = "99",
    number = "6",
    pages = "064003",
    year = "2019"
}

@article{Guo:2020sdu,
    author = "Guo, Hong and Kiorpelidi, Stella and Kuang, Xiao-Mei and Papantonopoulos, Eleftherios and Wang, Bin and Wu, Jian-Pin",
    title = "{Spontaneous holographic scalarization of black holes in Einstein-scalar-Gauss-Bonnet theories}",
    eprint = "2006.10659",
    archivePrefix = "arXiv",
    primaryClass = "hep-th",
    doi = "10.1103/PhysRevD.102.084029",
    journal = "Phys. Rev. D",
    volume = "102",
    number = "8",
    pages = "084029",
    year = "2020"
}

@article{Brihaye:2019gla,
    author = "Brihaye, Yves and Herdeiro, Carlos and Radu, Eugen",
    title = "{Black Hole Spontaneous Scalarisation with a Positive Cosmological Constant}",
    eprint = "1910.05286",
    archivePrefix = "arXiv",
    primaryClass = "gr-qc",
    doi = "10.1016/j.physletb.2020.135269",
    journal = "Phys. Lett. B",
    volume = "802",
    pages = "135269",
    year = "2020"
}

@article{Brihaye:2019dck,
    author = "Brihaye, Yves and Hartmann, Betti and Aprile, Nath{\'a}lia Pio and Urrestilla, Jon",
    title = "{Scalarization of asymptotically anti{\textendash}de Sitter black holes with applications to holographic phase transitions}",
    eprint = "1911.01950",
    archivePrefix = "arXiv",
    primaryClass = "gr-qc",
    doi = "10.1103/PhysRevD.101.124016",
    journal = "Phys. Rev. D",
    volume = "101",
    number = "12",
    pages = "124016",
    year = "2020"
}

@article{Myung:2018vug,
    author = "Myung, Yun Soo and Zou, De-Cheng",
    title = {{Instability of Reissner{\textendash}Nordstr{\"o}m black hole in Einstein-Maxwell-scalar theory}},
    eprint = "1808.02609",
    archivePrefix = "arXiv",
    primaryClass = "gr-qc",
    doi = "10.1140/epjc/s10052-019-6792-6",
    journal = "Eur. Phys. J. C",
    volume = "79",
    number = "3",
    pages = "273",
    year = "2019"
}

@article{Doneva:2022ewd,
    author = "Doneva, Daniela D. and Ramazano{\u{g}}lu, Fethi M. and Silva, Hector O. and Sotiriou, Thomas P. and Yazadjiev, Stoytcho S.",
    title = "{Spontaneous scalarization}",
    eprint = "2211.01766",
    archivePrefix = "arXiv",
    primaryClass = "gr-qc",
    doi = "10.1103/RevModPhys.96.015004",
    journal = "Rev. Mod. Phys.",
    volume = "96",
    number = "1",
    pages = "015004",
    year = "2024"
}

@article{Dima:2020yac,
    author = "Dima, Alexandru and Barausse, Enrico and Franchini, Nicola and Sotiriou, Thomas P.",
    title = "{Spin-induced black hole spontaneous scalarization}",
    eprint = "2006.03095",
    archivePrefix = "arXiv",
    primaryClass = "gr-qc",
    doi = "10.1103/PhysRevLett.125.231101",
    journal = "Phys. Rev. Lett.",
    volume = "125",
    number = "23",
    pages = "231101",
    year = "2020"
}

@article{Herdeiro:2020wei,
    author = "Herdeiro, Carlos A. R. and Radu, Eugen and Silva, Hector O. and Sotiriou, Thomas P. and Yunes, Nicol{\'a}s",
    title = "{Spin-induced scalarized black holes}",
    eprint = "2009.03904",
    archivePrefix = "arXiv",
    primaryClass = "gr-qc",
    doi = "10.1103/PhysRevLett.126.011103",
    journal = "Phys. Rev. Lett.",
    volume = "126",
    number = "1",
    pages = "011103",
    year = "2021"
}

@article{Doneva:2021tvn,
    author = "Doneva, Daniela D. and Yazadjiev, Stoytcho S.",
    title = "{Beyond the spontaneous scalarization: New fully nonlinear mechanism for the formation of scalarized black holes and its dynamical development}",
    eprint = "2107.01738",
    archivePrefix = "arXiv",
    primaryClass = "gr-qc",
    doi = "10.1103/PhysRevD.105.L041502",
    journal = "Phys. Rev. D",
    volume = "105",
    number = "4",
    pages = "L041502",
    year = "2022"
}

@article{Pombo:2023lxg,
    author = "Pombo, Alexandre M. and Doneva, Daniela D.",
    title = "{Effects of mass and self-interaction on nonlinear scalarization of scalar-Gauss-Bonnet black holes}",
    eprint = "2310.08638",
    archivePrefix = "arXiv",
    primaryClass = "gr-qc",
    doi = "10.1103/PhysRevD.108.124068",
    journal = "Phys. Rev. D",
    volume = "108",
    number = "12",
    pages = "124068",
    year = "2023"
}

@article{Belkhadria:2023ooc,
    author = "Belkhadria, Zakaria and Pombo, Alexandre M.",
    title = "{Mixed scalarization of charged black holes: From spontaneous to nonlinear scalarization}",
    eprint = "2311.15850",
    archivePrefix = "arXiv",
    primaryClass = "gr-qc",
    doi = "10.1103/PhysRevD.110.044014",
    journal = "Phys. Rev. D",
    volume = "110",
    number = "4",
    pages = "044014",
    year = "2024"
}

@article{Herdeiro:2018wub,
    author = "Herdeiro, Carlos A. R. and Radu, Eugen and Sanchis-Gual, Nicolas and Font, Jos{\'e} A.",
    title = "{Spontaneous Scalarization of Charged Black Holes}",
    eprint = "1806.05190",
    archivePrefix = "arXiv",
    primaryClass = "gr-qc",
    doi = "10.1103/PhysRevLett.121.101102",
    journal = "Phys. Rev. Lett.",
    volume = "121",
    number = "10",
    pages = "101102",
    year = "2018"
}

@article{Zhang:2021nnn,
    author = "Zhang, Cheng-Yong and Chen, Qian and Liu, Yunqi and Luo, Wen-Kun and Tian, Yu and Wang, Bin",
    title = "{Critical Phenomena in Dynamical Scalarization of Charged Black Holes}",
    eprint = "2112.07455",
    archivePrefix = "arXiv",
    primaryClass = "gr-qc",
    doi = "10.1103/PhysRevLett.128.161105",
    journal = "Phys. Rev. Lett.",
    volume = "128",
    number = "16",
    pages = "161105",
    year = "2022"
}

@article{Liu:2022fxy,
    author = "Liu, Yunqi and Zhang, Cheng-Yong and Chen, Qian and Cao, Zhoujian and Tian, Yu and Wang, Bin",
    title = "{Critical scalarization and descalarization of black holes in a generalized scalar-tensor theory}",
    eprint = "2208.07548",
    archivePrefix = "arXiv",
    primaryClass = "gr-qc",
    doi = "10.1007/s11433-023-2160-1",
    journal = "Sci. China Phys. Mech. Astron.",
    volume = "66",
    number = "10",
    pages = "100412",
    year = "2023"
}

@article{Herdeiro:2019yjy,
    author = "Herdeiro, Carlos A. R. and Radu, Eugen",
    title = "{Black hole scalarization from the breakdown of scale invariance}",
    eprint = "1901.02953",
    archivePrefix = "arXiv",
    primaryClass = "gr-qc",
    doi = "10.1103/PhysRevD.99.084039",
    journal = "Phys. Rev. D",
    volume = "99",
    number = "8",
    pages = "084039",
    year = "2019"
}

@article{Zhang:2024bfu,
    author = "Zhang, Lina and Pan, Qiyuan and Myung, Yun Soo and Zou, De-Cheng",
    title = "{Spontaneous scalarization of Bardeen black holes}",
    eprint = "2409.11669",
    archivePrefix = "arXiv",
    primaryClass = "gr-qc",
    doi = "10.1103/PhysRevD.110.124036",
    journal = "Phys. Rev. D",
    volume = "110",
    number = "12",
    pages = "124036",
    year = "2024"
}

@article{Zhang:2025msi,
    author = "Zhang, Lina and Zou, De-Cheng and Myung, Yun Soo",
    title = "{New scalarization of the Einstein{\textendash}Euler{\textendash}Heisenberg black hole}",
    eprint = "2510.07954",
    archivePrefix = "arXiv",
    primaryClass = "gr-qc",
    doi = "10.1140/epjc/s10052-025-15232-4",
    journal = "Eur. Phys. J. C",
    volume = "85",
    number = "12",
    pages = "1463",
    year = "2025"
}

@article{Belkhadria:2025lev,
    author = "Belkhadria, Zakaria and Mignemi, Salvatore",
    title = "{New model of spontaneous scalarization of black holes induced by curvature and matter}",
    eprint = "2506.12137",
    archivePrefix = "arXiv",
    primaryClass = "gr-qc",
    doi = "10.1103/ln23-nhpn",
    journal = "Phys. Rev. D",
    volume = "112",
    number = "4",
    pages = "044015",
    year = "2025"
}

@article{Breitenlohner:1982jf,
    author = "Breitenlohner, Peter and Freedman, Daniel Z.",
    title = "{Stability in Gauged Extended Supergravity}",
    reportNumber = "Print-82-0500 (MIT)",
    doi = "10.1016/0003-4916(82)90116-6",
    journal = "Annals Phys.",
    volume = "144",
    pages = "249",
    year = "1982"
}

@article{Zou:2022hjj,
    author = "Zou, De-Cheng and Meng, Bo and Zhang, Ming and Li, Sheng-Yuan and Lai, Meng-Yun and Myung, Yun Soo",
    title = "{Analytical Approximate Solutions for Scalarized AdS Black Holes}",
    eprint = "2301.04784",
    archivePrefix = "arXiv",
    primaryClass = "gr-qc",
    doi = "10.3390/universe9010026",
    journal = "Universe",
    volume = "9",
    number = "1",
    pages = "26",
    year = "2023"
}

@article{Hod:2020jjy,
    author = "Hod, Shahar",
    title = "{Onset of spontaneous scalarization in spinning Gauss-Bonnet black holes}",
    eprint = "2006.09399",
    archivePrefix = "arXiv",
    primaryClass = "gr-qc",
    doi = "10.1103/PhysRevD.102.084060",
    journal = "Phys. Rev. D",
    volume = "102",
    number = "8",
    pages = "084060",
    year = "2020"
}

@article{Guo:2025flg,
    author = "Guo, Hong and Liu, Hang and Myung, Yun Soo",
    title = "{Scalar-hairy AdS black hole in the Einstein{\textendash}Maxwell-scalar theory: first-order phase transition with a critical point}",
    eprint = "2512.22433",
    archivePrefix = "arXiv",
    primaryClass = "gr-qc",
    reportNumber = "CTPU-PTC-25-42",
    doi = "10.1140/epjc/s10052-026-15407-7",
    journal = "Eur. Phys. J. C",
    volume = "86",
    number = "2",
    pages = "204",
    year = "2026"
}

@article{Guo:2021zed,
    author = "Guo, Guangzhou and Wang, Peng and Wu, Houwen and Yang, Haitang",
    title = "{Scalarized Einstein{\textendash}Maxwell-scalar black holes in anti-de Sitter spacetime}",
    eprint = "2102.04015",
    archivePrefix = "arXiv",
    primaryClass = "gr-qc",
    reportNumber = "CTP-SCU/2021003",
    doi = "10.1140/epjc/s10052-021-09614-7",
    journal = "Eur. Phys. J. C",
    volume = "81",
    number = "10",
    pages = "864",
    year = "2021"
}

\end{document}